\documentclass[11pt]{article}
\usepackage{jheppub}

\usepackage{epsfig,multicol}
\usepackage{mathrsfs}
\usepackage{amsmath,amssymb,amsbsy,mathtools}
\usepackage{graphicx,picinpar,epsfig}
\usepackage{tikz}\usetikzlibrary{decorations.markings, arrows.meta}
\usepackage{calc}

\newcommand\fverb{\setbox\fverbbox=\hbox\bgroup\verb}
\newcommand\fverbdo{\egroup\medskip\noindent%
			\fbox{\unhbox\fverbbox}\ }
\newcommand\fverbit{\egroup\item[\fbox{\unhbox\fverbbox}]}
\newbox\fverbbox

\newcommand{\pslash}{p\kern-1ex /}
\newcommand{\qslash}{q\kern-1ex /}
\newcommand{\lslash}{l\kern-1ex /}
\newcommand{\sslash}{s\kern-1ex /}
\newcommand{\kaslash}{k_a\kern-2ex /}
\newcommand{\kbslash}{k_b\kern-2ex /}
\newcommand{\Dslash}{\mathcal{D}\kern-1.5ex /}

\newcommand{\beqa}{\begin{eqnarray}}
\newcommand{\eeqa}{\end{eqnarray}}

\newcommand{\ba}{\begin{eqnarray}}
\newcommand{\ea}{\end{eqnarray}}
\newcommand{\be}{\begin{equation}}

\numberwithin{equation}{section}

\title{Running coupling and non-perturbative corrections for O$(N)$ free energy
and for disk capacitor}

\author{
Zolt\'an Bajnok,\footnote{\tt bajnok.zoltan@wigner.hu}\ \  
J\'anos Balog,\footnote{\tt balog.janos@wigner.hu}\ \ 
\'Arp\'ad Heged\H us\footnote{\tt hegedus.arpad@wigner.hu}\ \ and \
Istv\'an Vona\footnote{\tt vona.istvan@wigner.hu}
\\
\vskip 1ex
{\it Holographic QFT Group, Institute for Particle and Nuclear Physics,\\
Wigner Research Centre for Physics} \\
{\it H-1525 Budapest 114, P.O.B. 49, Hungary}\\
}

\abstract{
We reconsider the complete solution of the linear TBA equation describing the energy density of finite density states in the $O(N)$ nonlinear sigma models by the Wiener-Hopf method. We keep all perturbative and non-perturbative contributions and introduce a running coupling in terms of which all asymptotic series appearing in the problem can be represented as pure power series without logs. We work out the first non-perturbative contribution in the $O(3)$ case and show that (presumably because of the instanton corrections) resurgence theory fails in this example. Using the relation of the $O(3)$ problem to the coaxial disks capacitor problem we work out the leading
non-perturbative terms for the latter and show that (at least to this order) resurgence theory, in particular the median resummation prescription, gives the correct answer. We demonstrate this by comparing the Wiener-Hopf results to the high precision numerical solution of the original integral equation.
}

\begin{document}


\newcommand{\con}{\,\star\hspace{-3.7mm}\bigcirc\,}
\newcommand{\convu}{\,\star\hspace{-3.1mm}\bigcirc\,}
\newcommand{\Eps}{\Epsilon}
\newcommand{\gM}{\mathcal{M}}
\newcommand{\dD}{\mathcal{D}}
\newcommand{\gG}{\mathcal{G}}
\newcommand{\pa}{\partial}
\newcommand{\eps}{\epsilon}
\newcommand{\La}{\Lambda}
\newcommand{\De}{\Delta}
\newcommand{\nonu}{\nonumber}
\newcommand{\beq}{\begin{eqnarray}}
\newcommand{\eeq}{\end{eqnarray}}
\newcommand{\ka}{\kappa}
\newcommand{\ee}{\end{equation}}
\newcommand{\an}{\ensuremath{\alpha_0}}
\newcommand{\bn}{\ensuremath{\beta_0}}
\newcommand{\dn}{\ensuremath{\delta_0}}
\newcommand{\al}{\alpha}
\newcommand{\bm}{\begin{multline}}
\newcommand{\fm}{\end{multline}}
\newcommand{\de}{\delta}
\newcommand{\dpd}{\int {\rm d}^d p}
\newcommand{\dqd}{\int {\rm d}^d q}
\newcommand{\dxd}{\int {\rm d}^d x}
\newcommand{\dyd}{\int {\rm d}^d y}
\newcommand{\dud}{\int {\rm d}^d u}
\newcommand{\dzd}{\int {\rm d}^d z}
\newcommand{\dpp}{\int \frac{{\rm d}^d p}{p^2}}
\newcommand{\dqq}{\int \frac{{\rm d}^d q}{q^2}}


\maketitle

\section{Introduction}

Quantum chromodynamics (QCD), the theory of the strong interaction
is asymptotically free in perturbation theory and there is a dynamically
generated mass scale. The nature of the perturbative expansion is
asymptotic, which manifests itself in the factorially growing coefficients.
This factorial growth can be tamed by switching to the Borel plane,
i.e. by analysing the perturbative coefficients divided by the appropriate
factorials. The perturbative expansion of the various observables
on the Borel plane then has a finite radius of convergence and exhibits
singularities typically on the real line. Singularities on the positive
real line prevent Borel summability, leading to non-perturbative ambiguities.
These ambiguities can originate from instantons and/or renormalons
\cite{Beneke:1998ui,Bauer:2011ws}. The full knowledge of the singularities,
in principle, enables one to formulate an ambiguity free trans-series
for each observable. It would be ideal to proceed along these lines
for QCD, however, only a very limited number of perturbative terms
are at our disposal. Thus, exactly soluble models with properties similar 
to QCD have been becoming more and more important. 

There are various toy models in two dimensions that share many physically
relevant aspects with QCD, but are nevertheless exactly soluble. They
provide ideal testing grounds of non-perturbative physics. Particularly
interesting are the asymptotically free theories with a dynamically
generated mass gap. By coupling an external field to one of the conserved
charges the free energy can be calculated in two alternative ways:
from perturbation theory based on the UV Lagrangean and also from
the thermodynamic limit of the Bethe ansatz, which uses the IR degrees
of freedom, the masses and scatterings of particles. 

In the UV description the singular point of the renormalized running
coupling defines a non-perturbative scale, $\Lambda$, which appears
in the perturbative expansion. On the IR side the expansion is based
on the Wiener-Hopf technique and includes the physical mass of the
particles. By matching the two expansions the highly non-trivial $m/\Lambda$
mass gap relation can be established. This calculation was first performed
in the $O(N)$ models \cite{Hasenfratz:1990ab,Hasenfratz:1990zz}
and later it was extended for various other two dimensional integrable
models: for the Gross-Neveu model \cite{Forgacs:1991rs,Forgacs:1991ru},
for the chiral $SU(N)\times SU(N)$ model \cite{Balog:1992cm}, for
their supersymmetric extensions \cite{Evans:1994sv,Evans:1994sy}
and also for the sine-Gordon model \cite{Zamolodchikov:1995xk}. This
calculation involves only the first few perturbative coefficients.
It is notoriously difficult to expand the TBA equations to higher orders
and the next order result, decisive for the AdS/CFT correspondence,
was obtained in the $O(N)$ models by direct perturbative calculations
\cite{Bajnok:2008it}. 

A breakthrough in this field was achieved by Volin \cite{Volin:2009wr,Volin:2010cq},
who managed in the $O(N)$ models to transform the expansion of TBA
into algebraic relations and to calculate the first
26 perturbative coefficients.
These were enough to see the factorial growth and to locate the leading
singularities on the Borel plane. His approach was extended to other
relativistic \cite{Marino:2019eym} and non-relativistic models \cite{Marino:2019fuy,Marino:2020dgc,Marino:2020ggm,Reichert:2020ymc}.
By focusing on the $O(3)$ and $O(4)$ models and performing the calculations
numerically, one can even obtain $336/2000$ perturbative terms, respectively
\cite{Abbott:2020mba,Abbott:2020qnl,Bajnok:2021zjm}. Alternatively,
one can make a systematic large $N$ expansion of the TBA equations,
which can be matched to large $N$ renormalon diagrams \cite{DiPietro:2021yxb,Marino:2021six}. 

Having enough perturbative coefficients fixes the leading factorial
and sub-factorial growths, which determine the nature and locations
of the closest Borel singularities. These singularities give rise
to non-perturbative corrections. The theory which connects the asymptotic
behaviour of the perturbative series to the non-perturbative corrections
is called resurgence, see \cite{Aniceto:2018bis,Dorigoni:2014hea}
for modern reviews and references therein. In its strongest version
it implies that the perturbative coefficients determine all the non-perturbative
corrections. By taking into account all the non-perturbative corrections
the physical observables can be written into trans-series forms. A
trans-series is a series containing all exponentially suppressed non-perturbative
corrections, where each term is multiplied with an (asymptotic) perturbative
series. It is understood to be Borel resummed laterally and the various
non-perturbative terms ensure that it is ambiguity free. The first
few terms of the trans-series were determined and matched to the numerical
solution of the TBA equation for the $O(4)$ model in \cite{Abbott:2020mba,Abbott:2020qnl},
while the leading order results were analytically proven in \cite{Bajnok:2021dri}.
Particularly important are the bridge equations which could relate
the perturbative expansions of the various non-perturbative terms
to each other. A big leap in this direction was the systematic expansion
of the TBA equation, which provided pertubative expansion of the
various non-perturbative contributions in parallel \cite{Marino:2021dzn}.
The aim of our paper is to extend and to elaborate this approach for
the $O(N)$ models in many respects. 

One of our main results is the introduction of the running coupling
on the TBA side in the $O(N)$ models, similarly to how it appeared for
the Gross-Neveu model in \cite{Forgacs:1991rs,Forgacs:1991ru}. In
the TBA formulation, the external field, $h$, forces the negatively
charged particles to condense into the $h$-dependent rapidity interval
$[-B,B]$. The rapidity density can be determined from the TBA integral
equation based on the scattering matrix, and provides the density
$\rho$ and energy density $\epsilon$. Large $h$ corresponds to
large $B$ and the systematic expansion goes in $B^{-1}$ multiplied
with a polynomial in $\log B$. Since $\epsilon/\rho^{2}$ is related
to the free energy, it can be expanded purely in powers of the Lagrangean's
running coupling, $\alpha$. This is, however, not expected either
from $\rho$ or from $\epsilon$ and one of our main result is the
introduction such a running coupling, $v$, in which they can
be expanded as a power series, i.e. without $\log v$. The main advantage
of this running coupling becomes obvious when we analyse models, which
are mathematically equivalent, but physically different from the $O(N)$
sigma models. In particular, for $N=3$ the integral equation is the
same which determines the groundstate energy in the Lieb-Liniger model
\cite{lieb1963exact}. It is also equivalent to the Love equation,
which describes the circular plate capacitor \cite{Love:1949}. 

The Maxwell-Kirchhoff disk capacitor problem is to calculate the capacitance
of two, oppositely charged, infinitely thin coaxial conducting disks
of radius $a$ and distance $d$. The leading order behaviour for
small $\kappa=d/a$ can be calculated easily, but subleading corrections
are very involved and challenged many prominent physicists. The expansion
involves powers and logarithms of $\kappa$ and is very difficult
to obtain systematically. The exact analytical description was provided
by Love and the kernel of his integral equation is the same which
appears in the O(3) sigma model. Adaptation of Volin's method allowed
to generate many perturbative terms \cite{Marino:2019fuy,Reichert:2020ymc},
but non-perturbative corrections have not been analyzed yet although the
first non-perturbative correction was given in \cite{Marino:2019fuy} for the
closely related Lieb-Liniger model. Our aim is to connect the capacitance
to the observables of the O(3) sigma model and analyze its non-perturbative
corrections. 

The O(3) model is interesting in its own right as the leading exponentially
suppressed (in perturbation theory) terms are not related to the asymptotics
of the perturbative coefficients \cite{Marino:2021dzn,Bajnok:2021zjm}.
We revisit also this model and, by exploiting our simplified equations,
we investigate this anomalous behaviour in detail. 

The paper is organized as follows: in the rest of this section we
first recall the interesting history of the capacitor problem. We
then introduce the form of the integral equations we are dealing with
together with various observables of relevance and establish a relationship
between them. In particular, the capacitance of the circular plate
capacitor will be related to the density and energy density of the
$O(3)$ sigma model. In section 2 we recall the Wiener-Hopf method
developed to solve the integral equation and present also a useful
simplification, which leads to the introduction of the running coupling.
By separating the singularities of the integral kernel we present
a form that allows systematic calculations of the various non-perturbative
corrections. We then comment on the structure of these non-perturbative
corrections and the resurgence properties of the perturbative expansion.
As the $O(3)$ kernel is drastically different from the $N\geq4$
cases we analyse it more carefully. Section 3 is devoted to the calculation
of the leading exponential corrections in the $O(3)$ sigma model,
while Section 4 focuses on the non-perturbative analysis of the circular
plate capacitor. Our conclusions are summarized in section 5. The
paper is closed by several Appendices. In Appendix A we introduce
our notations and the building blocks for the Wiener-Hopf analysis.
Appendix B provides details on the integral equation when the source
is a constant function. In Appendix C we calculate explicitly the
first two orders based on the relationship between the various observable
we found earlier and by introducing a new analytical tool. Here we
elaborate on how the Laplace transformation can be used to solve explicitly
the expansion of the integral equation. Finally, in Appendix D, we
explain how Volin's algorithm can be used to calculate the perturbative
expansion in our running coupling. 

The contribution of one of the authors of Ref. \cite{Balog:1992cm}, Ferenc
Niedermayer, was essential in obtaining the analytic solutions presented in
appendix C. We devote this paper to the memory of Ferenc.

\subsection{The Maxwell-Kirchhoff disk capacitor problem}


\begin{figure}
\leavevmode
\centerline{\includegraphics[scale=2.0]{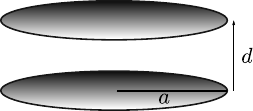} }
\caption{{\footnotesize
Geometry of the parallel disks capacitor.}}
\label{GEO}
\end{figure}


The problem is to calculate the capacity of two, infinitely thin coaxial
conducting disks of radius $a$ and distance $d$ (see Fig. \ref{GEO}). If
the disks are kept at fixed potential $\pm U$ (equal and opposite) the task is
to solve this boundary value problem of electrostatics, determine the electric
field around the two disks and the charge distribution on the disks and finally
calculate the capacity $C_{\rm phys}$ of the arrangement.

The solution is not available in closed form but there are results in the limit
of small separation of the disks. If
\begin{equation}
\kappa=\frac{d}{a}
\end{equation}  
is small, the problem reduces to the high-school problem of infinite parallel
conducting plates, the electric field becomes constant between the plates and
the capacity per unit surface area becomes
\begin{equation}
\frac{C_{\rm phys}}{a^2\pi}=\frac{\epsilon_0}{d},  
\end{equation}  
where $\epsilon_0$ is the vacuum permittivity.

If we want to go beyond this approximation, the effect of the edges of the disks
has to be taken into account. This legendary problem of electrostatics
attracted the attention of many prominent physicists in the last one and a half
century. The saga started by Maxwell \cite{maxwell} who was  the first
to study the edge effects.

Introducing the dimensionless quantity
\begin{equation}
{\cal C}=\frac{1}{4\pi\epsilon_0 a}C_{\rm phys}
\end{equation}  
its $\kappa\to0$ asymptotic expansion can be written as  
\begin{equation}
{\cal C}=\frac{1}{4\kappa}+\frac{\mathsf{L}-1}{4\pi}
+\frac{\kappa}{16\pi^2}({\mathsf{L}}^2-2)+
O(\kappa^2),  
\end{equation}  
where $\mathsf{L}=\ln(16\pi/\kappa)$. Here the first term
is the high-school result
and the next (NLO) term was found by Kirchhoff \cite{kirchhoff1877theorie}
by a heuristic derivation
(based on earlier result of Clausius and Helmholtz). Much later
Ignatowsky \cite{ignatowsky1931kreisscheibenkondensator} claimed the
constant in the NLO term should be replaced as
\begin{equation}
\ln\frac{16\pi}{e}\rightarrow\ln\frac{8}{\sqrt{e}}.
\end{equation}  
This turned out to be incorrect, however, curiously,
P\'olya and Szeg\H o \cite{polya1945inequalities}
proved that the sum of the first two terms with this modified constant gives
an exact lower bound\footnote{This exact lower bound was used in the numerical
studies of \cite{norgren2009capacitance}.} for the capacity!

Love \cite{Love:1949} derived the integral equation (\ref{Loveeq})
describing the problem and
Hutson \cite{hutson1963circular} proved rigorously from the Love equation
that Kirchhoff's NLO
result was correct. Actually, Love's equation already appeared in a much earlier
paper \cite{hafen1910studien}. Sneddon \cite{sneddon1966mixed}
simplified the derivation of Love's equation and
proved the existence and uniqueness of its solution. (For a recent elementary
derivation, see \cite{Felderhof:2013}.) 

Nearly a century after Kirchhoff's result the NNLO $\ln^2$ term was
calculated \cite{Leppington:1970}. The full NNLO calculation of
\cite{shaw1970circular} was
improved and corrected by \cite{wigglesworth1972comments,chew1982microstrip}.

The Lieb-Liniger model describes the system of one-dimensional bosons
interacting with a repulsive $\delta$-potential \cite{lieb1963exact}.
This non-relativistic
system is described by Lieb's integral equation.
Gaudin \cite{gaudin1971boundary} pointed out
first that Love's and Lieb's integral equations are identical and used potential
theory to calculate the NLO term in the free energy of the Lieb-Liniger model.
For a comprehensive review of the mathematics and physical applications of the
Love-Lieb integral equation, see \cite{Farina:2020zlr}.

It was noted in \cite{Hasenfratz:1990zz} that the TBA integral equation
describing the free energy of the $O(3)$ non-linear sigma model in an external
field is also closely related to Love's equation. Thus these three completely
distinct physical systems are all described by the same mathematics.

Systematic asymptotic expansion of the integral equations (following Maxwell's
original ideas) was based (using the disk language) on the method of
\lq\lq matched'' asymptotic expansions of the potential near and away from the
edges. The first few terms in the asymptotic expansion were calculated in
\cite{lieb1963exact,popov1977theory,hutson1963circular}.
This method was elegantly extended to all orders, for the
$O(N)$ free energy, by Volin \cite{Volin:2009wr,Volin:2010cq}. Volin's method
was adapted to the disk problem in \cite{Marino:2019fuy}.
In \cite{Reichert:2020ymc} the small
$\kappa$ expansion of ${\cal C}$ was given explicitly up to $O(\kappa^7)$.

Numerical studies of the integral equation were initiated in 
\cite{nystrom1930praktische,nomura,cooke1956solution,cooke1958coaxial}
and continued with ever greater precision 
\cite{norgren2009capacitance,paffuti2017numerical,paffuti2019galerkin}.

In this paper we study the disk problem using its relation to the free energy
of the $O(3)$ model and introduce a running coupling $\beta$ by
\begin{equation}
\frac{2\pi}{\kappa}=\frac{1}{\beta}+\ln\beta-\ln\frac{8}{e}
\end{equation}  
in terms of which the capacity can be compactly written as an asymptotic power
series, see (\ref{C0beta}).


\subsection{Integral equations and observables}

In this paper we analyze linear integral equations of the form 
\begin{equation}
\chi_{i}(\theta)-\int_{-B}^{B}d\theta' K(\theta-\theta')\chi_{i}(\theta')=r_{i}(\theta),
\qquad\quad -B\leq\theta\leq B,  
\label{TBA}
\end{equation}
where the kernel $K(\theta)$ and the sources $r_{i}(\theta)$ are
symmetric functions of $\theta$, implying the same for the solutions,
too. These solutions $\chi_{i}(\theta)$ depend implicitly on $B$,
but we suppress this $B$-dependence in the notation.

Observe that
the $B$-derivative of $\chi_{i}(\theta)$, which we denote by a dot,
$\partial\chi_{i}(\theta)/\partial B=\dot{\chi}_{i}(\theta)$, satisfies
the same type of equation 
\begin{equation}
\dot{\chi}_{i}(\theta)-\int_{-B}^{B} d\theta' K(\theta-\theta')\dot{\chi}_{i}(\theta')=R_{i}(\theta)
\end{equation}
with the source term 
\begin{equation}
R_{i}(\theta)=\left[K(\theta-B)+K(\theta+B)\right]\chi_{i}(B),
\end{equation}
which is again symmetric in $\theta$. 

We are interested in observables of the form
\begin{equation}
O_{ij}=\int_{-B}^{B}\frac{d\theta}{2\pi}\chi_{i}(\theta)r_{j}(\theta),
\end{equation}
where we again suppressed the $B$-dependence of $O_{ij}$ and will
do so if it does not lead to any confusion. The symmetry of the kernel
and the integral equation implies that $O_{ij}$ is symmetric in $i,j$.
Actually, not all $O_{ij}$ are independent as they can be expressed
in terms of $\chi_{i}(B)$ and integration constants, which follows
from the key formula
\begin{equation}
\dot{O}_{ij}\equiv\frac{\partial O_{ij}}{\partial B}=\frac{1}{\pi}\chi_{i}(B)\chi_{j}(B)\label{eq:master}.
\end{equation}
This nice expression can be derived by observing that 
\begin{align}
\dot{O}_{ij} & =\frac{1}{\pi}\chi_{i}(B)r_{j}(B)+\int_{-B}^{B}\frac{d\theta}{2\pi}\dot{\chi}_{i}(\theta)r_{j}(\theta)
\end{align}
and by exploiting that the second term is of the form of $O_{ij}$.
Its symmetry implies that $\int_{-B}^{B}\frac{d\theta}{2\pi}\dot{\chi}_{i}(\theta)r_{j}(\theta)=\int_{-B}^{B}\frac{d\theta}{2\pi}\chi_{j}(\theta)R_{i}(\theta)$
thus we can recognize the appearance of $\chi_{j}(B)=\chi_{j}(-B)$
leading to the required conclusion (\ref{eq:master}). 

In this paper we analyze the two dimensional $O(N)$ sigma model in
a magnetic field. The kernel
\begin{equation}
K(\theta)=\frac{1}{2\pi i}\partial_{\theta}\log S(\theta)
\end{equation}
is the logarithmic derivative of the scattering matrix 
\begin{equation}
S(\theta)=-\frac{\Gamma(\frac{1}{2}-\frac{i\theta}{2\pi})\Gamma(\Delta-\frac{i\theta}{2\pi})\Gamma(1+\frac{i\theta}{2\pi})\Gamma(\frac{1}{2}+\Delta+\frac{i\theta}{2\pi})}{\Gamma(\frac{1}{2}+\frac{i\theta}{2\pi})\Gamma(\Delta+\frac{i\theta}{2\pi})\Gamma(1-\frac{i\theta}{2\pi})\Gamma(\frac{1}{2}+\Delta-\frac{i\theta}{2\pi})},
\end{equation}
where the parameter $\Delta$ is related to $N$ as $1/\Delta=N-2$. 

We are interested in two problems: In the first, the source of the
integral equation is $r_{c}(\theta)=\cosh\theta$, while in the second
it is simply $r_{1}(\theta)=1$. The physical meaning of $m\chi_{c}(\theta)$
is the rapidity density of particles in the groundstate. The density
and energy density are simply 
\begin{equation}
\rho=mO_{c1}=m\int_{-B}^{B}\frac{d\theta}{2\pi}\chi_{c}(\theta)\quad;\qquad
\epsilon=m^{2}O_{cc}=m^{2}\int_{-B}^{B}\frac{d\theta}{2\pi}\chi_{c}(\theta)\cosh\theta,
\end{equation}
which defines a one-to-one correspondence between $\rho$ and $B$.
They are fixed by the magnetic field, through $h=\partial_{\rho}\epsilon=\dot{\epsilon}/\dot{\rho}$,
which follows from minimising the free energy over $\rho$. Alternatively,
one can determine directly the free energy ${\cal F}(h)=m^2O_{cc}-hmO_{c1}$
as a function of $h$, where $B$ is expressed with $h$ from $h\chi_{1}(B)=m\chi_{c}(B)$. 

We will pay particular attention to the case $\Delta=1$, i.e. to
the $O(3)$ model, where the kernel is a simple rational function
$K(\theta)=1/(\theta^{2}+\pi^{2})$. As we discussed above, the corresponding
integral
equation shows up in two unrelated but interesting problems, namely
in the Lieb-Liniger model and in the circular disk capacitor. In the capacitor
problem the relevant observable is the capacity (as a function of $B$),
which is simply related to 
\begin{equation}
O_{11}=\int_{-B}^{B}\frac{d\theta}{2\pi}\chi_{1}(\theta).
\end{equation}

In view of (\ref{eq:master}), the first and the second problems with
solutions $\chi_{c}(\theta)$ and $\chi_{1}(\theta)$ are not independent.
The knowledge of any of them determines the 3 independent integrals up to integration
constants, e.g. 
\begin{equation}
\dot{O}_{11}=\frac{\dot O_{c1}^{2}}{\dot O_{cc}}
\label{O11}  
\end{equation}



\section{Wiener-Hopf method and running coupling}

In this section we present a solution of the TBA integral equations (\ref{TBA})
following the Wiener-Hopf method. We restrict our consideration to the case of
the $O(N)$ non-linear sigma models and in particular the $O(3)$ model.
Our derivation is very similar to that of
\cite{Hasenfratz:1990zz,Hasenfratz:1990ab,Marino:2021dzn}
but here we concentrate on the introduction and use of
running couplings, which are useful in presenting perturbative series.

Many of the technical details and various definitions and results are collected
in appendix~\ref{appX}.

\subsection{Derivation of the Wiener-Hopf equations}

Let us temporarily suppress the $c,1$ indices and extend the TBA equation
(\ref{TBA}) to all $\theta$ as follows.
\begin{equation}
\chi(\theta)-\int_{-\infty}^\infty K(\theta-\theta')\chi(\theta')d\theta'=
r(\theta)+R(\theta)+L(\theta).
\label{TBA1}
\end{equation}
Here we extended $\chi(\theta)$ and $r(\theta)$ by defining
\begin{equation}
\chi(\theta)=r(\theta)=0,\qquad\quad \vert\theta\vert>B  
\end{equation}
and added $R(\theta)$ and $L(\theta)$ satisfying
\begin{equation}
R(\theta)=0,\quad \theta\leq B;\qquad\qquad L(\theta)=R(-\theta).   
\end{equation}
Next we use the Fourier transformed functions $\tilde\chi(\omega)$,
$\tilde r(\omega)$, $\tilde R(\omega)$ to define
\begin{equation}
X_+(\omega)=e^{-i\omega B}\tilde R(\omega),\quad  
r_\pm(\omega)=e^{\pm i\omega B}\tilde r(\omega),\quad  
\chi_\pm(\omega)=e^{\pm i\omega B}\tilde \chi(\omega),\quad
X_-(\omega)=X_+(-\omega).
\end{equation}
$r_\pm$ and $\chi_\pm$ are analytic everywhere and vanish for large
$\vert\omega\vert$ in the upper/lower complex half-planes $H_\pm$,
whereas $X_\pm$ are analytic in $H_\pm$ and
vanish there for large $\vert\omega\vert$.
\begin{equation}
Q_\pm(\omega)=G_\pm(\omega)X_\pm(\omega),\qquad\quad Q_-(\omega)=Q_+(-\omega)  
\end{equation}
have similar properties, see (\ref{Gplmi}) and (\ref{Gplus}).

Going to Fourier space the extended TBA equation (\ref{TBA1}) becomes
\begin{equation}
[1-\tilde K(\omega)]\tilde\chi(\omega)=\tilde r(\omega)+\tilde R(\omega)+
\tilde L(\omega),  
\label{TBAFour}
\end{equation}
which can be rearranged as ($\alpha(\omega)$ is defined by (\ref{alphadef}))
\begin{equation}
\frac{\chi_+}{G_+}=G_- r_++\alpha Q_+ + Q_-.  
\end{equation}
Decomposing it into negative and positive parts, we get the following two
equations
\begin{equation}
Q_-+(\alpha Q_+)^{(-)}+(G_- r_+)^{(-)}=0,
\end{equation}
\begin{equation}
\frac{\chi_+}{G_+}=(\alpha Q_+)^{(+)}+(G_- r_+)^{(+)}.
\end{equation}
We now spell out the first of these for $\omega=-2i\xi$ and use $Q_-(\omega)=
Q_+(-\omega)$. We obtain
\begin{equation}
Q_+(2i\xi)+\frac{1}{2\pi}\int_{-\infty}^\infty \frac{\alpha(\omega)Q_+(\omega)}
{2\xi-i\omega}d\omega  
+\frac{1}{2\pi}\int_{-\infty}^\infty \frac{G_-(\omega)r_+(\omega)}
{2\xi-i\omega}d\omega=0.  
\label{WH1}
\end{equation}
Similarly, the second equation for $\omega=i\kappa$ becomes
\begin{equation}
\chi_+(i\kappa)=G_+(i\kappa)\left\{
\frac{1}{2\pi}\int_{-\infty}^\infty \frac{\alpha(\omega)Q_+(\omega)}
{\kappa+i\omega}d\omega  
+\frac{1}{2\pi}\int_{-\infty}^\infty \frac{G_-(\omega)r_+(\omega)}
{\kappa+i\omega}d\omega\right\}.  
\label{WH2}
\end{equation}
It is sufficient to know $\chi_+$ along the positive imaginary axis, since the
two density integrals
\begin{equation}
O_{.c}=\frac{1}{2\pi}\int_{-\infty}^\infty \chi(\theta)\cosh\theta d\theta
\quad{\rm and}\quad
O_{.1}=\frac{1}{2\pi}\int_{-\infty}^\infty \chi(\theta) d\theta
\end{equation}
are given by
\begin{equation}
O_{.c}=\frac{e^B}{2\pi}\chi_+(i)\qquad{\rm and}\qquad
O_{.1}=\frac{1}{2\pi}\chi_+(0)  
\label{213}
\end{equation}
and using the Cauchy integral
\begin{equation}
\chi_+(i\kappa)=\frac{1}{2\pi i}\int_{-\infty}^\infty \frac{\chi_+(\omega)d\omega}
{\omega-i\kappa} 
\end{equation}
the boundary value $\chi(B)$ can also be obtained from
\begin{equation}
\chi(B)=\lim_{\kappa\to\infty}\kappa\chi_+(i\kappa).  
\end{equation}

So far we reviewed the standard way \cite{Hasenfratz:1990zz,Hasenfratz:1990ab,
Marino:2021dzn} of applying the Wiener-Hopf method to the $O(N)$ nonlinear
sigma model. In the rest of this section we will follow a novel way of analyzing
the problem by slightly rearranging the terms in the integral equation
(\ref{WH1}) allowing us to introduce running couplings already at an early
stage of the calculation.

Technically, it is very difficult to extract the particle density $O_{c1}$
using (\ref{213}) as the kernels are singular at the origin.
An alternative way to obtain the particle density $O_{c1}$ is to go back to
the original TBA equation (\ref{TBAFour}) for the $\chi_c$ problem and use the
small $\vert\omega\vert$ expansions
\begin{equation}
1-\tilde K(\omega)=\pi\Delta\vert\omega\vert+O(\omega^2),\quad
\tilde\chi_c(\omega)=2\pi O_{c1}+O(\omega^2),\quad  
\tilde r_c(\omega)=2\sinh B+O(\omega^2),  
\end{equation}
together with the small $\vert\omega\vert$ expansion of $\tilde R(\omega)$,
which is analytic in the upper half plane only:
\begin{equation}
\tilde R(\omega)=a-ib\omega-ic\omega\ln(-i\omega)+O(\vert\omega\vert^2),
\label{Romega}
\end{equation}
where $a$, $b$, $c$ are some constants.
Here, and also throughout this paper, the symbol $O(x^p)$ includes possible
logarithmic terms of the form $x^p(\ln x)^q$.
Using $\tilde L(\omega)=\tilde R(-\omega)$, for real $\omega$ we find
\begin{equation}
\tilde R(\omega)+\tilde L(\omega)=2a-\pi c\vert\omega\vert+O(\vert\omega\vert^2)
\end{equation}
and using (\ref{TBAFour}) we can determine 
\begin{equation}
a=-\sinh B,\qquad\quad c=-2\pi\Delta O_{c1}.
\end{equation}
Thus the value of the particle density $O_{c1}$ is hidden in the small
$\omega$ expansion of $\tilde R$. 

Since it is difficult to perform this small $\omega$ expansion or to evaluate
(\ref{WH2}) at the singular point $\kappa=0$, we will solve both problems
($\chi_c$ and $\chi_1$) and calculate the densities from
\begin{equation}
O_{cc}=\frac{e^B}{2\pi}\chi_{c+}(i),\qquad\quad O_{c1}=O_{1c}=\frac{e^B}{2\pi}
\chi_{1+}(i).  
\end{equation}
Finally, $O_{11}$ will be obtained (up to an integration constant) using
(\ref{O11}). This way we only use (\ref{WH2}) for $\kappa=1$ and
$\kappa=\infty$, which are well defined.

\subsection{The $\chi_c$ problem}

In this case
\begin{equation}
r_{c+}(\omega)=\frac{1}{2}e^{2i\omega B}\left(\frac{e^B}{1+i\omega}
+\frac{e^{-B}}{i\omega-1}\right)
+\frac{1}{2}\left(\frac{e^B}{1-i\omega}
-\frac{e^{-B}}{1+i\omega}\right).
\label{rc+}
\end{equation}
It is useful to introduce the new unknown $q_+(\omega)$ by
\begin{equation}
Q_+(\omega)+\frac{G_+(\omega)}{2}\left(\frac{e^B}{1+i\omega}
+\frac{e^{-B}}{i\omega-1}\right)=\frac{e^B G_+(i)}{2}q_+(\omega)  
\end{equation}
because then the first term in (\ref{rc+}) can be absorbed into the integral
containing $\alpha(\omega)$ and the contribution of the second term of
(\ref{rc+}) can be explicitly integrated in (\ref{WH1}) by closing the
integration contour at infinity in $H_-$. We obtain
\begin{equation}
q_+(2i\xi)+\frac{1}{2\pi}\int_{-\infty}^\infty \frac{\alpha(\omega)q_+(\omega)}
{2\xi-i\omega}d\omega=\frac{1}{1-2\xi}.  
\label{WH1a}
\end{equation}
Manipulating (\ref{WH2}) similarly, we get
\begin{equation}
O_{cc}=\frac{e^{2B}G_+^2(i)}{8\pi}W,\qquad\quad \chi_c(B)=\frac{e^B G_+(i)}{2}w,  
\label{Occ}
\end{equation}
where
\begin{equation}
W=1+\frac{1}{\pi}\int_{-\infty}^\infty \frac{\alpha(\omega)q_+(\omega)}{1+i\omega}
d\omega,\qquad\qquad w=1+\frac{1}{2\pi}\int_{-\infty}^\infty \alpha(\omega)
q_+(\omega)d\omega.
\label{Ww}
\end{equation}
We note that using (\ref{Romega}) and (\ref{Gplus}) we can derive the small
$\vert\omega\vert$ expansion
\begin{equation}
\frac{e^B G_+(i)}{2}q_+(\omega)=\frac{\sqrt{-i\omega}}{\sqrt{\pi\Delta}}[
 b_1+c\ln(-i\omega)+O(\vert\omega\vert)],\qquad b_1=b-B\sinh B+\cosh B.
\label{qplusexp}
\end{equation}

The introduction of the new variable $q_+$ simplifies the equations but the
price to pay is that $q_+$ is no longer analytic in $H_+$: it has a pole at
$\omega=i$. Sometimes it is useful to remove this pole term and write
\begin{equation}
q_+(2i\xi)=k_+(2i\xi)+\frac{1}{1-2\xi},  
\end{equation}
where $k_+$ is analytic in $H_+$.

\begin{figure}
\leavevmode
\centerline{\includegraphics[scale=0.7]{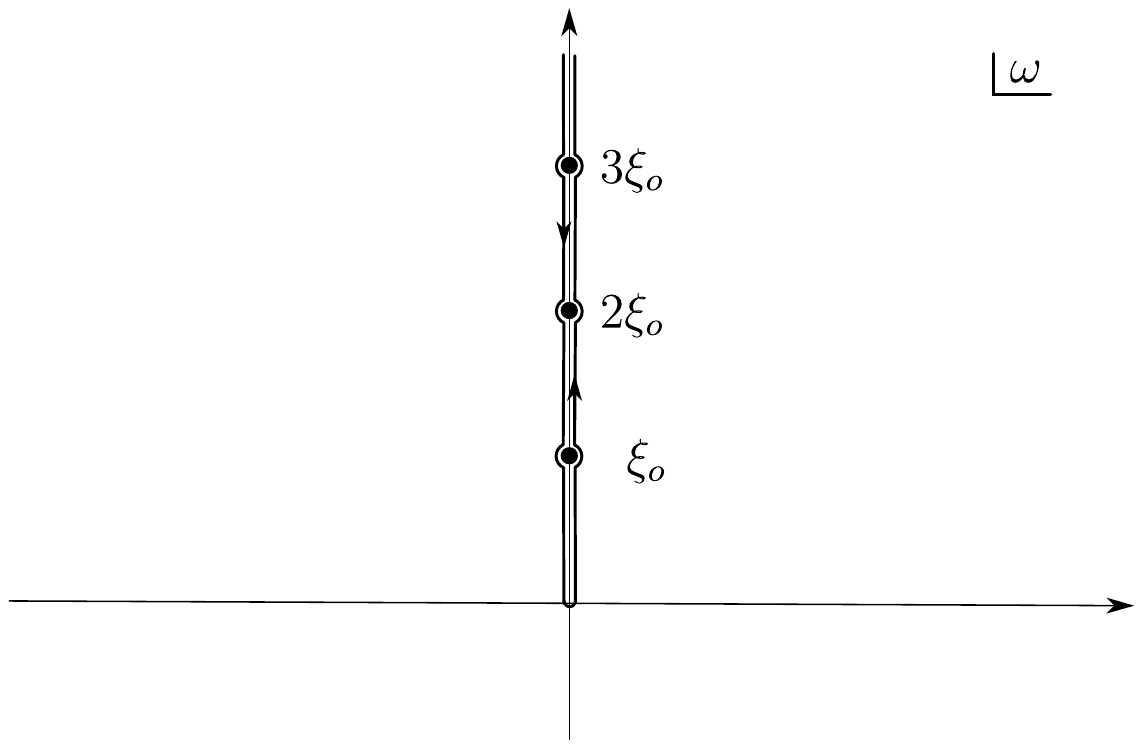} }
\vspace{-11.0cm}
\caption{{\footnotesize
The ${\cal C}_{LR}$ contour.}}
\label{CLR}
\end{figure}

The next step in the Wiener-Hopf analysis is the deformation of the contour
of the $\omega$ integration. Using the fact that $\alpha q_+$ is analytic in
$H_+$ except on the imaginary axis, we deform the $\omega$ contour to
${\cal C}_{LR}$, which is defined as coming down from infinity to zero
slightly to the left along the imaginary axis and then going up to
infinity slightly to right of the imaginary axis (see Fig. \ref{CLR}):
\begin{equation}
q_+(2i\xi)+\frac{i}{2\pi}\int_{{\cal C}_{LR}} \frac{\alpha(2i\xi')q_+(2i\xi')}
{\xi+\xi'}d\xi'=\frac{1}{1-2\xi}.  
\label{WH1b}
\end{equation}
The contribution of the poles at $\xi=\xi_\ell$ (see appendix~\ref{appX}) to
this integral cancel due to the opposite signs in (\ref{poles}) for the left
and right integrals. In this region the integration along ${\cal C}_{LR}$
reduces to the principal value integral of the discontinuity.
The pole of $q_+$ at $\xi=1/2$ only contributes for $N=3$, since for larger
$N$ the function $\alpha$ vanishes at this point.
Using (\ref{alphabeta}), (\ref{WH1b}) simplifies to
\begin{equation}
q_+(2i\xi)+\frac{\delta_{N,3}}{e}\,\frac{e^{-2B}}{1+2\xi}+\frac{1}{\pi}{\cal P}  
\int_0^\infty\frac{e^{-4B\xi'}\beta(\xi')q_+(2i\xi')}{\xi+\xi'}d\xi'
=\frac{1}{1-2\xi}.
\label{WH1c}
\end{equation}
For the $W$ integral in (\ref{Ww}) the contribution of the $\xi=\xi_\ell$ poles
also cancels, but the region around $\xi=1/2$ must be treated carefully.
Using the results (\ref{half}-\ref{half2}) we obtain
\begin{equation}
W=1+2\Lambda_oe^{-2B}+\frac{4}{\pi}{\cal P}\int_0^\infty\frac
{e^{-4B\xi}\beta(\xi)q_+(2i\xi)}{1-2\xi}d\xi,
\label{W1}    
\end{equation}
where
\begin{equation}
\Lambda_o=\left\{\begin{matrix}-\frac{1}{2}\sin\frac{\pi\gamma_1}{2}\tilde h_1
\qquad\quad & N\geq4,\\
\frac{1}{e}(2B+k_+(i)+1-\gamma_E-\ln2)\qquad\quad &N=3.\end{matrix}\right.
\end{equation}
Here $\tilde h_1$ is defined by (\ref{half2}) and for $N=3$ 
\begin{equation}
k_+(i)=-\frac{1}{2e}e^{-2B}-\frac{2}{\pi}{\cal P}\int_0^\infty\frac
{e^{-4B\xi}\beta(\xi)q_+(2i\xi)}{1+2\xi}d\xi.
\label{k+}    
\end{equation}
Finally evaluating $w$ in (\ref{Ww}) gives
\begin{equation}
w=1+\frac{\delta_{N,3}}{e}e^{-2B}+\frac{2}{\pi}{\cal P}\int_0^\infty
e^{-4B\xi}\beta(\xi)q_+(2i\xi) d\xi.
\label{w1}    
\end{equation}
Following \cite{Marino:2021dzn} we now introduce the integration along the
contour ${\cal C}_+$, which goes from $0$ to $\infty$ slightly to the left
of the imaginary axis (in the variable $2i\xi$). In terms of $\xi$ this is
slightly above the real axis. The reason to use this integration path is that
it resembles the $+$ lateral Borel resummation.
\newcommand{\inC}{\int_{{\cal C}_+}}
We introduce this new integration using the rule
\begin{equation}
\inC=-i\pi\sum_{{\rm res}}+{\cal P}\int.
\label{integrationcontour}
\end{equation}
Here in front of the sum of residues term we have the factor $-i\pi$ because
the integral near the poles is only a half-circle and its orientation is
clock-wise. (\ref{WH1c}) becomes
\begin{equation}
q_+(2i\xi)+\frac{\delta_{N,3}}{e}\,\frac{\nu}{1+2\xi}+i\sum_{\ell=1}^\infty
\frac{H_\ell q_\ell}{\xi+\ell\xi_o}\nu^{2\ell\xi_o}+\frac{1}{\pi}\inC
\frac{e^{-4B\xi'}\beta(\xi')q_+(2i\xi')}{\xi+\xi'}d\xi'=\frac{1}{1-2\xi}.
\label{WH1d}
\end{equation}
Here we introduced the notations
\begin{equation}
\nu=e^{-2B},\qquad\quad q_\ell=q_+(2i\ell\xi_o)
\end{equation}
and the definition of $\xi_o$ and $H_\ell$ is given by (\ref{xi0def}) and
(\ref{poles}), respectively.
Next we introduce the variable $Q(x)$ with rescaled argument:
\begin{equation}
Q(x)=q_+(ivx),  
\end{equation}
where the parameter $v$ will be fixed later. At this point the
Wiener-Hopf equation becomes
\begin{equation}
Q(x)+\frac{\delta_{N,3}\nu}{e(1+vx)}+2i\sum_{\ell=1}^\infty
\frac{H_\ell q_\ell}{vx+2\ell\xi_o}\nu^{2\ell\xi_o}+\frac{1}{\pi}\inC
\frac{e^{-2Bvy}\beta(vy/2)Q(y)}{x+y}d y=\frac{1}{1-vx}.
\label{WH1e}
\end{equation}
We can write down the equation determining the parameters $q_s$:
\begin{equation}
q_s+\frac{\delta_{N,3}\nu}{e(1+2s\xi_o)}+\frac{i}{\xi_o}\sum_{\ell=1}^\infty
\frac{H_\ell q_\ell}{s+\ell}\nu^{2\ell\xi_o}+\frac{v}{\pi}\inC
\frac{e^{-2Bvx}\beta(vx/2)Q(x)}{2s\xi_o+vx}d x=\frac{1}{1-2s\xi_o}.
\label{qs}
\end{equation}
The physical quantities $W$, $w$ and $k_+(i)$ (for $N=3$) can also be rewritten:
\begin{equation}
W=1+(i\beta_o+2\Lambda_o)\nu+4i\sum_{\ell=1}^\infty
\frac{H_\ell q_\ell}{1-2\ell\xi_o}\nu^{2\ell\xi_o}+\frac{2v}{\pi}\inC
\frac{e^{-2Bvx}\beta(vx/2)Q(x)}{1-vx}d x,
\label{W2}
\end{equation}
\begin{equation}
w=1+\frac{\delta_{N,3}}{e}\nu+2i\sum_{\ell=1}^\infty
H_\ell q_\ell \nu^{2\ell\xi_o}+\frac{v}{\pi}\inC
e^{-2Bvx}\beta(vx/2)Q(x) d x,
\label{w2}
\end{equation}
\begin{equation}
k_+(i)=-\frac{\nu}{2e}-2i\sum_{\ell=1}^\infty
\frac{H_\ell q_\ell}{1+2\ell}\nu^{2\ell}-\frac{v}{\pi}\inC
\frac{e^{-2Bvx}\beta(vx/2)Q(x)}{1+vx}d x.
\label{k+2}
\end{equation}
The definition of $\beta_o$ is given by (\ref{bet0def}).
\subsection{Running coupling}

Following the idea of \cite{Forgacs:1991rs}, where a running coupling has been
introduced for fermionic models, we now fix the value of the rescaling parameter
$v$, thereby promoting it to the status of a running coupling. First we note
that (see appendix~\ref{appX})
\begin{equation}
\beta(\xi)=e^{2\gamma_1\xi\ln2\xi}A(\xi),\qquad\quad A(\xi)=\cos(\pi\gamma_1\xi)\,
H_p(\xi),\qquad\quad \gamma_1=2\Delta-1.  
\end{equation}
It is easy to see that if we define $v$ so that it satisfies the relation
\begin{equation}
2B=\frac{1}{v}+\gamma_1\ln v +L,  
\end{equation}
where $L$ is some constant\footnote{This constant should not be
confused with the
logarithmic variable $\mathsf{L}$ used in the disk capacitor case.}, then
\begin{equation}
e^{-2Bvx}\beta(vx/2)=e^{-x}{\cal A}(x),  
\end{equation}
where
\begin{equation}
{\cal A}(x)=e^{vx(\gamma_1\ln x-L)}A(vx/2).
\end{equation}
The advantage of using the running coupling $v$ instead of $B$ itself is that
${\cal A}(x)$ can be expanded around $v=0$ in a power series:
\begin{equation}
{\cal A}(x)=\sum_{k=0}^\infty (vx)^k L_k(\ln x),  
\end{equation}
where $L_k$ is a polynomial of degree $k$ in its argument. Since
\begin{equation}
A(vx/2)=1+\alpha_1 vx+O(v^2),
\end{equation}
the leading expansion coefficients are
\begin{equation}
L_0(z)=1,\qquad,\qquad L_1(z)=\gamma_1 z+\alpha_1-L.
\end{equation}
Later we will see that (most of) the physical quantities can also be expanded
in $v$ as a power series. The choice of the value of the constant $L$ is a
matter of convenience and does not change the above conclusion. This is true
since a given running coupling can be power expanded in terms of a running
coupling defined by a different $L$ value.

\subsection{Exact equations}

Using the perturbatively expandable function ${\cal A}(x)$, the equations of the
Wiener-Hopf method can be rewritten as follows.
\begin{equation}
Q(x)+\frac{\delta_{N,3}\nu}{e(1+vx)}+2i\sum_{\ell=1}^\infty
\frac{H_\ell q_\ell}{vx+2\ell\xi_o}\nu^{2\ell\xi_o}+\frac{1}{\pi}\inC
\frac{e^{-y}{\cal A}(y)Q(y)}{x+y}d y=\frac{1}{1-vx},
\label{WH1f}
\end{equation}
\begin{equation}
q_s+\frac{\delta_{N,3}\nu}{e(1+2s\xi_o)}+\frac{i}{\xi_o}\sum_{\ell=1}^\infty
\frac{H_\ell q_\ell}{s+\ell}\nu^{2\ell\xi_o}+\frac{v}{\pi}\inC
\frac{e^{-x}{\cal A}(x)Q(x)}{2s\xi_o+vx}d x=\frac{1}{1-2s\xi_o},
\label{qs1}
\end{equation}
\begin{equation}
W=1+M\nu+4i\sum_{\ell=1}^\infty
\frac{H_\ell q_\ell}{1-2\ell\xi_o}\nu^{2\ell\xi_o}+\frac{2v}{\pi}\inC
\frac{e^{-x}{\cal A}(x)Q(x)}{1-vx}d x,
\label{W3}
\end{equation}
\begin{equation}
w=1+\frac{\delta_{N,3}}{e}\nu+2i\sum_{\ell=1}^\infty
H_\ell q_\ell \nu^{2\ell\xi_o}+\frac{v}{\pi}\inC
e^{-x}{\cal A}(x)Q(x) d x,
\label{w3}
\end{equation}
\begin{equation}
(N=3)\quad k_+(i)=-\frac{\nu}{2e}-2i\sum_{\ell=1}^\infty
\frac{H_\ell q_\ell}{1+2\ell}\nu^{2\ell}-\frac{v}{\pi}\inC
\frac{e^{-x}{\cal A}(x)Q(x)}{1+vx}d x.
\label{k+3}
\end{equation}
Here
\begin{equation}
M=\left\{\begin{matrix}
  -2e\left(\frac{\Delta}{e}\right)^{2\Delta}\,\frac{\Gamma(1-\Delta)}
{\Gamma(1+\Delta)}e^{i\pi\Delta}\qquad\quad &N\geq4,\\  
-\frac{i\pi}{e}+\frac{2}{e}[2B+1-\gamma_E-\ln2+k_+(i)]\qquad\quad &N=3.
\end{matrix}\right.
\label{nagyM}
\end{equation}

\subsection{Trans-series and perturbative expansion, $N\geq4$}

The form of the exact equation (\ref{WH1f}) suggests that the solution can be 
written in a trans-series form
\begin{equation}
Q(x)=\sum_{m=0}^\infty Q_m(x)\nu^{2m\xi_o}.  
\end{equation}
Here the coefficient functions $Q_m(x)$ still depend on $B$, but in the above
representation we temporarily treat $\nu$ as an independent expansion
parameter and its value $\nu=e^{-2B}$ will be used only at the end of the
calculation.

In the same spirit we write
\begin{equation}
q_s=\sum_{m=0}^\infty q_{s,m}\nu^{2m\xi_o},  
\end{equation}
\begin{equation}
W=W_0+M\nu+\sum_{m=1}^\infty W_m\nu^{2m\xi_o},  
\label{Wexp}
\end{equation}
\begin{equation}
w=\sum_{m=0}^\infty w_m\nu^{2m\xi_o}.  
\end{equation}
The leading and next-to-leading coefficients satisfy
\begin{equation}
Q_0(x)+\frac{1}{\pi}\inC\frac{e^{-y}{\cal A}(y)Q_0(y)}{x+y}d y=\frac{1}{1-vx},  
\end{equation}
\begin{equation}
Q_1(x)+\frac{1}{\pi}\inC\frac{e^{-y}{\cal A}(y)Q_1(y)}{x+y}d y
=-\frac{2iH_1q_{1,0}}{2\xi_o+vx},  
\end{equation}
where
\begin{equation}
q_{1,0}=\frac{1}{1-2\xi_o}-\frac{v}{\pi}\inC\frac{e^{-x}{\cal A}(x)Q_0(x)}
{2\xi_o+vx}d x.  
\end{equation}
The leading and subleading coefficients of $W$ and $w$ are
\begin{equation}
W_0=1+\frac{2v}{\pi}\inC
\frac{e^{-x}{\cal A}(x)Q_0(x)}{1-vx}d x,
\label{W0x}
\end{equation} 
\begin{equation}
W_1=\frac{4iH_1q_{1,0}}{1-2\xi_o}+\frac{2v}{\pi}\inC
\frac{e^{-x}{\cal A}(x)Q_1(x)}{1-vx}d x,
\label{W1x}
\end{equation}
\begin{equation}
w_0=1+\frac{v}{\pi}\inC
e^{-x}{\cal A}(x)Q_0(x)d x,
\label{w0x}
\end{equation}
\begin{equation}
w_1=2iH_1q_{1,0}+\frac{v}{\pi}\inC
e^{-x}{\cal A}(x)Q_1(x)d x.
\label{w1x}
\end{equation}

Next we expand the coefficient functions perturbatively in the running
coupling $v$. For the leading term $Q_0(x)$ we define
\begin{equation}
{\cal A}(x)Q_0(x)=\psi(x)=\sum_{k=0}^\infty v^k\psi_k(x).  
\end{equation}
All $B$-dependence is given through the running coupling $v$, the component
functions $\psi_k$ are $B$-independent. $\psi(x)$ satisfies the integral
equation
\begin{equation}
\frac{\psi(x)}{{\cal A}(x)}+\frac{1}{\pi}\inC \frac{e^{-y}\psi(y)}
{x+y}dy=\frac{1}{1-vx}.
\label{psi}
\end{equation}
Here are the LO and NLO equations of the
perturbative expansion:
\begin{equation}
\psi_0(x)+\frac{1}{\pi}\int_0^\infty\frac{e^{-y}\psi_0(y)}{x+y}d y=1,  
\label{psi0}
\end{equation}
\begin{equation}
\psi_1(x)+\frac{1}{\pi}\int_0^\infty\frac{e^{-y}\psi_1(y)}{x+y}d y
=x+xL_1(\ln x)\psi_0(x).  
\label{psi1}
\end{equation}
For the physical quantities we need the moments of the component functions
defined by
\begin{equation}
\Omega_{k,s}=\frac{1}{\pi}\int_0^\infty e^{-x}x^s\psi_k(x)dx,\qquad\quad
\Omega_k=\Omega_{k,0}.  
\label{Omegaks}
\end{equation}
Expressed in terms of these moments,
\begin{equation}
W_0=1+2\sum_{k=0}^\infty \overline{\Omega}_k v^{k+1},\qquad\quad
\overline{\Omega}_k=\sum_{r=0}^k\Omega_{k-r,r},  
\label{W0pert}
\end{equation}
\begin{equation}
w_0=1+\sum_{k=0}^\infty \Omega_k v^{k+1}.
\label{w0pert}
\end{equation}
In appendix \ref{appA} we give the solution of the LO and NLO perturbative
problems and calculate the corresponding moments by explicitly solving the
integral equations (\ref{psi0}) and (\ref{psi1}). It is extremely difficult
to go to higher orders with this method. Luckily, using Volin's method
\cite{Volin:2009wr,Volin:2010cq} it is possible to obtain the \lq\lq
perturbative'' coefficients $\overline{\Omega}_k$ to very high order
algebraically. Although Volin's algorithm was worked out originally for
the $1/B$ expansion coefficients, it is possible to modify this
algorithm so that it gives directly the series (\ref{W0pert}). We can look
at (\ref{W0pert}) and (\ref{w0pert}) as structural results, proving that these
physical quantities have perturbative expansions in terms of the running
coupling $v$.

The LO and NLO results are
\begin{equation}
\overline{\Omega}_0=\Omega_0=\frac{1}{4},\qquad\quad
\overline{\Omega}_1=\Omega_1+\Omega_{0,1}=\frac{3}{16}+\frac{3\Delta}{4}
-\frac{r_o}{2}-\frac{L}{4},
\end{equation}
where
\begin{equation}
r_o=\left(2\Delta-\frac{1}{2}\right)\ln2-\Delta\ln\Delta.
\end{equation}
If one is using the $L=-2r_o$ scheme, $\ln2$ and $\ln\Delta$ disappear from
the NLO coefficient. We found that this feature is preserved also at higher
orders.

\subsection{The $N=3$ case}

For $N=3$ the trans-series is of the form
\begin{equation}
Q(x)=\sum_{m=0}^\infty Q_m(x)\nu^m.  
\end{equation}
Note that this trans-series contains powers of $\nu$, not just powers of
$\nu^{2\xi_o}$ as in the general case ($\xi_o=1$ for $N=3$).
Similarly, we write
\begin{equation}
q_s=\sum_{m=0}^\infty q_{s,m}\nu^m,\qquad  
W=\sum_{m=0}^\infty W_m\nu^m,\qquad  
w=\sum_{m=0}^\infty w_m\nu^m,\qquad  
M=\sum_{m=0}^\infty M_m\nu^m. 
\end{equation}
The leading coefficient $Q_0(x)$ satisfies the same equation and has the same
type of perturbative expansion as in the general $N\geq4$ case. $W_0$ and $w_0$
respectively are given by the same formulas as (\ref{W0x}) and (\ref{W0pert}),
(\ref{w0x}) and (\ref{w0pert}), respectively. On the other hand, we have
$O(\nu)$ terms here, which are absent (except for a constant M in (\ref{Wexp}))
for $N\geq4$. The corresponding equations are
\begin{equation}
Q_1(x)+\frac{1}{\pi}\inC\frac{e^{-y}{\cal A}(y)Q_1(y)}{x+y}d y=
-\frac{1}{e}  \,\frac{1}{1+vx},  
\end{equation}
\begin{equation}
W_1=M_0+\frac{2v}{\pi}\inC\frac{e^{-x}{\cal A}(x)Q_1(x)}{1-vx}d x,
\end{equation}
\begin{equation}
w_1=\frac{1}{e}+\frac{v}{\pi}\inC e^{-x}{\cal A}(x)Q_1(x) d x,
\end{equation}
where
\begin{equation}
M_0=-\frac{i\pi}{e}+\frac{2}{e}\left(2B+1-\gamma_E-\ln2
-\frac{v}{\pi}\inC\frac{e^{-x}{\cal A}(x)Q_0(x)}{1+vx}d x\right).
\end{equation}
It is useful to introduce
\begin{equation}
{\cal A}(x)Q_1(x)=-\frac{1}{e}\,\phi(x),\qquad\quad \phi(x)=\sum_{k=0}^\infty
v^k\phi_k(x).  
\end{equation}
It satisfies an integral equation similar to (\ref{psi}):
\begin{equation}
\frac{\phi(x)}{{\cal A}(x)}+\frac{1}{\pi}\inC \frac{e^{-y}\phi(y)}
{x+y}dy=\frac{1}{1+vx}.
\label{phi}
\end{equation}
We also introduce the moments of $\phi_k$:
\begin{equation}
\Sigma_{k,s}=\frac{1}{\pi}\int_0^\infty e^{-x}x^s\phi_k(x)dx,\qquad\quad
\Sigma_k=\Sigma_{k,0}.
\label{Sigmaks}
\end{equation}
It is easy to express $w_1$ with these moments:
\begin{equation}
w_1=\frac{1}{e}\left(1-\frac{v}{\pi}\inC e^{-x}\phi(x)dx\right)=
\frac{1}{e}\left(1-\sum_{k=0}^\infty \Sigma_k v^{k+1}\right).  
\end{equation}
The formula for $W_1$ is more complicated:
\begin{equation}
\begin{split}  
W_1&=-\frac{i\pi}{e}+\frac{2}{e}\left(2B+1-\gamma_E-\ln2
-\frac{v}{\pi}\inC\frac{e^{-x}\psi(x)}{1+vx}dx  
-\frac{v}{\pi}\inC\frac{e^{-x}\phi(x)}{1-vx}dx\right)\\  
&=-\frac{i\pi}{e}+\frac{2}{e}\left(\frac{1}{v}+\ln v+L+1-\gamma_E-\ln2
-\sum_{k=0}^\infty Z_k v^{k+1}\right),  
\end{split}
\end{equation}
where
\begin{equation}
Z_k=\sum_{r=0}^k[\Sigma_{k-r,r}+(-1)^r\Omega_{k-r,r}].
\label{Zk}
\end{equation}
We see that $W_1$ contains a $\ln v$ term (coming from the $2B$ contribution),
but all higher corrections are perturbative.

The expression (\ref{Zk}) can be written as
\begin{equation}
Z_k=\Sigma_k+\Omega_k+Z_{kx},  
\end{equation}
where $\Omega_k$ and $\Sigma_k$ already appeared in $w_0$ and $w_1$,
respectively, but we also need the extra contributions $Z_{kx}$ to
calculate higher corrections. Since
\begin{equation}
\psi_0(x)=\phi_0(x),\qquad\quad \Sigma_{0,r}=\Omega_{0,r},  
\end{equation}
we have
\begin{equation}
Z_{0x}=Z_{1x}=0,\qquad\quad Z_{2x}=\Sigma_{1,1}-\Omega_{1,1}+2\Omega_{0,2}.
\label{Zmx}
\end{equation}
Here
\begin{equation}
\Sigma_{1,1}-\Omega_{1,1}=\frac{1}{\pi}\int_0^\infty e^{-x}xk_1(x)dx,\qquad\quad
k_1(x)=\phi_1(x)-\psi_1(x).  
\end{equation}
$k_1$ satisfies the integral equation
\begin{equation}
k_1(x)+\frac{1}{\pi}\int_0^\infty \frac{e^{-y}k_1(y)}
{x+y}dy=-2x.
\label{k1}
\end{equation}
This is related to the NLO problem solved in appendix \ref{appA} and we find
\begin{equation}
Z_{2x}=-\frac{3}{64}.
\end{equation}

\subsection{Solution of the $\chi_1$ problem}

The treatment of the $\chi_1$ problem is completely analogous to the $\chi_c$
case, although technically a little more challenging. The details are given in
appendix~\ref{appY}, here we only summarize the main results.

$O_{1c}$ and $\chi_1(B)$ are of the form
\begin{equation}
O_{1c}=\frac{e^BG_+(i)}{4\pi\sqrt{v\Delta}}\hat\rho,  
\label{O1crho}
\end{equation}
where
\begin{equation}
\hat\rho=\hat\rho_0-\frac{\delta_{N,3}}{e}\hat\rho_1\nu+O(\nu^{2\xi_o}),
\end{equation}
\begin{equation}
\hat\rho_0=1+\sum_{k=1}^\infty R_k v^k,\qquad\quad
\hat\rho_1=1+\sum_{k=1}^\infty \tilde R_k v^k
\end{equation}
and
\begin{equation}
\chi_1(B)=\frac{1}{2\sqrt{v\Delta}}u,\qquad
u=\sum_{m=0}^\infty \hat u_m\nu^{2m\xi_0},\qquad
\hat u_0=1+\sum_{k=1}^\infty r_k v^k.  
\label{udefs}
\end{equation}
Comparing (\ref{qplusexp}) and (\ref{O1crho}) we see that $\hat\rho$ is also
encoded in the small $x$ expansion of $Q(x)$:
\begin{equation}
Q(x)=\sqrt{x}[G+\tilde G\ln x]+O(x^{3/2}),\qquad\quad \tilde G=-\frac{1}
{\sqrt{\pi}}\hat\rho.
\end{equation}
At leading (perturbative) order $Q(x)$ reduces to $Q_0(x)$ but this has the
same leading small $x$ expansion coefficients as $\psi(x)$ and we find
\begin{equation}
\psi(x)=\sqrt{x}[g+\tilde g\ln x]+O(x^{3/2}),\qquad\quad \tilde g=-\frac{1}
{\sqrt{\pi}}\hat\rho_0.
\label{psiasy}
\end{equation}




\providecommand{\tabularnewline}{\\}



\subsection{The structure of the non-perturbative corrections}

In this subsection we summarise the structure of the non-perturbative
corrections and comment on the leading behaviour. 

The non-perturbative terms originate from various sources. They have
their seeds in the pole singularities of the integral kernel $\alpha(\omega)$,
in the source term of the integral equation, $\cosh\theta$, affecting
observables of the form $O_{c .}$, as well as in the observable, i.e.
in $O_{. c}$ via $\chi_{+}(i)$. 

The integral kernel gives rise to non-perturbative corrections of
the form 
\begin{equation}
e^{-4Bl\xi_{o}}\quad;\quad l=1,2,\dots\qquad,\qquad\xi_{o}=\begin{cases}
\begin{array}{c}
k-1\\
2k-1
\end{array} & \begin{array}{c}
N=2k\\
N=2k+1
\end{array}\end{cases}
\end{equation}
Observe that we have ``twice'' as many terms in the even case than
in the odd case. This gives a different non-perturbative structure
for even and odd models, a fact, which was not pointed out explicitly
in \cite{Marino:2021dzn}. 

The source, $\cosh\theta$, produces an extra non-perturbative term
in the integral equation for the $O(3)$ model of order $\nu=e^{-2B}$
, which further generates terms of order $\nu^{1+2l\xi_{o}}$ and
makes the $N=3$ case drastically different from the $N\geq4$ cases.
We thus first focus on the simpler $N\geq4$ cases, where this term
is absent. 

Nevertheless, the observable $O_{cc}=e^{2B}WG_{+}(i)^{2}/(8\pi)$
even in these models gets an exceptional (leading) non-perturbative
correction of order $e^{-2B}$ in $W$. Clearly, this term is a constant
for $O_{cc}$ and disappears from $\dot{O}_{cc}$ and from $\chi_{c}(B)$.
The non-perturbative corrections indicate a trans-series form of the
various observables. In particular for $W$, which is related to the
energy of the system, we obtain 
\begin{equation}
W=W_{0}+M\nu+\sum_{l=1}^{\infty}W_{l}\nu^{2l\xi_{o}}
\end{equation}
where each term $W_{i}$ has an (asymptotic) expansion in the running
coupling $v$. (A similar trans-series of the observable $O_{c1}$
misses the $M\nu$ term.) It is plausible that this trans-series is a
resurgent one and the asymptotic behaviour of the perturbative coefficients of
$W_{0}$ carries all information on the expansion of the higher
terms $W_{i>0}$. We have verified this assertion in the $N=4$ case
\cite{Abbott:2020qnl} up to $4^{\rm th}$ order in $\nu$.

Let us now investigate the term $M$, which should determine the leading
asymptotics of the perturbative series. It is given by (\ref{nagyM}) and
it behaves for $N\geq4$ as :
\begin{equation}
M=\mathrm{real\times}e^{i\pi\Delta}\quad;\quad\Delta^{-1}=N-2
\end{equation}
where we indicated merely the reality properties. 
\begin{itemize}
\item For $N=4$ the term $M$ is purely imaginary and was completely recovered
from the asymptotics in \cite{Abbott:2020qnl}. Actually, this model is the
$SU(2)$ principal chiral model at the same time, so we see behaviour typical
for that class of models, see \cite{DiPietro:2021yxb,Marino:2021dzn} for
details about the principal chiral models. 
\item For $N>4$ the term $M$ is complex and only the imaginary
part was recovered from the asymptotics \cite{Marino:2021dzn}. The real part,
which is not related to the asymptotics, was also tested for various
$N$ by comparing to the difference between the numerical solution
of the integral equation and the median Borel resummation of the truncated
perturbative series. In \cite{DiPietro:2021yxb} the authors attributed this
mismatch of the real part to the different definitions of the free energy in
the perturbative and in the TBA approach. In particular, they differ
in an $h$-independent piece, and we have seen that $M$ corresponds
exactly to a constant term in $O_{cc}$. The perturbative expansion
uses vanishing free energy for $h=0$, while in the TBA approach the
free energy vanishes for $h=m$. The difference between the two is
the bulk energy constant \cite{Marino:2021dzn}
\begin{equation}
\epsilon_{\mathrm{bulk}}=\frac{m^{2}}{8}\cot(\pi\Delta).
\end{equation}
This is similar to what happens in the sine-Gordon model
\cite{Zamolodchikov:1995xk,Samaj:2013yva}. It is also analogous to the
mismatch in the ground-state energy between the conformal perturbation
theory and the TBA \cite{Zamolodchikov:1989cf}.
The same term, actually,  was observed for $N=6$ in analysing the cusp
anomalous dimension \cite{Basso:2009gh} as in this case the low energy
behaviour is controlled by the $O(6)$ model.

\item In the large $N$ limit the expression is purely real and cannot be
obtained from the asymptotics. This bulk energy constant can, however,
be tested against the contribution of the renormalon diagrams at large
$N$, see \cite{Marino:2021six}.
\end{itemize}
For other observables such as for $O_{c1}$ or for $\chi_{1}(B)$
we do not have this exceptional leading term, and most probably the
trans-series is a resurgent one. This was explicitly checked for the
leading term in the $O(4)$ model in \cite{Abbott:2020qnl}.

Let us focus now on the exceptional $O(3)$ model. The observable
$W$ has the trans-series expansion of the form\footnote{Note that $W_l$ already
contains the contribution coming from $M_{l-1}$, too.}
\begin{equation}
W=\sum_{l=0}^{\infty}W_{l}\nu^{l}=W_{0}+\nu W_{1}+\dots
\end{equation}
where 
\begin{equation}
W_{1}=-\frac{i\pi}{e}+\mathrm{real}(v)
\end{equation}
and again, we merely indicated the reality structure. The result has
a constant imaginary part, and a highly non-trivial running coupling-dependent
real contribution. As we observed in \cite{Bajnok:2021zjm} the imaginary part
can be recovered from the asymptotics of the perturbative expansion of
$W_{0}$, but the real part can not. This also follows from the fact
that the perturbative coefficients are all real, so their leading
non-perturbative contributions must be purely imaginary \cite{Aniceto:2013fka}. 

In the following section we analyze the real part of $W$ and various
other observables in the $O(3)$ model in more detail. 



\newcommand{\galphaminusone}{2.0000(1 \! \pm \! 5)}
\newcommand{\galphaln}{1.000000(7 \! \pm \! 1)}
\newcommand{\galphanull}{-1.0429515(9 \! \pm \! 2)}
\newcommand{\galphaone}{0.4999999(4\! \pm \! 5)}
\newcommand{\galphatwo}{-0.000000(2\!\pm\! 4)}
\newcommand{\galphathree}{0.01230(6 \! \pm \! 4)}
\newcommand{\galphafour}{0.0420(1 \!\pm\! 3)}
\newcommand{\galphafive}{0.090(2\! \pm \! 1)}

\newcommand{\bvnull}{1.00000000000   (02\! \pm \! 15 )}
\newcommand{\rbvnull}{1.5\times 10^{-12}}
\newcommand{\pbvnull}{\text{10$\%$}}
\newcommand{\bvone}{2.8294415416   (81\! \pm \! 27 )}
\newcommand{\rbvone}{9.6\times 10^{-12}}
\newcommand{\pbvone}{\text{6$\%$}}
\newcommand{\bvtwo}{2.610691541   (6\! \pm \! 4 )}
\newcommand{\rbvtwo}{1.6\times 10^{-10}}
\newcommand{\pbvtwo}{\text{10$\%$}}
\newcommand{\bvthree}{0.1335131   (70\! \pm \! 19 )}
\newcommand{\rbvthree}{1.4\times 10^{-7}}
\newcommand{\pbvthree}{\text{4$\%$}}
\newcommand{\bvfour}{-1.009192   (66\! \pm \! 35 )}
\newcommand{\rbvfour}{3.4\times 10^{-7}}
\newcommand{\pbvfour}{\text{20$\%$}}
\newcommand{\bvfive}{0.45939   (1\! \pm \! 4 )}
\newcommand{\rbvfive}{8.5\times 10^{-6}}
\newcommand{\pbvfive}{\text{4$\%$}}
\newcommand{\bvsix}{2.64409\pm 0.00006}
\newcommand{\rbvsix}{2.2\times 10^{-5}}
\newcommand{\pbvsix}{\text{2$\%$}}
\newcommand{\bvseven}{6.50490\pm 0.00031}
\newcommand{\rbvseven}{4.8\times 10^{-5}}
\newcommand{\pbvseven}{\text{40$\%$}}
\newcommand{\bveight}{25.343\pm 0.005}
\newcommand{\rbveight}{1.9\times 10^{-4}}
\newcommand{\pbveight}{\text{50$\%$}}
\newcommand{\bvnine}{123.652\pm 0.029}
\newcommand{\rbvnine}{2.3\times 10^{-4}}
\newcommand{\pbvnine}{\text{70$\%$}}
\newcommand{\bvten}{656.1\pm 0.6}
\newcommand{\rbvten}{9.8\times 10^{-4}}
\newcommand{\pbvten}{\text{80$\%$}}
\newcommand{\bunull}{1.00000000000   (03\! \pm \! 22 )}
\newcommand{\rbunull}{2.2\times 10^{-12}}
\newcommand{\pbunull}{\text{10$\%$}}
\newcommand{\buone}{2.5225887222   (4\! \pm \! 4 )}
\newcommand{\rbuone}{1.5\times 10^{-11}}
\newcommand{\pbuone}{\text{10$\%$}}
\newcommand{\butwo}{2.303838722   (2\! \pm \! 6 )}
\newcommand{\rbutwo}{2.5\times 10^{-10}}
\newcommand{\pbutwo}{\text{5$\%$}}
\newcommand{\buthree}{0.5806790   (83\! \pm \! 28 )}
\newcommand{\rbuthree}{4.7\times 10^{-8}}
\newcommand{\pbuthree}{\text{9$\%$}}
\newcommand{\bufour}{-0.338059   (3\! \pm \! 4 )}
\newcommand{\rbufour}{1.3\times 10^{-6}}
\newcommand{\pbufour}{\text{20$\%$}}
\newcommand{\bufive}{0.49658   (3\! \pm \! 6 )}
\newcommand{\rbufive}{1.2\times 10^{-5}}
\newcommand{\pbufive}{\text{4$\%$}}
\newcommand{\busix}{3.18624\pm 0.00008}
\newcommand{\rbusix}{2.4\times 10^{-5}}
\newcommand{\pbusix}{\text{10$\%$}}
\newcommand{\buseven}{11.4120\pm 0.0004}
\newcommand{\rbuseven}{3.9\times 10^{-5}}
\newcommand{\pbuseven}{\text{30$\%$}}
\newcommand{\bueight}{46.492\pm 0.006}
\newcommand{\rbueight}{1.4\times 10^{-4}}
\newcommand{\pbueight}{\text{40$\%$}}
\newcommand{\bunine}{220.05\pm 0.04}
\newcommand{\rbunine}{1.9\times 10^{-4}}
\newcommand{\pbunine}{\text{40$\%$}}
\newcommand{\buten}{1163.0\pm 1.0}
\newcommand{\rbuten}{8.7\times 10^{-4}}
\newcommand{\pbuten}{\text{70$\%$}}
\newcommand{\bwnull}{1.00000000000   (0\! \pm \! 5 )}
\newcommand{\rbwnull}{5.2\times 10^{-12}}
\newcommand{\pbwnull}{\text{4$\%$}}
\newcommand{\bwone}{1.7500000000   (2\! \pm \! 9 )}
\newcommand{\rbwone}{5.2\times 10^{-11}}
\newcommand{\pbwone}{\text{20$\%$}}
\newcommand{\bwtwo}{1.53125000   (03\! \pm \! 17 )}
\newcommand{\rbwtwo}{1.1\times 10^{-9}}
\newcommand{\pbwtwo}{\text{20$\%$}}
\newcommand{\bwthree}{1.2895635   (0\! \pm \! 6 )}
\newcommand{\rbwthree}{4.3\times 10^{-8}}
\newcommand{\pbwthree}{\text{20$\%$}}
\newcommand{\bwfour}{1.892612   (3\! \pm \! 8 )}
\newcommand{\rbwfour}{4.2\times 10^{-7}}
\newcommand{\pbwfour}{\text{10$\%$}}
\newcommand{\bwfive}{4.432524\pm 0.000013}
\newcommand{\rbwfive}{3.0\times 10^{-6}}
\newcommand{\pbwfive}{\text{2$\%$}}
\newcommand{\bwsix}{13.26959\pm 0.00029}
\newcommand{\rbwsix}{2.2\times 10^{-5}}
\newcommand{\pbwsix}{\text{30$\%$}}
\newcommand{\bwseven}{47.5311\pm 0.0014}
\newcommand{\rbwseven}{2.9\times 10^{-5}}
\newcommand{\pbwseven}{\text{10$\%$}}
\newcommand{\bweight}{197.729\pm 0.014}
\newcommand{\rbweight}{7.1\times 10^{-5}}
\newcommand{\pbweight}{\text{20$\%$}}
\newcommand{\bwnine}{934.15\pm 0.17}
\newcommand{\rbwnine}{1.8\times 10^{-4}}
\newcommand{\pbwnine}{\text{7$\%$}}
\newcommand{\bwten}{4927.9\pm 3.5}
\newcommand{\rbwten}{7.2\times 10^{-4}}
\newcommand{\pbwten}{\text{40$\%$}}
\newcommand{\cvnull}{0.999999999   (6\! \pm \! 8 )}
\newcommand{\rcvnull}{7.9\times 10^{-10}}
\newcommand{\pcvnull}{\text{60$\%$}}
\newcommand{\cvone}{2.579441541   (80\! \pm \! 13 )}
\newcommand{\rcvone}{5.0\times 10^{-11}}
\newcommand{\pcvone}{\text{90$\%$}}
\newcommand{\cvtwo}{2.32944154   (00\! \pm \! 33 )}
\newcommand{\rcvtwo}{1.4\times 10^{-9}}
\newcommand{\pcvtwo}{\text{50$\%$}}
\newcommand{\cvthree}{0.7470019   (21\! \pm \! 21 )}
\newcommand{\rcvthree}{2.8\times 10^{-8}}
\newcommand{\pcvthree}{\text{80$\%$}}
\newcommand{\cvfour}{-0.775631   (45\! \pm \! 14 )}
\newcommand{\rcvfour}{1.8\times 10^{-7}}
\newcommand{\pcvfour}{\text{600000000$\%$}}
\newcommand{\cvfive}{-0.15008   (63\! \pm \! 34 )}
\newcommand{\rcvfive}{2.3\times 10^{-5}}
\newcommand{\pcvfive}{\text{4000000$\%$}}
\newcommand{\cvsix}{3.11819\pm 0.00011}
\newcommand{\rcvsix}{3.4\times 10^{-5}}
\newcommand{\pcvsix}{\text{3000000$\%$}}
\newcommand{\cvseven}{11.197\pm 0.004}
\newcommand{\rcvseven}{3.4\times 10^{-4}}
\newcommand{\pcvseven}{\text{300000$\%$}}
\newcommand{\cveight}{37.62\pm 0.04}
\newcommand{\rcveight}{1.1\times 10^{-3}}
\newcommand{\pcveight}{\text{90000$\%$}}
\newcommand{\cvnine}{175.1\pm 0.6}
\newcommand{\rcvnine}{3.4\times 10^{-3}}
\newcommand{\pcvnine}{\text{30000$\%$}}
\newcommand{\cvten}{-897.\pm 9.}
\newcommand{\rcvten}{9.5\times 10^{-3}}
\newcommand{\pcvten}{\text{10000$\%$}}
\newcommand{\cveleven}{60739.\pm 120.}
\newcommand{\rcveleven}{2.0\times 10^{-3}}
\newcommand{\pcveleven}{\text{50000$\%$}}
\newcommand{\cunull}{0.99999999   (94\! \pm \! 11 )}
\newcommand{\rcunull}{1.1\times 10^{-9}}
\newcommand{\pcunull}{\text{50$\%$}}
\newcommand{\cuone}{2.272588722   (40\! \pm \! 22 )}
\newcommand{\rcuone}{9.7\times 10^{-11}}
\newcommand{\pcuone}{\text{70$\%$}}
\newcommand{\cutwo}{2.02258872   (0\! \pm \! 5 )}
\newcommand{\rcutwo}{2.6\times 10^{-9}}
\newcommand{\pcutwo}{\text{50$\%$}}
\newcommand{\cuthree}{1.1078654   (73\! \pm \! 35 )}
\newcommand{\rcuthree}{3.2\times 10^{-8}}
\newcommand{\pcuthree}{\text{70$\%$}}
\newcommand{\cufour}{0.159218   (93\! \pm \! 21 )}
\newcommand{\rcufour}{1.3\times 10^{-6}}
\newcommand{\pcufour}{\text{70000000$\%$}}
\newcommand{\cufive}{0.49126   (6\! \pm \! 6 )}
\newcommand{\rcufive}{1.2\times 10^{-5}}
\newcommand{\pcufive}{\text{8000000$\%$}}
\newcommand{\cusix}{3.86889\pm 0.00015}
\newcommand{\rcusix}{3.8\times 10^{-5}}
\newcommand{\pcusix}{\text{3000000$\%$}}
\newcommand{\cuseven}{17.017\pm 0.006}
\newcommand{\rcuseven}{3.4\times 10^{-4}}
\newcommand{\pcuseven}{\text{300000$\%$}}
\newcommand{\cueight}{70.85\pm 0.06}
\newcommand{\rcueight}{8.7\times 10^{-4}}
\newcommand{\pcueight}{\text{100000$\%$}}
\newcommand{\cunine}{311.7\pm 0.9}
\newcommand{\rcunine}{2.9\times 10^{-3}}
\newcommand{\pcunine}{\text{30000$\%$}}
\newcommand{\cuten}{-1457.\pm 15.}
\newcommand{\rcuten}{1.0\times 10^{-2}}
\newcommand{\pcuten}{\text{10000$\%$}}
\newcommand{\cueleven}{106016.\pm 283.}
\newcommand{\rcueleven}{2.7\times 10^{-3}}
\newcommand{\pcueleven}{\text{40000$\%$}}
\newcommand{\cwnull}{0.99999999   (8\! \pm \! 4 )}
\newcommand{\rcwnull}{3.8\times 10^{-9}}
\newcommand{\pcwnull}{\text{60$\%$}}
\newcommand{\cwone}{1.500000000   (4\! \pm \! 8 )}
\newcommand{\rcwone}{5.0\times 10^{-10}}
\newcommand{\pcwone}{\text{50$\%$}}
\newcommand{\cwtwo}{1.2499999   (93\! \pm \! 16 )}
\newcommand{\rcwtwo}{1.3\times 10^{-8}}
\newcommand{\pcwtwo}{\text{40$\%$}}
\newcommand{\cwthree}{1.599459   (29\! \pm \! 12 )}
\newcommand{\rcwthree}{7.5\times 10^{-8}}
\newcommand{\pcwthree}{\text{50$\%$}}
\newcommand{\cwfour}{2.819320   (4\! \pm \! 6 )}
\newcommand{\rcwfour}{2.3\times 10^{-7}}
\newcommand{\pcwfour}{\text{400000000$\%$}}
\newcommand{\cwfive}{6.571694\pm 0.000021}
\newcommand{\rcwfive}{3.2\times 10^{-6}}
\newcommand{\pcwfive}{\text{30000000$\%$}}
\newcommand{\cwsix}{19.22654\pm 0.00032}
\newcommand{\rcwsix}{1.7\times 10^{-5}}
\newcommand{\pcwsix}{\text{6000000$\%$}}
\newcommand{\cwseven}{68.751\pm 0.016}
\newcommand{\rcwseven}{2.3\times 10^{-4}}
\newcommand{\pcwseven}{\text{400000$\%$}}
\newcommand{\cweight}{290.84\pm 0.16}
\newcommand{\rcweight}{5.4\times 10^{-4}}
\newcommand{\pcweight}{\text{200000$\%$}}
\newcommand{\cwnine}{1529.4\pm 2.8}
\newcommand{\rcwnine}{1.8\times 10^{-3}}
\newcommand{\pcwnine}{\text{50000$\%$}}
\newcommand{\cwten}{-6313\pm 61}
\newcommand{\rcwten}{9.7\times 10^{-3}}
\newcommand{\pcwten}{\text{10000$\%$}}
\newcommand{\cweleven}{(4.055\pm 0.014)\times 10^5}
\newcommand{\rcweleven}{3.4\times 10^{-3}}
\newcommand{\pcweleven}{\text{30000$\%$}}

\section{The $O(3)$ sigma model revisited}

As an application of our method, we proceed by calculating further
perturbative coefficients for the leading exponential correction of
the standard ratio of the energy density $\epsilon = m^2  O_{cc}$
and the square of the number density $\rho = m O_{1c}$; that is $\epsilon/\rho^{2} =  O_{cc}/O_{1c}^{2}$.
Starting from the solution of the $O_{cc}$ problem up to the leading
exponential correction, i.e. $W_{0},W_{1}$ - it is possible to reconstruct
the series $\hat{\rho}_{1}$ for $O_{c1}$ by using \eqref{eq:master}.
The two relevant equations are:
\begin{align}
\dot{O}_{1c} & =\frac{1}{\pi}\chi_{1}(B)\chi_{c}(B) & \dot{O}_{cc} & =\frac{1}{\pi}\chi_{c}^{2}(B)\label{eq:derivativeOij}
\end{align}
where $\chi_{1}(B)$ is the only expression, which does not contain a term of order $e^{-2B}$: 
\begin{align}
O_{cc} & =\frac{e^{2B}G_{+}^{2}(i)}{8\pi}\left(W_{0}+e^{-2B}W_{1}+\mathcal{O}(e^{-4B})\right); & \chi_{c}(B) & =\frac{e^{B}G_{+}(i)}{2}\left(w_{0}+e^{-2B}w_{1}+\mathcal{O}(e^{-4B})\right);\nonumber\\
O_{1c} & =\frac{e^{B}G_{+}(i)}{4\pi\sqrt{v}}\left(\hat{\rho}_{0}-e^{-1}e^{-2B}\hat{\rho}_{1}+\mathcal{O}(e^{-4B})\right); & \chi_{1}(B) & =\frac{1}{2\sqrt{v}}\left(\hat{u}_{0}+\mathcal{O}(e^{-4B})\right).
\end{align}
 After expanding the relations (\ref{eq:derivativeOij}), one gets
two equalities for the energy density part - by matching $\mathcal{O}(1)$
and $\mathcal{O}(e^{-2B})$ terms:
\begin{align}
W_{0}+\frac{1}{2}\dot{W}_{0} & =w_{0}^{2}\ ; & \dot{W}_{1} & =4w_{0}w_{1}\label{eq:wcond}
\end{align}
and another two for the number density problem:
\begin{align}
\hat{\rho}_{0}+\dot{\hat{\rho}}_{0}-\frac{\dot{v}}{2v}\hat{\rho}_{0} & =\hat{u}_{0}w_{0}\ ; & -\hat{\rho}_{1}+\dot{\hat{\rho}}_{1}-\frac{\dot{v}}{2v}\hat{\rho}_{1} & =-e\hat{u}_{0}w_{1}.\label{eq:ucond}
\end{align}
After expressing the ratio $w_{1}/w_{0}$ from both sets of equations
(\ref{eq:wcond}) and (\ref{eq:ucond}), and switching to $v$-derivatives,
one can easily reconstruct the series $\hat{\rho}_{1}$ from $\hat{\rho}_{0},W_{0}$
and $W_{1}$. 

The conditions (\ref{eq:wcond}) are also sufficient to establish some
relations among the moments $\Sigma_{k,s}$ and $\Omega_{k,s}$ defined
in (\ref{Omegaks}) and (\ref{Sigmaks}). In the end, one only has to calculate $\Omega_{0,2}$
and the linear combination $\Sigma_{1,1}-\Omega_{1,1}$ analytically
(see Appendix \ref{appA}), all the other moments needed to calculate
$W_{1}$ up to the $\mathcal{O}(v^{3})$ term can be obtained from
$W_{0}$. To go beyond these orders, some increasingly harder problems must be solved
via the method presented in Appendix \ref{appA}, thus we stopped at this
- feasible - order.

The missing pieces are then the moments appearing in $W_{0}$ and
$\hat{\rho}_{0}$. Both quantities are in the perturbative sector,
thus Volin's algorithm is suitable to generate their expansion. A
slight modification of the algorithm is
needed to produce the result in terms of the running coupling instead
of a $B^{-1},\ln B$ trans-series expansion.\footnote{We used the same trick to calculate the capacitance immediately as
a power series in $v$ (see Section \ref{sec:capacitance}).}

At this point, we also have to comment on the running coupling 
\begin{equation}
2B=\frac{1}{v}+\ln v+L\label{eq:runningcouplingwithL}
\end{equation}
being used here. The original definition based on perturbation theory
of the $\alpha$ running coupling in \cite{Bajnok:2008it} - see (\ref{eq:volinalpha}) - 
can be rewritten as 
\begin{equation}
2B=\frac{2}{\alpha}+\ln v-\ln\left(\frac{8}{e}\right)-\ln\hat{\rho}^{2}(v).
\end{equation}

If we choose $L=-\ln\left(\frac{8}{e}\right)$ (let us denote this special coupling by $\beta$ instead of the generic $v$),
then the two couplings - up to a trivial factor of two - differ only
in $\mathcal{O}(\beta^{3})$ terms from each other:\footnote{Note that $\hat{\rho}(\beta)$ is an exact quantity here, it also contains
exponential, non-perturbative corrections. It is necessary to take
the $\mathcal{O}(e^{-1/\beta})$ term into account when calculating (\ref{eq:chiO3}).}
\begin{equation}
2B=\frac{1}{\beta}+\ln \beta-\ln\left(\frac{8}{e}\right)\quad\Rightarrow\quad\alpha=\frac{2\beta}{1+\beta\ln\hat{\rho}^{2}(\beta)}.\label{eq:specialL}
\end{equation}
 Also empirically, the coefficients in the above set of series of
the densities are then free of any $\ln2$-s; it seems to be advisable
then to pick this gauge for the intermediate analytical calculations.
However, we are rather presenting our results for the $O(3)$ sigma
model in terms of $\alpha$:
\begin{align}
	& \chi  = \frac{\epsilon}{\pi\rho^{2}}  =S_{+}\left(\alpha+\frac{\alpha^{2}}{2}+\frac{\alpha^{3}}{2}+\mathcal{O}(\alpha^{4})\right)+\label{eq:chiO3}\\
 & +\frac{32}{e^{2}}e^{-2/\alpha}S_+\left\{ -\frac{i\pi}{2}+\frac{2}{\alpha}+\left(3-\gamma_{E}-5\ln2\right)+\ln\alpha+\frac{\alpha}{2}+0\cdot\alpha^{2}+\left(\frac{1}{8}-\frac{3}{32}\zeta_{3}\right)\alpha^{3}+\mathcal{O}(\alpha^{4})\right\}\nonumber\\
	& + \mathcal{O}(e^{-4/\alpha}),\nonumber
\end{align}
as we would like to compare them to \cite{Marino:2021dzn} and  \cite{Bajnok:2021zjm}. Our
new results are the coefficients of the $\alpha^{2}$ and $\alpha^{3}$
terms of the non-perturbative sector; as one can see, the first turned
out to be zero. The notation $S_{+}$ here means the lateral Borel
resummation of the perturbative part \cite{Bajnok:2021zjm}. Its
leading imaginary term is $+i\frac{16\pi}{e^{2}}e^{-2/\alpha}$, which comes
from a simple pole on the Borel plane. The Borel transform of $\chi$ is defined here as
\begin{equation}
	B(\chi) = \sum_{n=1}^\infty c_n s^{n-1}, \quad c_n = \frac{\chi^{(0)}_n 2^{n-1} }{\Gamma(n)} ,
	\label{eq:BorelO3}
\end{equation}
where $\chi^{(0)}_n$-s are the perturbative coefficients of $\chi$; and this singularity is at $s = 1$. This pole term drops out between the resummation and the explicit second term calculated by our method in (\ref{eq:chiO3}). 

Let us now turn to the numerical test of the above result. As it
was noted in \cite{Bajnok:2021zjm}, originally we found the leading exponentially
suppressed contribution in (\ref{eq:chiO3}) numerically, and it turned
out to be inexplainable by the resurgence of the perturbative coefficients.
While for $N\geq4$ strong resurgence (as defined in \cite{DiPietro:2021yxb}), i.e. that the full trans-series can be reconstructed from perturbation
theory up to overall constants, seems to be at work, here the leading non-perturbative correction is of a different origin. It can be attributed to
instanton effects \cite{Marino:2022ykm}.

To be able to compare the Borel-resummation of the perturbative sector
to the exact solution, we solved the TBA integral equation
for $\chi_{c}(\theta)$ at some $B$ values in a suitable range, 
$9\leq B\leq15$. The precision of this solution was sufficient
for comparing terms of the order of magnitude $10^{-14}\lesssim e^{-2B}\lesssim 10^{-8}$.
By expanding $\chi_{c}(\theta)$ in rapidity space over the basis
of even Chebyshev polynomials up to some high order (e.g. we used $n_\text{cut} = 352$), one can reformulate
the problem (\ref{TBA}) as a matrix inversion. The precision
can be improved by increasing the polynomial order at which one cuts
the basis \cite{Abbott:2020qnl}.\footnote{The error estimate of the TBA solution was around $10^{-75}$ for the largest $B$ values.}

On the other hand, we used the perturbative sector of the quantity
$\chi$ in (\ref{eq:chiO3}) generated by Volin's algorithm \cite{Volin:2010cq},
to analyze its asymptotic behavior, and calculate the lateral Borel resummation. We were able to obtain the first 336 of these coefficients, by optimizing the algorithm - and also, instead of calculating with exact symbolic values, we used high precision numerics (a few thousand digits) from the start.

The coefficients $c_n$ in (\ref{eq:BorelO3}) have the following asymptotic structure:
\begin{equation}
	c_n = \frac{8}{e^2} + \frac{n}{2^{n-1}}\bigg\lbrace a_0 + \frac{a_1}{n}+ \frac{a_2}{n(n-1)}+\ldots+\frac{a_k}{n(n-1)\ldots(n-k+1)}+\mathcal{O}(1/n^{k+1}) \bigg\rbrace
	\label{eq:O3asymp}
\end{equation}
As explained in \cite{Bajnok:2021zjm} it is possible to acquire 20 of these asymptotic coefficients, i.e. the perturbative coefficients in the expression of the first alien derivative \cite{Abbott:2020qnl} of the $\chi(\alpha)$ function at the closest logarithmic cut on its Borel plane starting from $s=2$. This alien derivative gives the imaginary part of the Borel resummation, which is defined as:
\begin{equation}
	S_+(\chi) = 2\int_0^{e^{i\varphi} \infty} B(\chi) e^{-2 s/\alpha} \mathrm{d}s,
\label{eq:lateralBorel}
\end{equation}
$\varphi$ being the angle of the integration contour and the x-axis.\footnote{\label{foot:Borel}Here we evaluated the integral at $\varphi = 1/2$, which was chosen such that the line avoids the spurious poles of the Pad\'e-approximant of the series $B(\chi)$ in (\ref{eq:BorelO3}). For the same reason, an extra conformal transformation was used as well \cite{Bajnok:2021zjm}. The error in the numerical evaluation of the integral was estimated to be around $10^{-30}$.} 

The coefficients in (\ref{eq:O3asymp}) are also growing factorially:
\begin{equation}
	\frac{a_{n-1}}{a_0} = n! \left(b_0+ \frac{b_1}{n}+\mathcal{O}(n^{-2})\right),
\end{equation}
and the leading singularity on their Borel plane is at the same point $s=2$. Here $b_0$ and $b_1$ are some asymptotic coefficients; we could not estimate the further ones reliably by our numerics. 
The theory of resurgence \cite{Abbott:2020qnl} then would predict a difference of the median
resummation - which we think of as the trans-series expansion of the
exact solution $\chi_{\mathrm{TBA}}$ of the integral equation - and
the real part of the lateral Borel resummation (\ref{eq:lateralBorel}). According to the theory - see also (\ref{eq:Smeddiff}) - the magnitude of this difference should be: 
\begin{equation}
S_{\mathrm{med}}(\chi)-\mathrm{Re}\,S_{+}(\chi)\propto a_0 b_0 e^{-8/\alpha}.
\end{equation}
What one finds instead, is that the difference is much bigger, i.e. $\propto e^{-2/\alpha}$; and
the analytical results in (\ref{eq:chiO3}) are in good agreement
with our numerical findings (see Table \ref{tab:gs}). Namely, we could fit\footnote{As the series multiplying the exponential factor in (\ref{eq:realChi}) is asymptotic, we fitted its coefficients order by order. That is, we subtracted the already known terms (preferably the exact value), and determined the leading coefficient at each step. The latter was done by fitting a series of polynomials with an increasing degree to the remaining part; the leading coefficient then stabilized around the correct value.} the new $g_{2},g_{3}$ coefficients up to 4 digits, where the $g$-s are defined as:
\begin{equation}
\chi_{\mathrm{TBA}}-\mathrm{Re}\,S_{+}(\chi)=\frac{32}{e^{2}}e^{-2/\alpha}\left\{ g_{-1}\alpha^{-1}+g_{\ln}\ln\alpha+g_{0}+g_{1}\alpha+g_{2}\alpha^{2}+g_{3}\alpha^{3}+\mathcal{O}(\alpha^{4})\right\} .
\label{eq:realChi}
\end{equation}
\begin{table}[h]
\centering{}%
\begin{tabular}{ll}
\hline 
exact value & numerical estimate\tabularnewline
\hline 
\hline 
$g_{-1}=2$ & $\quad\galphaminusone$\tabularnewline
$g_{\ln}=1$ & $\quad\galphaln$\tabularnewline
$g_{0}=3-\gamma_{E}-5\ln2 =-1.04295157\ldots  $ & $\,\galphanull$\tabularnewline
$g_{1}=\frac{1}{2}$ & $\quad\galphaone$\tabularnewline
$g_{2}=0$ & $\,\galphatwo$\tabularnewline
$g_{3}=\frac{1}{8}-\frac{3}{32}\zeta_{3} =  0.012307\ldots$ & $\quad\galphathree$\tabularnewline
$g_{4}=?$ & $\quad\galphafour$\tabularnewline
$g_{5}=?$ & $\quad\galphafive$\tabularnewline
\hline 
\end{tabular}\caption{The coefficients of the series multiplying the leading exponential
term in the difference of the numerical TBA solution and the lateral
	Borel resummation. The validity of the fits is shown in Figure \ref{fig:realpartO3}.}
\label{tab:gs}
\end{table}

\begin{figure}[t]
	\centering
	\includegraphics[width=.49\textwidth]{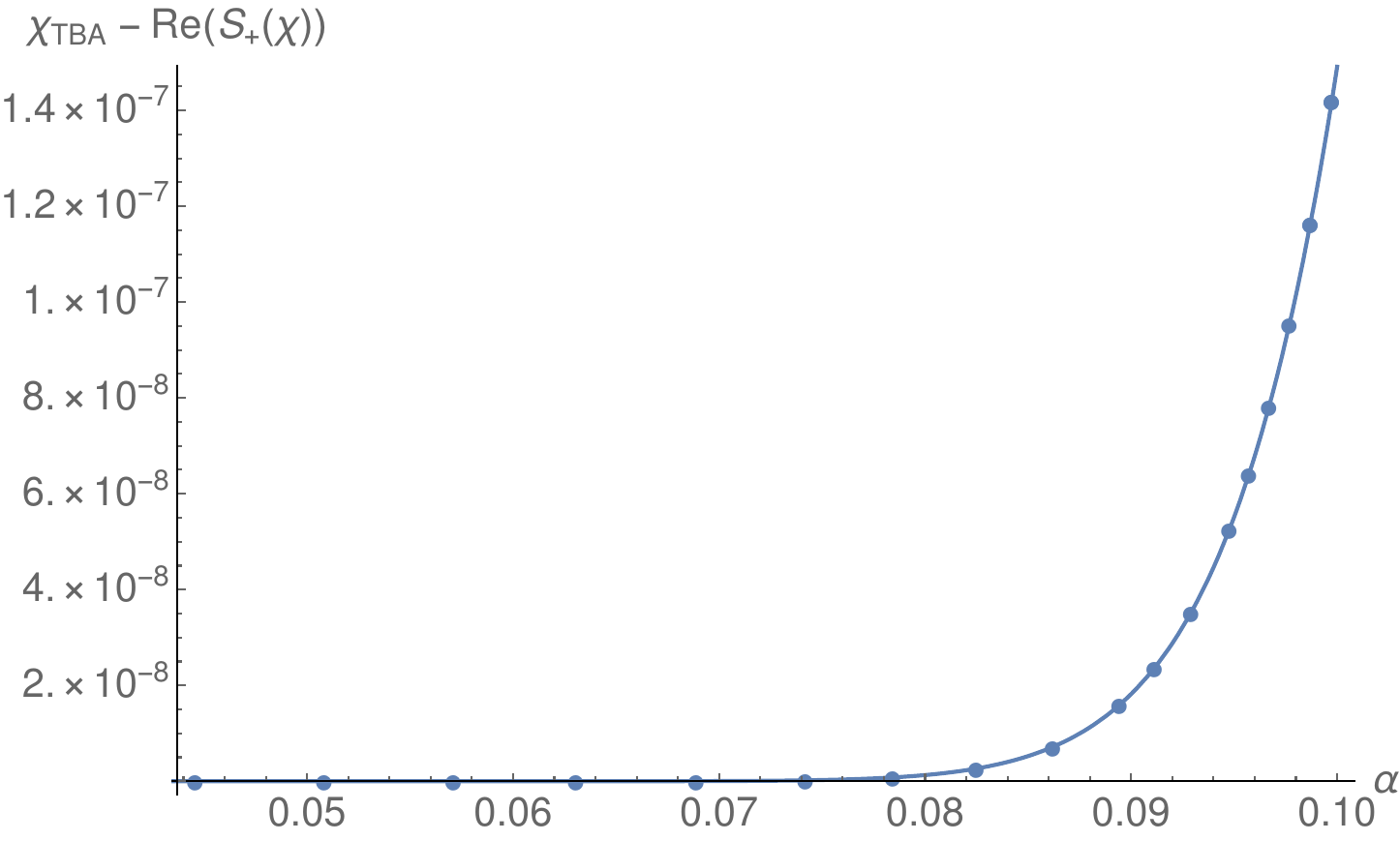}
	\includegraphics[width=.49\textwidth]{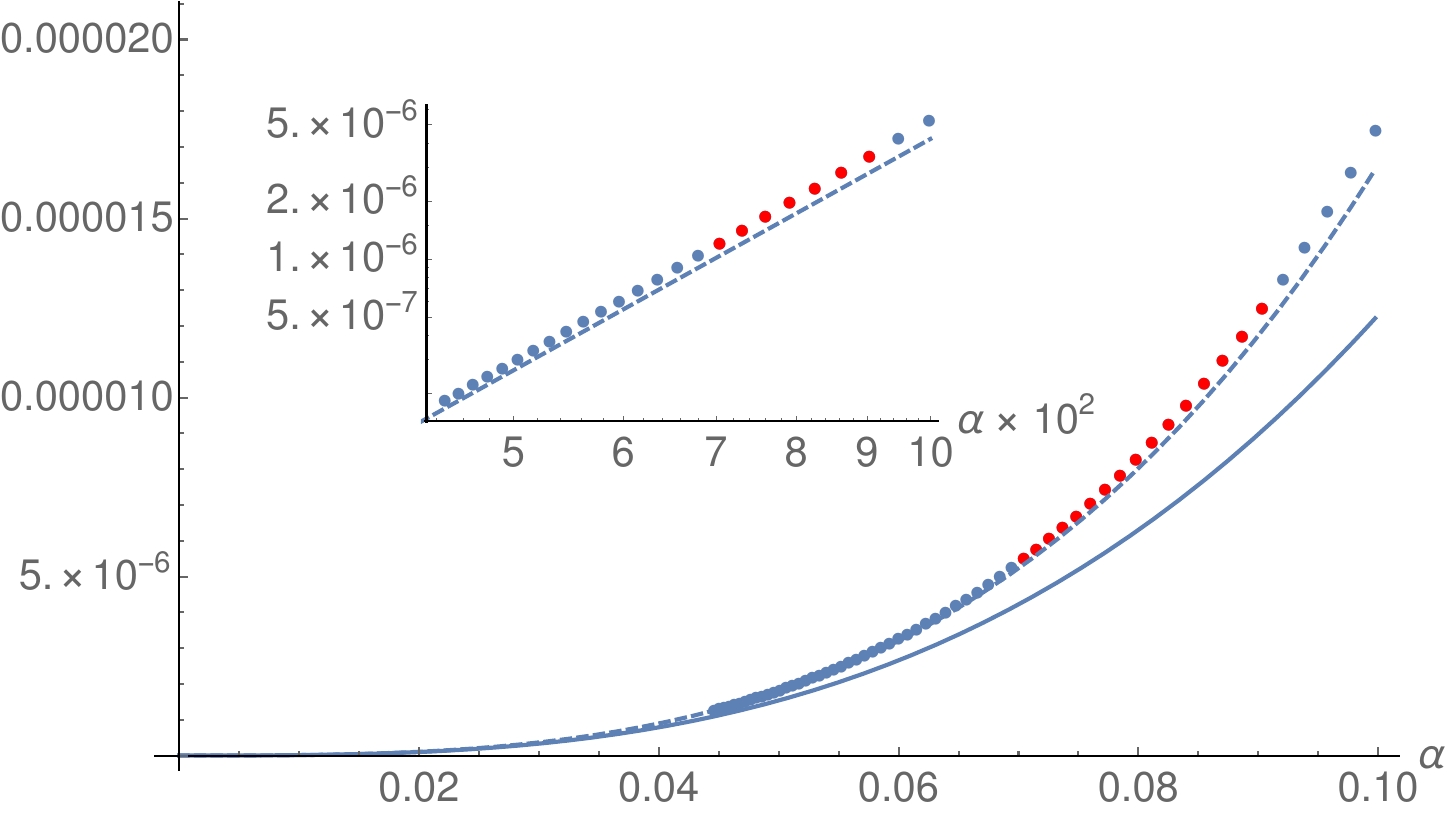}
	\caption{ On the left hand side, one can see the difference between the TBA result and the real part of the lateral Borel resummation. The points show the numerics, while the solid line is nothing but the formula (\ref{eq:realChi}) truncated at $g_3$ in the expansion. The comparison of the so far unknown coefficients $g_2$ and $g_3$  in Table \ref{tab:gs} to the subtracted numerical data $e^2/32  e^{2/\alpha} \left(\chi_\text{TBA}-S_+(\chi)\right) - \left(2/\alpha + (3-\gamma_E -5\ln 2) + \ln\alpha + \alpha/2 \right)$ is shown on the right. The points in red indicate the actual region over which the fit was performed (this was restricted, due to a substantial improvement in the error estimates of the coefficients). The continuous line is simply $\left(1/8-3\zeta_3/32\right)\alpha^3$, while the dashed line also contains the next term (the value of $g_4$ is numerical here): $g_3 \alpha^3 + g_4\alpha^4$. The inset shows the difference between the numerical data and the $g_3 \alpha^3$ term on a log-log scale; the steepness of this line is consistent with the next term $g_4 \alpha^4$ (dashed line).
In order to maintain visibility we plotted only every fifth data point in the
inset.}
	\label{fig:realpartO3}
\end{figure}

\section{Disk capacitor problem}
\label{sec:capacitance}

As explained in the introduction, we deal with the capacitance $C_{\text{phys}}$
of two oppositely charged, thin coaxial disks of radius $a$ at distance
$d$ from each other. This quantity goes to
\begin{equation}
C_{\text{phys}}\to\epsilon_{0}\frac{\pi a^{2}}{d}
\end{equation}
for $d\ll a$, i.e. to the well-known formula for parallel-plate capacitors;
where $\epsilon_{0}$ is the vacuum permittivity. However, for finite
$d$, the dependence on the distance is not so trivial. The problem
can be solved exactly by Love's equation (1949) \cite{Love:1949} (see a particularly simple derivation in \cite{Felderhof:2013}):
\begin{equation}
f(x,\kappa)-\frac{\kappa}{\pi}\int_{-1}^{1}\mathrm{d}y\frac{f(y,\kappa)}{\kappa^{2}+(y-x)^{2}}=1,
\label{Loveeq}
\end{equation}
where $\kappa\equiv d/a$, and the (dimensionless) capacitance can
be expressed as an integral of the solution to this equation over
the given domain:
\begin{equation}
C\equiv\frac{C_{\text{phys}}}{4\epsilon_{0}d},\qquad C=\frac{1}{2\kappa}\int_{-1}^{1}\mathrm{d}x\,f(x,\kappa).
\end{equation}
Note that the integral equation above is nothing else, but the $\chi_{1}$
problem for the $O(3)$ kernel up to some identifications, and thus
the capacitance is equivalent to the density $O_{11}$:
\begin{equation}
B\equiv\frac{\pi}{\kappa},\quad\chi_{1}(\theta)\equiv f(\kappa\,\theta/\pi,\kappa),\quad C\equiv O_{11}.
\end{equation}

More recently, effective algorithms have been developed
\cite{Marino:2019fuy,Reichert:2020ymc} for producing a
large- and small-$\kappa$ expansion for the
capacitance. The latter one is based on a modification of Volin's
method \cite{Volin:2010cq} for the $r_{1}(\theta)=1$ source. The first nine
coefficients of this expansion were given in
\cite{Reichert:2020ymc} analytically,
however, to be able to perform an asymptotic analysis, one needs a
lot more of them.

By using the relation (\ref{O11}) it is possible to reconstruct the
perturbative coefficients of the capacitance from the particle- and
energy-density of the $O(3)$ sigma model. As the running coupling
proved to be handy for eliminating logarithms, we make use of it from
the start. After using the relation among the derivatives (\ref{O11}) and
the expressions for the corresponding $O_{cc}$ and $O_{1c}$ 
(\ref{Occ}), (\ref{O1crho}) one gets 
\begin{equation}
	C'(v)=-\frac{1}{8\pi v^{3}}\frac{\left[ \hat{\rho}(v)-2v^{2}\hat{\rho}'(v)\right]^2}{(1-v)W(v)-v^{2}W'(v)}, 
\end{equation}
where we also had to exchange the variables via $\partial_B = 2 v^2/(v-1) \partial_v$. 

The above formula is exact, but we only used it to obtain the perturbative sector of the capacitance, which we denote by $C_0$. One has to integrate the series expansion of $C_0'(v)$ formally term-by-term, and account for the loss of a constant term $c_\mathrm{o} = (L-L_0)^2 + 2(L-L_0) - 1 $ - we fixed this integration constant
from the NNLO result derived first in \cite{wigglesworth1972comments,chew1982microstrip}. Here we introduced $L_0 \equiv -\ln(8/e)$, just to show that for $L=L_0$ - which corresponds to the choice in (\ref{eq:specialL}) - some terms disappear. The expansion itself looks as:
\begin{equation}
C_0'(v) = -\frac{1}{8\pi}\bigg\lbrace v^{-3}+\left(L-L_0 \right) v^{-2} - \left[\left(L-L_0\right) +\frac{1}{2}\left(L-L_0\right)^2+\frac{1}{4}-\frac{3}{4} \zeta_3 \right]+\mathcal{O}(v) \bigg\rbrace  
\end{equation}
and remarkably, the $1/v$ term is not present; thus there will be no $\ln(v)$ term in $C_0$:
\begin{align}
	C_0(v) =  \frac{1}{16\pi}\bigg\lbrace & v^{-2}+ 2\left(L-L_0 \right) v^{-1} + c_\mathrm{o} +\nonumber\\
	 &+ \left[2\left(L-L_0\right) + \left(L-L_0\right)^2+\frac{1}{2}-\frac{3}{2} \zeta_3 \right] v + \mathcal{O}(v^2) \bigg\rbrace
\end{align}
As a check, and to demonstrate its effectiveness, we may rewrite the same perturbative result shown
in \cite{Reichert:2020ymc} in terms of the special running coupling (\ref{eq:specialL}),
leading to a formula, which is completely free of logarithms:
\begin{align}
	C_0(\beta)=\frac{1}{16\pi}&\bigg\lbrace  \frac{1}{\beta^{2}}-1+\beta\left(\frac{1}{2}-\frac{3}{2}\zeta_{3}\right)+\beta^{2}\left(\frac{7}{6}-\frac{5}{2}\zeta_{3}\right)+\beta^{3}\left(\frac{1}{2}\zeta_{3}-\frac{135}{32}\zeta_{5}+\frac{11}{48}\right)+\nonumber\\
 & +\beta^{4}\left(\frac{73}{8}\zeta_{3}+\frac{81}{16}\zeta_{3}^{2}-\frac{819}{32}\zeta_{5}-\frac{67}{120}\right)+\nonumber\\
 & +\beta^{5}\left(\frac{25}{8}\zeta_{3}+\frac{705}{16}\zeta_{3}^{2}-\frac{1179}{32}\zeta_{5}-\frac{14175}{256}\zeta_{7}-\frac{23}{80}\right)+\nonumber\\
 & +\beta^{6}\left(-\frac{4651}{192}\zeta_{3}+\frac{433}{4}\zeta_{3}^{2}+\frac{17919}{128}\zeta_{5}+\frac{27135}{128}\zeta_{3}\zeta_{5}-\frac{161109}{256}\zeta_{7}+\frac{9449}{20160}\right)+\nonumber\\
 & +O(\beta^{7})\bigg\rbrace.
\label{C0beta}
\end{align}

Making use of the above method, we were able to ``recycle'' our
336 coefficients obtained for the $O(3)$ sigma model, and perform
an asymptotic analysis for the perturbative coefficients of the capacitance.
To be able to do this, we had to implement a slight modification (see Appendix \ref{app:Volin}) of the original algorithm described in Appendix E.3 of \cite{Volin:2010cq}. Otherwise, the numerical methods used in the asymptotic analysis, the lateral Borel resummation, and the exact solution of the TBA  were very similar to the case of the $O(3)$ non-linear sigma model. 

Let us then define the normalized quantity 
\begin{equation}
\Phi(\beta)  =16\pi \beta^{2}C_{0}(\beta)=\sum_{n=0}^{\infty}a_{n}\beta^{n}\label{eq:gammaseries}
\end{equation}
whose expansion starts with $a_{0}=1$, then its asymptotics looks like 
\begin{equation}
a_{n}=\frac{n!}{2^{n}}\left\{ A_{0}+\frac{A_{1}}{n}+\frac{A_{2}}{n(n-1)}+\frac{A_{3}}{n(n-1)(n-2)}+\mathcal{O}(n^{-4})\right\} \label{eq:capAsy}
\end{equation}
where the leading coefficient is\footnote{A similar analysis was performed
in \cite{Marino:2019fuy} for the Lieb-Liniger model.}
\begin{equation}
A_{0}=-\frac{2^{6}}{\pi e^{4}},
\end{equation}
and the asymptotic coefficients relative to this latter are:
\begin{equation}
\frac{A_{n}}{A_{0}}=1,\frac{7}{2},\frac{49}{8},\frac{149}{48}+6\zeta_{3},-\frac{6143}{384}+\frac{77}{2}\zeta_{3},\ldots.
\end{equation}
As in \cite{Abbott:2020qnl,Bajnok:2021zjm} we found these results numerically\footnote{With the help of successive Richardson transformations, as explained in \cite{Abbott:2020qnl}.},
then searched for a plausible expression on the basis of odd zeta-s
over the field of rationals. Similar to the $O(3)$ case, there were around 20 of these asymptotic coefficients, of which we could have a reliable numerical estimate. Then the ambiguity (the imaginary part of the lateral Borel resummation\footnote{Compared to the evaluation of the similar integral in the $O(3)$ case (\ref{eq:lateralBorel}), here we even made a Pad\'e-approximation after the conformal transformation mentioned in Footnote \ref{foot:Borel}, since without this step, the conformal transformation introduced spurious oscillations of unknown source in $S_+(\Phi)$ as a function of the coupling $\beta$.}) looks as:
\begin{equation}
	\mathrm{Im}\, S_{+}(\Phi)=-\frac{2^{7}}{e^{4}\beta}e^{-2/\beta}\left\{ b_{0}+b_{1}\beta+b_{2}\beta^{2}+b_{3}\beta^{3}+\mathcal{O}(\beta^4)\right\}  
\end{equation}
where the coefficients $b_n$ are in the following relation with the
asymptotic ones:
\begin{equation}
	b_{n}=\frac{A_{n}}{A_{0}}2^{-n},\quad b_{0}=1,\;b_{1}=\frac{7}{4},\; b_{2}=\frac{49}{32},\quad\ldots,\label{eq:bcoeffs}
\end{equation}
and we will show in the following, that eventually, this ambiguity seems to
get canceled by the leading exponential imaginary term obtained from
the Wiener-Hopf method: at least up to $b_{2}$, we could derive the
same coefficients analytically. 

To show this, we start again from the (\ref{eq:master}) relation for $O_{11}$,
and notice immediately, that since the leading exponential term in
$\chi_{1}(B)$ is of the order of $e^{-4B}$:
\begin{equation}
\dot{O}_{11}=\frac{1}{\pi}\chi_{1}^{2}(B),\quad\chi_{1}(B)=\frac{1}{2\sqrt{v}}\left(\hat{u}_{0}-ie^{-2}\hat{u}_{1}e^{-4B}+\mathcal{O}(e^{-8B})\right),
\end{equation}
therefore the capacitance itself has a similar structure: 
\begin{equation}
C=C_{0}+e^{-4B}C_{1}+\mathcal{O}(e^{-8B}),
\end{equation}
and also that $C_{1}$ must be imaginary. By matching the $\mathcal{O}(1)$
parts, in terms of the running coupling (\ref{eq:runningcouplingwithL}) we can express $C_{0}$ as:
\begin{equation}
\frac{\mathrm{d}C_{0}}{\mathrm{d}v}=-\frac{1-v}{8\pi v^{3}}\hat{u}_{0}^{2},
\end{equation}
and similarly, after matching the $e^{-4B}$ terms, one can reconstruct
$C_{1}$ from:
\begin{equation}
C_{1}+\frac{v^{2}}{2(1-v)}\frac{\mathrm{d}C_{1}}{\mathrm{d}v}=\frac{i}{8\pi e^{2}v}\hat{u}_{0}\hat{u}_{1}.
\end{equation}
We can calculate the perturbative part $\hat{u}_{0}$
by using the left equations of (\ref{eq:wcond}) and (\ref{eq:ucond}):
\begin{equation}
\hat{u}_{0}=1+\frac{1}{2}\beta+\frac{3}{8}\beta^{2}+\mathcal{O}(\beta^{3}).\label{eq:u0cap}
\end{equation}
Here, and throughout of this section, we use the special coupling (\ref{eq:specialL}).
Finally, the non-perturbative part $\hat{u}_{1}$ 
is given by (\ref{u1expansion}), (\ref{a12}), and (\ref{f1res}). The necessary
input is just the above expansion coefficients
$r_{1}=\frac{1}{2},\;r_{2}=\frac{3}{8}$ of $\hat{u}_{0}$ from (\ref{eq:u0cap}),
and $\Omega_{1}=\frac{13}{32}$  from the expansion of $w_{0}$ (\ref{w0pert}).
Then up to the same order in $\beta$: 
\begin{equation}
\hat{u}_{1}=1+\frac{3}{4}\beta+\frac{9}{32}\beta^{2}+\mathcal{O}(\beta^{3}).
\end{equation}
Thus the reconstructed $C_{1}$ is 
\begin{equation}
C_{1}=\frac{i}{8\pi e^{2}\beta}\left\{ 1+\frac{7}{4}\beta+\frac{49}{32}\beta^{2}+\mathcal{O}(\beta^{3})\right\} ,
\end{equation}
and as one can see the coefficients here match those in (\ref{eq:bcoeffs}).
Let us note that calculating one further coefficient still seems to
be feasible by the methods described in this paper.

The above findings support the claim, that the method originally developed in \cite{Marino:2021dzn} and reformulated in this paper  
provides us the physical value of the given quantity
since here we saw that the lateral Borel resummation of the perturbative
part - which we got by expanding the integrals over the contour $\mathcal{C}_{+}$
in (\ref{integrationcontour}) - together with the explicit residue terms is at least
free of the leading ambiguity. Let us emphasize that this fact - and the one that for the capacitance resurgence seems to explain also the leading real exponential contribution in the trans-series (see later) - supports the strong version of resurgence theory, namely that the perturbative coefficients contain all information about the trans-series expansion. This is in contrast to the $O(3)$ sigma model's energy density, where parts of the trans-series are missed by the median resummation.\footnote{Note that in both cases, we tested against only this specific type of resummation, which means an even stronger statement for the capacitance (that is, median resummation gives the correct answer); while some other recipe for extracting information from the perturbative sector might still apply for the $O(3)$ case in the realm of strong resurgence. It is also not entirely clear yet, how exactly the Wiener-Hopf method in \cite{Marino:2021dzn} will correspond to the median resummation - if at all - for sub-leading exponential terms in the trans-series. }

To complete the numerical analysis, we compared the lateral Borel
resummation above the imaginary axis to a high-precision numerical
solution of the integral equation as for the $O(3)$ case: 
\begin{equation}
\Phi_{\mathrm{TBA}}-\mathrm{Re}\,S_{+}(\Phi)=-\frac{2^{12}}{e^{8}\beta^{3}}e^{-4/\beta}\left\{ c_{0}+c_{1}\beta+c_{2}\beta^{2}+c_{3}\beta^{3}+\mathcal{O}(\beta^{4})\right\} \label{eq:gammaReal}
\end{equation}
where the $c_{n}$ coefficients are measured to be those shown in Table \ref{tab:realpartcapacitance}.
\begin{table}[t]
\centering{}%
\begin{tabular}{ll}
\hline 
exact value & numerical estimate\tabularnewline
\hline
\hline 
$c_{0}=1$ & $\;\; \;\; \;\; \;\;\;\;\cwnull$\tabularnewline
$c_{1}=\frac{3}{2}$ & $\;\; \;\; \;\; \;\;\;\;\cwone$\tabularnewline
$c_{2}=\frac{5}{4}$ & $ \;\;\;\;\;\;\;\;\;\;\cwtwo$\tabularnewline
$c_{3}\overset{?}{=}\frac{3}{4}\zeta_{3}+\frac{67}{96} = 1.59945934\ldots $ & $\;\; \;\; \;\; \;\;\;\;\cwthree$\tabularnewline
$c_{4}=?$ & $\;\; \;\; \;\; \;\;\;\;\cwfour$\tabularnewline
$c_{5}=?$ & $\;\; \;\; \;\; \;\;\;\;\cwfive$\tabularnewline
\hline
\end{tabular}\caption{The coefficients of the exponentially suppressed discrepancy
	between the TBA numerics and the real part of the lateral Borel resummation. The agreement of the recognized coefficients to the numerical data can be seen in Figure \ref{fig:realpartcapacitance}. Note that the exact value of $c_3$ is only an educated guess. }
\label{tab:realpartcapacitance}
\end{table}

\begin{figure}[t]
	\centering
	\includegraphics[width=.49\textwidth]{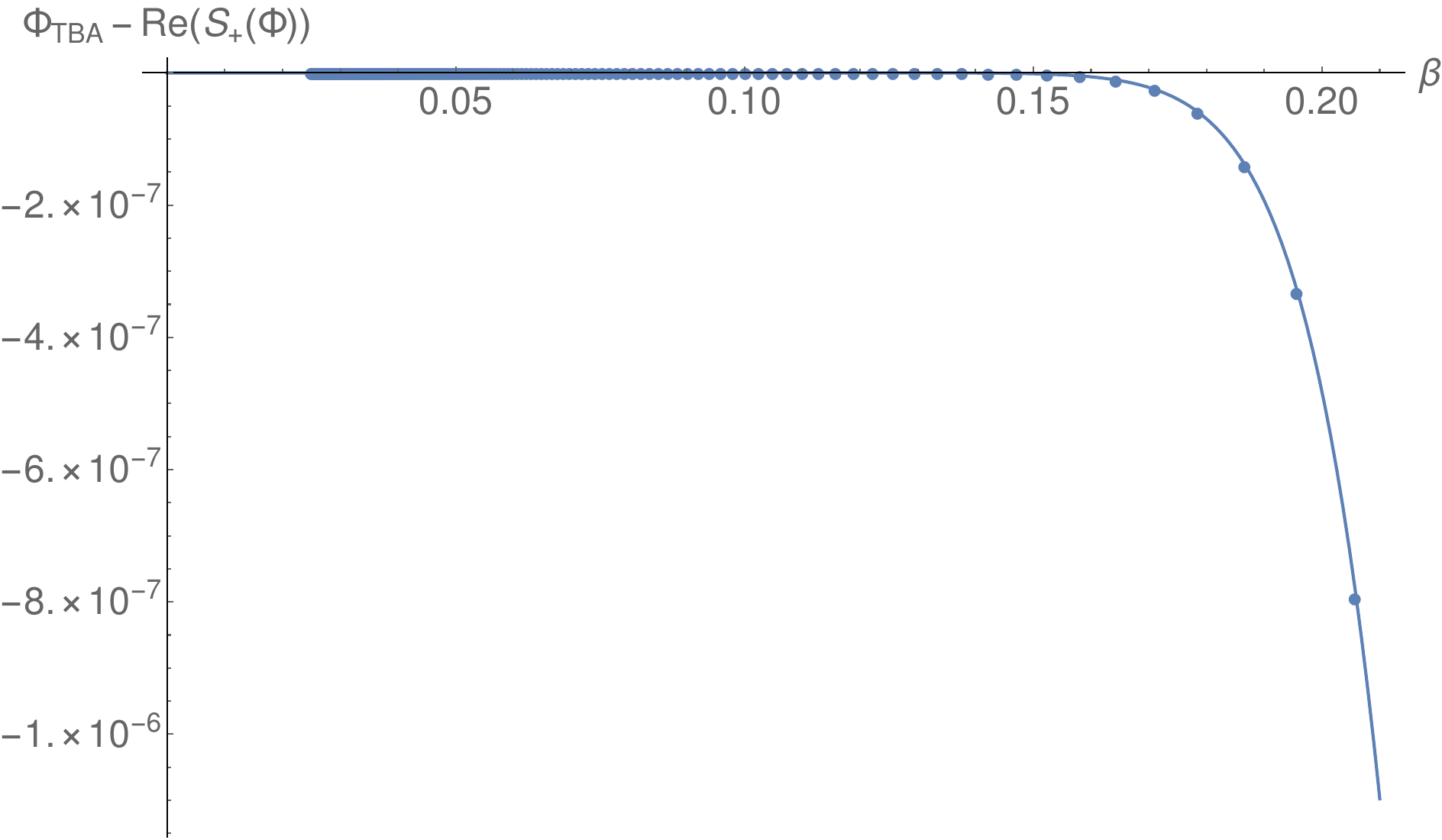}
	\includegraphics[width=.49\textwidth]{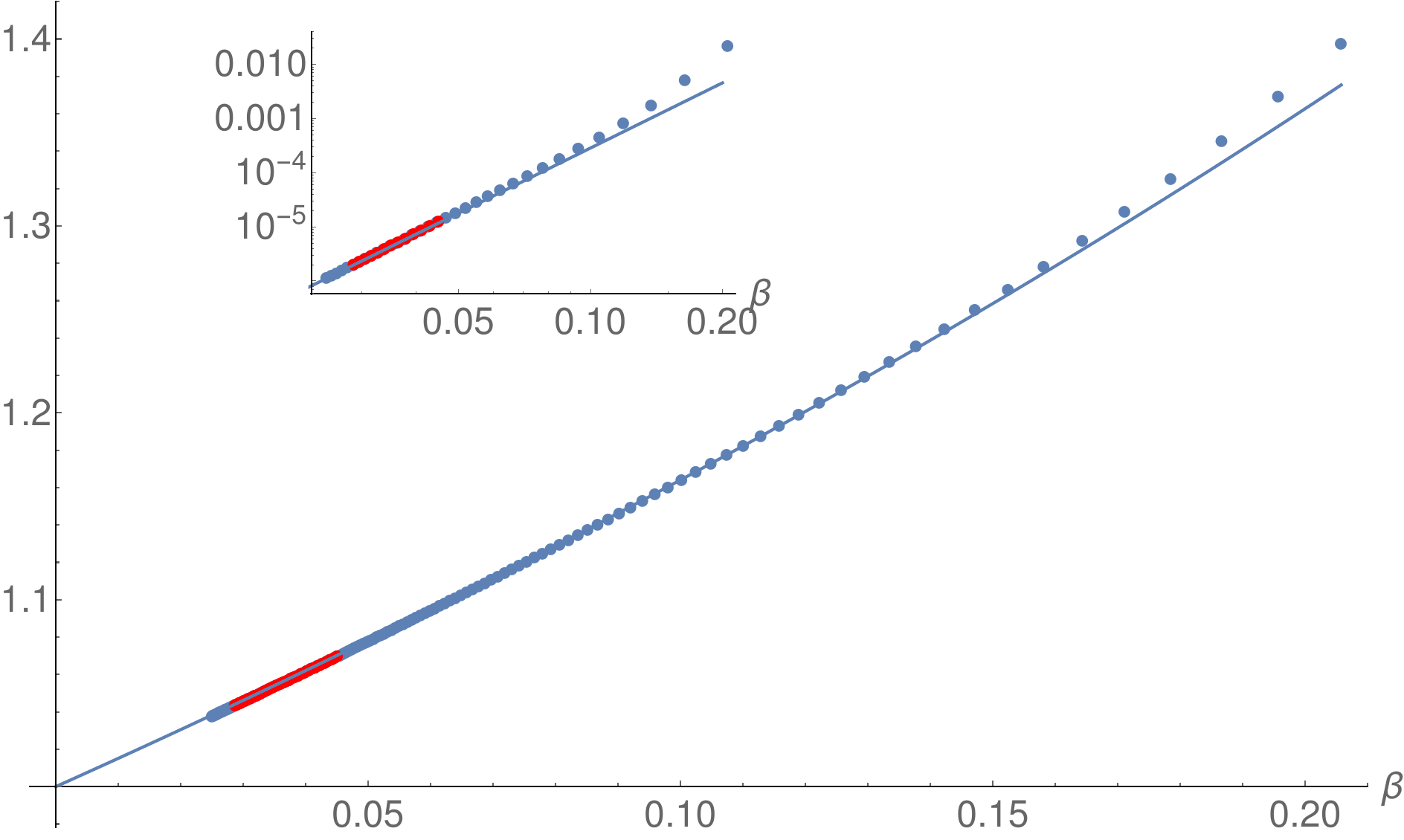}
	\caption{On the left, the points show the numerics, that is, the l.h.s of (\ref{eq:gammaReal}), while the solid line is nothing but the formula on the r.h.s. up to $c_3$ in the expansion. On the right, the comparison of the same coefficients in Table \ref{tab:realpartcapacitance} to the numerical data $-e^8 \beta^3/2^{12} e^{4/\beta} \left(\Phi_\text{TBA}-S_+(\Phi)\right)$ is shown.  The red points show the fitting region again.  The continuous line is the exact polynomial $1 + 3/2 \beta + 5/4 \beta^2 +(3/4 \zeta_3+67/96) \beta^3$. The inset shows the difference between the numerical data and the above third-order polynomial on a log-log scale; its steepness is consistent with the next term being proportional to $\beta^4$ in the expansion.}
	\label{fig:realpartcapacitance}
\end{figure}

The number of perturbative coefficients was not sufficient enough
to obtain a reliable estimate of the second alien derivative of $\Phi(\beta)$
at the closest singularity on its Borel plane, which is assumed to
be proportional to the above difference (\ref{eq:gammaReal}). Thus, here we only try to support the leading asymptotic coefficient by resurgence theory. 
That is, we expect the asymptotic structure of the $A_{n}$-s themselves in (\ref{eq:capAsy})
to look like:
\begin{equation}
	\frac{A_{n-1}}{A_{0}}=n!\left\{ \mathcal{A}_{0}+\frac{\mathcal{A}_{1}}{n}+\frac{\mathcal{A}_{2}}{n(n-1)}+\frac{\mathcal{A}_{3}}{n(n-1)(n-2)}+\mathcal{O}(n^{-4})\right\} ,\label{eq:asyasy}
\end{equation}
where the numerical estimate for the leading asymptotic coefficient
is consistent with $\mathcal{A}_{0}=\frac{16}{e^{4}\pi}$ up to two
digits.

We tested this number as follows. The
median resummation formula \cite{Abbott:2020qnl} is
\begin{equation}
S_{\mathrm{med}}(\Phi)-\mathrm{Re}\,S_{+}(\Phi)=-\frac{e^{-4/\beta}}{8}S_{+}(D_{2}^{2}\Phi+D_{1}D_{3}\Phi)+\mathcal{O}(e^{-6/\beta}),\label{eq:Smeddiff}
\end{equation}
where $D_{n}\Phi$ is the alien derivative at $s=n$ on the Borel
plane corresponding to the asymptotic series (\ref{eq:gammaseries})
and its Borel transform
\begin{equation}
B(\Phi)=\sum_{n=0}^{\infty}\frac{a_{n}}{n!}s^{n}.
\end{equation}
If we assume that $D_{1}D_{3}\Phi=0$, as we did for the $O(4)$
case  \cite{Abbott:2020qnl}, then the strength of the (\ref{eq:Smeddiff}) 
difference is governed by the second alien derivative $ D_{2}^{2}\Phi$ only.
After substituting the asymptotics (\ref{eq:capAsy})  and (\ref{eq:asyasy}), the structure of the function $B(\Phi)$ looks as:
\begin{align}
	B(\Phi) &= \text{ non-singular } + \frac{A_0}{1-s/2} -\ln\left(1-\frac{s}{2}\right)\bigg\{ \text{ non-singular } + \nonumber\\
	& + A_{0}\left[- \frac{2^{4}\, \mathcal{A}_{0}}{(s-4)^{3}}+\frac{2^{2}\,\mathcal{A}_{1}}{(s-4)^{2}}-\frac{2\,\mathcal{A}_{2}}{s-4}-\left(\mathcal{A}_{3}+\mathcal{O}(s-4)\right)\ln(2-s/2) \right]\bigg\},
\end{align}
and thus the leading contribution in $D^2_2\Phi$ comes from the $1/(s-4)^3$ term multiplying the logarithmic cut. The leading term in the second alien derivative can be calculated as
\begin{align}
	S_+(D^{2}_{2} \Phi) &= S_{+}(D_{2}\Phi)-S_{-}(D_{2}\Phi) =  \nonumber \\
	                    &=  2^{4}A_{0}\mathcal{A}_{0}\times\frac{1}{\beta}\int_{\mathcal{H}}\mathrm{d}s\, e^{-s/\beta}\ln\left(1-\frac{s}{2}\right)\left(\frac{1}{(s-4)^{3}}+\mathcal{O}\left((s-4)^{-2}\right)\right) =\nonumber\\
	& = 2^4 A_{0}\mathcal{A}_{0}\times\frac{(-2\pi i)^{2}}{\beta}\times e^{-4/\beta}\left( \frac{1}{2\beta^{2}}+\mathcal{O}\left(\beta^{-1}\right)\right), 
\end{align}
where one  $(-2\pi i)$ factor comes from the discontinuity of the logarithm, and another comes by Cauchy theorem when evaluating the Hankel-type contour $\mathcal{H}$ around the logarithmic cut.
This way (\ref{eq:Smeddiff}) and (\ref{eq:gammaReal}) are consistent with each other, at least up to the Stokes constant - the strength of the term - before the exponential\footnote{Here we suppress higher terms, both perturbative and non-perturbative.}:
\begin{equation}
S_{\mathrm{med}}(\Phi)-\mathrm{Re}\,S_{+}(\Phi)=-\frac{e^{-4/\beta}}{8 \beta^3}\times\left(-\pi^{2}2^{5}A_{0}\mathcal{A}_{0}\right)=-\frac{2^{12}}{e^{8}\beta^3}e^{-4/\beta}.
\end{equation}
Finally, we give the capacitance formula in terms of the original
variable $\kappa$, which is proportional to the distance of the plates: 
\begin{align}
	\mathcal{C} &= S_{+}\left(\frac{1}{4\kappa}+\frac{\mathsf{L}-1}{4\pi}+\frac{\kappa}{16\pi^{2}}\left(\mathsf{L}^{2}-2\right)+\mathcal{O}(\kappa^{2})\right)\nonumber\\
		    &+\frac{i}{4\pi e^{2}}e^{-4\pi/\kappa}S_{+}\left\{ 1+\left(\mathsf{L}+\frac{3}{4}\right)\frac{\kappa}{2\pi}+\left(\mathsf{L}+\frac{17}{32}\right)\frac{\kappa^{2}}{4\pi^{2}}+\mathcal{O}(\kappa^{3})\right\} +\mathcal{O}(e^{-8\pi/\kappa})
\end{align}
where $\mathsf{L}\equiv\ln\left(16\pi/\kappa\right)$, and $\mathcal{C} \equiv \kappa/\pi\, C$ is another possible definition of the dimensionless capacitance used in the introduction and in \cite{Reichert:2020ymc}. The difference between the real part of the lateral Borel resummation $S_{+}(\mathcal{C})$, and
the TBA data looks now as: 
\begin{equation}
    \mathcal{C}_{\text{TBA}}-\text{Re}\ S_{+}(\mathcal{C})=-\frac{1}{8\pi e^{4}}e^{-8\pi/\kappa}\left\{ 1+\left(\mathsf{L}+\frac{1}{2}\right)\frac{\kappa}{2\pi}+\left(\mathsf{L}+\frac{1}{4}\right)\frac{\kappa^{2}}{4\pi^{2}}+\mathcal{O}(\kappa^{3})\right\} .	
\end{equation}

\section{Conclusion}

In this paper we investigated integral equations with kernels related
to the $O(N)$ sigma models. We analyzed various observables including
the density $O_{c1}$ and energy density $O_{cc}$ of the groundstate
energy of the $O(N)$ sigma models in a magnetic field and the capacitance
of the circular plate capacitor $O_{11}$. We also established relations
between these observables, see (\ref{O11}).
We advanced in the systematic, Wiener-Hopf
type solution of this class of models considerably. We first simplified
the Wiener-Hopf equations to ensure that only one integral operator
appears. Based on the analytical structure of its kernel we could
introduce a running coupling and show that all observables can be
expanded solely in its powers without any logarithms of it. We also
formulated a form of the equations which handled the non-perturbative
terms systematically showing also only power-like dependence on the
running coupling. These simplified equations are easier to handle
than their previous versions and we also developed a method based
on Laplace transformation how the various orders could be calculated
explicitly. Combining the relations between the observables and the
already known perturbative expansions we could advance in the calculations
of new non-perturbative terms. We investigated the structure of the
non-perturbative corrections and determined their leading terms. 

As the $O(3)$ kernel is exceptional, we elaborated on this case separately.
We studied the behaviour of the free energy in the $O(3)$ sigma model,
where we compared the exactly determined non-perturbative terms to
the direct numerical solution of the integral equations and found
complete agreement. We also confronted the asymptotics of the perturbative
series with the leading non-perturbative corrections and confirmed
the previous findings that they do not match. 

The introduction of our running coupling simplified the perturbative
expansion of the coaxial disk capacitance drastically. By expressing
its derivatives in terms of the density and groundstate energy of
the $O(3)$ sigma model we were able to analyze also its leading non-perturbative
corrections. We then investigated how resurgence theory works and
we found that the asymptotics of the perturbative expansion contained
all information about the non-perturbative terms. We checked these
results against the numerical solution of the integral equation. 

Our simplified equations can be used to calculate the trans-series
representation of the various observables. This trans-series is a
sum of all the exponentially suppressed terms $\nu^{n}$ with
$\nu=e^{-1/v-L}/v$
multiplied by a typically asymptotic series in the running coupling
$v$.
We indicate in Table
\ref{table:NP} the structure of the non-perturbative corrections
for various observables in the $O(N)$ model. The terms in red are
not connected to the asymptotics of the perturbative series, while
the ones in black are. The anomalous (constant) term in the energy
density for the $O(N)$ sigma models for $N\geq5$ can be attributed
to the different definitions of the zero point energy in the two descriptions:
in perturbation theory the groundstate energy vanishes for zero magnetic
field, $h=0$, while in the TBA formulation it vanishes at $h=m$
\cite{Marino:2021dzn}. The biggest difference is in the $O(3)$ model,
where the real part of the leading non-perturbative corrections is
not related to the asymptotic perturbative series. We attributed this behaviour
to the presence of instantons \cite{Bajnok:2021zjm}, and this has been verified
in \cite{Marino:2022ykm}. This makes
the $O(3)$ theory exceptional among the other $O(N)$ models. 
Other non-perturbative corrections, which are related to the asymptotics
of the perturbative series were previously connected to renormalons
\cite{DiPietro:2021yxb,Marino:2021six}. 

\begin{table}
\begin{centering}
\begin{tabular}{|c|c||c|c|c|c|c|c|c|c|c|}
\hline 
 & $O_{cc}$ & ${\color{red}\nu}$ & $\nu^{2}$ & ${\color{red}\nu^{3}}$ & $\nu^{4}$ & ${\color{red}\nu}^{{\color{red}5}}$ & $\nu^{6}$ &  &  & $\nu^{l}$\tabularnewline
\cline{2-11} \cline{3-11} \cline{4-11} \cline{5-11} \cline{6-11} \cline{7-11} \cline{8-11} \cline{9-11} \cline{10-11} \cline{11-11} 
$O(3)$ & $O_{c1}$ & ${\color{red}\nu}$ & $\nu^{2}$ & ${\color{red}\nu^{3}}$ & $\nu^{4}$ & ${\color{red}\nu^{5}}$ & $\nu^{6}$ &  &  & $\nu^{l}$\tabularnewline
\cline{2-11} \cline{3-11} \cline{4-11} \cline{5-11} \cline{6-11} \cline{7-11} \cline{8-11} \cline{9-11} \cline{10-11} \cline{11-11} 
 & $O_{11}$ &  & $\nu^{2}$ &  & $\nu^{4}$ &  & $\nu^{6}$ &  &  & $\nu^{2l}$\tabularnewline
\hline 
\hline 
$O(4)$ & $O_{cc}$ & $\nu$ & $\nu^{2}$ &  & $\nu^{4}$ &  & $\nu^{6}$ &  &  & $\nu^{2l}$\tabularnewline
\cline{2-11} \cline{3-11} \cline{4-11} \cline{5-11} \cline{6-11} \cline{7-11} \cline{8-11} \cline{9-11} \cline{10-11} \cline{11-11} 
 & $O_{c1}$ &  & $\nu^{2}$ &  & $\nu^{4}$ &  & $\nu^{6}$ &  &  & $\nu^{2l}$\tabularnewline
\hline 
\hline 
$O(5)$ & $O_{cc}$ & ${\color{red}\nu}$ &  &  &  &  & $\nu^{6}$ &  &  & $\nu^{6l}$\tabularnewline
\cline{2-11} \cline{3-11} \cline{4-11} \cline{5-11} \cline{6-11} \cline{7-11} \cline{8-11} \cline{9-11} \cline{10-11} \cline{11-11} 
 & $O_{c1}$ &  &  &  &  &  & $\nu^{6}$ &  &  & $\nu^{6l}$\tabularnewline
\hline 
\hline 
$O(6)$ & $O_{cc}$ & ${\color{red}\nu}$ &  &  & $\nu^{4}$ &  &  &  &  & $\nu^{4l}$\tabularnewline
\cline{2-11} \cline{3-11} \cline{4-11} \cline{5-11} \cline{6-11} \cline{7-11} \cline{8-11} \cline{9-11} \cline{10-11} \cline{11-11} 
 & $O_{c1}$ &  &  &  & $\nu^{4}$ &  &  &  &  & $\nu^{4l}$\tabularnewline
\hline 
\hline 
$O(2k)$ & $O_{cc}$ & ${\color{red}\nu}$ &  &  &  &  &  & $\nu^{2(k-1)}$ &  & $\nu^{2l(k-1)}$\tabularnewline
\cline{2-11} \cline{3-11} \cline{4-11} \cline{5-11} \cline{6-11} \cline{7-11} \cline{8-11} \cline{9-11} \cline{10-11} \cline{11-11} 
 & $O_{c1}$ &  &  &  &  &  &  & $\nu^{2(k-1)}$ &  & $\nu^{2l(k-1)}$\tabularnewline
\hline 
\hline 
$O(2k+1)$ & $O_{cc}$ & ${\color{red}\nu}$ &  &  &  &  &  & $\nu^{2(2k-1)}$ &  & $\nu^{2l(2k-1)}$\tabularnewline
\cline{2-11} \cline{3-11} \cline{4-11} \cline{5-11} \cline{6-11} \cline{7-11} \cline{8-11} \cline{9-11} \cline{10-11} \cline{11-11} 
 & $O_{c1}$ &  &  &  &  &  &  & $\nu^{2(2k-1)}$ &  & $\nu^{2l(2k-1)}$\tabularnewline
\hline 
\end{tabular}
\par\end{centering}
\caption{Structure of the exponential corrections in $\nu=e^{-2B}$ for different
$O(N)$ models and observables. The non-perturbative terms in red
are not completely related to the asymptotics of the perturbative
series, although for $N\protect\geq5$ this mismatch can be attributed
to the different definitions of the $h=0$ bulk energy constant.}

\label{table:NP}
\end{table}

Our formulation allows a systematic calculation of the non-perturbative
terms. Beyond the leading order, however, several contributions will
appear to each trans-series term and a future analysis should work
out their details. Also it would be important to understand that to
which resummation prescription, median or not, our method will lead
to. The technique based on the Laplace transformation gives explicit
expressions to the various perturbative coefficients, but becomes more
and more complicated at higher orders. It would be very nice to adapt
Volin's method beyond the perturbative order and calculate non-perturbative
corrections in a similar fashion. In the $O(4)$ model we observed
very interesting relations between different trans-series terms \cite{Abbott:2020mba,Abbott:2020qnl,Bajnok:2021dri}.
It would be very insightful to derive those relations directly from
the integral equations, or derive similar relations for other trans-series
terms. The introduction of the running coupling simplified all our
calculations considerably. We believe they should also play a central
role in other asymptotically free models, too.

%
\section*{Acknowledgements}
\par\bigskip

Our work was supported by ELKH, with infrastructure provided by the
Hungarian Academy of Sciences. This work
was also supported by the NKFIH grant K134946.

\appendix

\section{Building blocks needed for the Wiener-Hopf analysis}
\label{appX}

In this appendix we collect a number of definitions, results and technical
tools which are used in the main text.

The Fourier transform of the kernel $K(\theta)$ is
\begin{equation}
\tilde K(\omega)=\int_{-\infty}^\infty d\theta e^{i\omega\theta} K(\theta)=
\frac{e^{-\pi\vert\omega\vert}+e^{-2\pi\Delta\vert\omega\vert}}
{1+e^{-\pi\vert\omega\vert}}.  
\end{equation}
We will define the Fourier transform of other functions analogously and
also indicate it by a tilde. Note that for $O(3)$ the Fourier transform of the
kernel simplifies to $e^{-\pi\vert\omega\vert}$.

In the Wiener-Hopf analysis an important role is played by the multiplicative
decomposition
\begin{equation}
G_+(\omega)G_-(\omega)=\frac{1}{1-\tilde K(\omega)},  
\end{equation}
where $G_\pm$ are analytic in $H_\pm$, the upper and lower complex half-planes,
respectively. Note that $\tilde K(0)=1$ and $G_\pm(\omega)$ are singular near
$\omega=0$ like $\vert\omega\vert^{-1/2}$. This is the typical behaviour
for bosonic models. (For fermionic models $G_\pm$ are regular at $\omega=0$ and
this makes their Wiener-Hopf analysis simpler.) $\tilde K(\omega)$ is real and
symmetric and therefore
\begin{equation}
G_-(\omega)=G_+(-\omega).
\label{Gplmi}  
\end{equation}
$G_+$ is given explicitly by
\begin{equation}
G_+(\omega)=\frac{1}{\sqrt{-i\Delta\omega}}\
\frac{\Gamma(1-i\omega\Delta)}{\Gamma(1/2-i\omega/2)}
\exp\left\{-i\omega\Delta[1-\ln(-i\omega\Delta)]+i\omega[1-\ln(-i\omega/2)]/2
\right\}.
\label{Gplus}
\end{equation}
For small real $\omega$
\begin{equation}
G_+(\omega)\approx\frac{1}{\sqrt{\pi\Delta\vert\omega\vert}}e^{\frac{i\pi}{4}
{\rm sign}(\omega)}    
\end{equation}
and for asymptotically large argument in $H_+$
\begin{equation}
G_+(i\kappa)=1+O(1/\kappa),\qquad\quad \kappa\to\infty.
\end{equation}
We also need
\begin{equation}
G_+(i)=\frac{1}{\sqrt{2\Delta}}\Gamma(1+\Delta)e^{\gamma_1/2-\Delta\ln\Delta},
\qquad\quad \gamma_1=2\Delta-1.
\end{equation}

Functions $f(\omega)$ vanishing for $\vert\omega\vert\to\infty$ can be
decomposed additively into $f^{(+)}$ and $f^{(-)}$, analytic in $H_+$ and $H_-$,
respectively:
\begin{equation}
f^{(\pm)}(\omega)
=\mp\frac{1}{2\pi i}\int_{-\infty}^\infty \frac{f(\omega')d\omega'}
{\omega-\omega'}.  
\end{equation}
For real $\omega$, $f^{(+)}(\omega)+f^{(-)}(\omega)=f(\omega)$. If $f(\omega)$ is
already analytic in $H_+$ and vanishes for large $\vert\omega\vert$ then
$f^{(+)}(\omega)=f(\omega)$ and $f^{(-)}(\omega)=0$,
and analogously for $f(\omega)$ analytic in $H_-$.

$G_+$ is analytic in $H_+$ by definition but it is also analytic in $H_-$,
except for the negative imaginary axis (where it is discontinuous and also has
poles). Similarly $G_-$ is everywhere analytic, except along the positive
imaginary axis. An important variable in our analysis is
\begin{equation}
\alpha(\omega)=e^{2i\omega B}\frac{G_-(\omega)}{G_+(\omega)},  
\label{alphadef}
\end{equation}
which is discontinuous along the positive imaginary axis and has poles there.
Explicitly, just on the left and right side of the imaginary axis,
\begin{equation}
\alpha(2i\xi\pm\varepsilon)=e^{-4B\xi}[\mp i\cos(\pi\gamma_1\xi)
+\sin(\pi\gamma_1\xi)]H(\xi),  
\end{equation}
where
\begin{equation}
H(\xi)=e^{2\gamma_1\xi\ln2\xi} H_p(\xi), 
\end{equation}
with
\begin{equation}
H_p(\xi)=\frac{\Gamma(1-2\Delta\xi)}{\Gamma(1+2\Delta\xi)}\,
\frac{\Gamma(1/2+\xi)}{\Gamma(1/2-\xi)}\,
e^{2\xi[2\Delta\ln\Delta+\ln2-\gamma_1]}.
\end{equation}
An important property of $H_p(\xi)$ is that it can be expanded around $\xi=0$
in powers of $\xi$:
\begin{equation}
H_p(\xi)=1+2\alpha_1\xi+O(\xi^2),   
\end{equation}
where
\begin{equation}
\alpha_1=\gamma_1(\gamma_E-1+\ln\Delta)+\ln(\Delta/2).
\end{equation}
The discontinuity is
\begin{equation}
\alpha(2i\xi+\varepsilon)-\alpha(2i\xi-\varepsilon)=-2ie^{-4B\xi}\beta(\xi),
\qquad\quad \beta(\xi)=\cos(\pi\gamma_1\xi)\, H(\xi).  
\label{alphabeta}
\end{equation}

$H(\xi)$ is meromorphic in $H_+$, it has poles (and zeroes) on the positive
imaginary axis. The poles are located at
\begin{equation}
\xi=\xi_\ell=\ell\xi_o,\qquad \ell=1,2,\dots\qquad
\xi_o=\left\{\begin{matrix}
\frac{N-2}{2}\qquad\quad &N\ {\rm even,}\\
N-2 \qquad\quad & N \ {\rm odd.}
\end{matrix}\right.
\label{xi0def}
\end{equation}
The zeroes are at
\begin{equation}
\xi=m+\frac{1}{2},\qquad\quad m=0,1,2\dots  
\end{equation}
with the exception (for odd $N$) of $m$ integers of the form
\begin{equation}
m=\frac{N-3}{2}+(N-2)k,\qquad\quad k=0,1,2\dots  
\end{equation}
We see that
\begin{equation}
H(1/2)=0,\qquad\quad {\rm except\ for\ }N=3.
\end{equation}
We also see that
\begin{equation}
\sin(\pi\gamma_1\xi_\ell)=0,\qquad\quad \cos(\pi\gamma_1\xi_\ell)=\varepsilon_\ell
\end{equation}
where
\begin{equation}
\varepsilon_\ell=\left\{\begin{matrix} (-1)^\ell\qquad\quad &N\ {\rm odd,}\\
(-1)^{N\ell/2}\qquad\quad & N\ {\rm even.}
\end{matrix}\right.
\end{equation}
Near $\xi=\xi_\ell$ if
\begin{equation}
H(\xi)\approx \frac{\tilde H_\ell}{\xi-\xi_\ell}  
\end{equation}
with some residue $\tilde H_\ell$ then
\begin{equation}
\alpha(2i\xi\pm\varepsilon)=\mp ie^{-4B\xi_\ell}\,\frac{H_\ell}{\xi-\xi_\ell},
\qquad\quad \beta(\xi)\approx \frac{H_\ell}{\xi-\xi_\ell},\qquad\quad
H_\ell=\varepsilon_\ell \tilde H_\ell.
\label{poles}
\end{equation}

We will also need the behaviour of $\alpha(2i\xi\pm\varepsilon)$
and $\beta(\xi)$ near $\xi=1/2$. We find
\begin{equation}
\alpha(2i\xi\pm\varepsilon)=e^{-2B}\left\{h_o+h_1^\pm(\xi-1/2)
+O((\xi-1/2)^2)\right\},  
\label{half}
\end{equation}
where
\begin{equation}
h_o=\frac{\delta_{N,3}}{e},\qquad\quad
h_1^\pm=\left\{\begin{matrix}
\left(\mp i\cos\frac{\pi\gamma_1}{2}+\sin\frac{\pi\gamma_1}{2}\right)
\tilde h_1\qquad\quad & N\geq4,\\
\tilde h_1-\frac{4B}{e}\pm\frac{i\pi}{e}\qquad\qquad & N=3,
\end{matrix}\right.
\label{half1}
\end{equation}
and
\begin{equation}
\tilde h_1=\left\{\begin{matrix}
-\frac{\Gamma(1-\Delta)}{\Gamma(1+\Delta)}\exp\{2\Delta\ln\Delta+\ln2-\gamma_1
\}\qquad\quad & N\geq4,\\
\frac{2}{e}(\gamma_E-1+\ln2)\qquad\quad & N=3.
\end{matrix}\right.
\label{half2}
\end{equation}
Again, we see that $\alpha(i)=0$, except for $N=3$.
On the other hand, $\beta(1/2)$ always vanishes:
\begin{equation}
\beta(\xi)=\beta_o(\xi-1/2)+O((\xi-1/2)^2), 
\end{equation}
where
\begin{equation}
\beta_o=\left\{\begin{matrix}\cos\left(\frac{\pi\gamma_1}{2}\right)
\tilde h_1\qquad\quad &
N\geq4,\\-\pi/e\qquad\quad &N=3.\end{matrix}\right.  
\label{bet0def}
\end{equation}

\section{The $\chi_1$ problem}
\label{appY}

The source term here is
\begin{equation}
r_{1+}(\omega)=\frac{e^{2i\omega B}}{i\omega}-\frac{1}{i\omega}.  
\end{equation}
We proceed analogously to the $\chi_c$ problem and absorb the first term
into the new unknown
\begin{equation}
\hat Q_+(\omega)=Q_+(\omega)+\frac{G_+(\omega)}{i\omega}.  
\label{Qhat}
\end{equation}
Again, the rest of the source term can be easily dealt with by closing the
contour of integration in $H_-$.

However, there is an extra difficulty in this problem since the new variable
(\ref{Qhat}) has a non-integrable singularity at the origin:
\begin{equation}
\hat Q_+(2i\xi)\approx \frac{-1}{2\xi\sqrt{2\xi\pi\Delta}}.  
\end{equation}
This means that we have to be careful with the derivation and regularize
integrals around the origin. This regularization can be removed at the end
of the calculation after applying appropriate subtractions to the integrals.
We again deform the original contour to the (regularized) ${\cal C}_{LR}$
contour and formulate the integral equation in terms of principal value
integration. No terms proportional to $e^{-2B}$ arise here and finally we
arrive at
\begin{equation}
\begin{split}
\xi\hat Q_+&(2i\xi)
-\frac{1}{\pi}{\cal P}\int_0^\infty\frac
{e^{-4B\xi'}\beta(\xi')\xi'\hat Q_+(2i\xi')}{\xi+\xi'}d\xi'\\
&+\frac{1}{\pi}{\cal P}\int_0^\infty\Big\{
e^{-4B\xi'}\beta(\xi')\hat Q_+(2i\xi')+\frac{1}{2\xi'\sqrt{2\xi'\pi\Delta}}
\Big\}d\xi'=0.
\end{split}
\label{hatQeq}
\end{equation}
We simplify the equation by introducing the rescaled variable
\begin{equation}
K(x)=\sqrt{v\pi\Delta}vx\hat Q_+(ivx)  
\end{equation}
and rewrite the integrals to go along the contour ${\cal C}_+$ (plus residue
terms):
\begin{equation}
\begin{split}
K(x)-&\frac{1}{\pi}\inC\frac{e^{-y}{\cal A}(y)K(y)}{x+y}dy+
\frac{ivx}{\xi_o}\sum_{\ell=1}^\infty \frac{H_\ell y_\ell}{\ell(2\ell\xi_o+vx)}
\nu^{2\ell\xi_o}\\
+&\frac{1}{\pi}\inC\Bigg\{e^{-y}{\cal A}(y)\frac{K(y)}{y}+\frac{1}{y\sqrt{y}}
\Bigg\}dy=0.
\end{split}
\end{equation}
Here
\begin{equation}
y_s=\sqrt{v\pi\Delta}2s\xi_o\hat Q_+(2is\xi_o)
\end{equation}
and it can be calculated from
\begin{equation}
\begin{split}
y_s-&\frac{v}{\pi}\inC\frac{e^{-x}{\cal A}(x)K(x)}{2s\xi_o+vx}dx+
\frac{is}{\xi_o}\sum_{\ell=1}^\infty \frac{H_\ell y_\ell}{\ell(s+\ell)}
\nu^{2\ell\xi_o}\\
+&\frac{1}{\pi}\inC\Bigg\{e^{-x}{\cal A}(x)\frac{K(x)}{x}+\frac{1}{x\sqrt{x}}
\Bigg\}dx=0.
\end{split}
\end{equation}

We again represent our variables as the trans-series
\begin{equation}
K(x)=\sum_{m=0}^\infty K_m(x)\,\nu^{2m\xi_o},\qquad\quad
y_s=\sum_{m=0}^\infty y_{s,m}\,\nu^{2m\xi_o}.
\end{equation}
Note that these are valid for both the $N\geq4$ and the $N=3$ cases.
The LO and NLO equations are
\begin{equation}
K_0(x)-\frac{1}{\pi}\inC\frac{e^{-y}{\cal A}(y)K_0(y)}{x+y}dy
+\frac{1}{\pi}\inC\Bigg\{e^{-y}{\cal A}(y)\frac{K_0(y)}{y}+\frac{1}{y\sqrt{y}}
\Bigg\}dy=0,
\end{equation}
\begin{equation}
\overline{K}_1(x)+\frac{1}{\pi}\inC\frac{e^{-y}{\cal A}(y)
\overline{K}_1(y)}{x+y}dy=-\frac{iv}{\xi_o}\,\frac{H_1 y_{1,0}}{2\xi_o+vx}.
\end{equation}
Here
\begin{equation}
\overline{K}_1(x)=\frac{K_1(x)}{x}
\end{equation}
and
\begin{equation}
y_{1,0}=\frac{v}{\pi}\inC\frac{e^{-y}{\cal A}(y)K_0(y)}{2\xi_o+vy}dy
-\frac{1}{\pi}\inC\Bigg\{e^{-y}{\cal A}(y)\frac{K_0(y)}{y}+\frac{1}{y\sqrt{y}}
\Bigg\}dy.
\label{y10}
\end{equation}

For the perturbative expansion of the leading coefficient we will use
\begin{equation}
{\cal A}(x)K_0(x)=\omega(x)=\sum_{k=0}^\infty v^k\omega_k(x)  
\end{equation}
and also define the ($B$-independent) functions $g_k$ by
\begin{equation}
\omega_k(x)=x g_k(x)-\frac{\delta_{k,0}}{\sqrt{x}}.  
\end{equation}
Later we will need their moments
\begin{equation}
\bar\gamma_{k,r}=\frac{1}{\pi}\int_0^\infty e^{-x}x^r g_k(x)dx,\qquad\quad
\bar\gamma_k=\bar\gamma_{k,0}  
\end{equation}
and we will also need
\begin{equation}
I(p)=\int_0^\infty\frac{e^{-px}-1}{x\sqrt{x}}dx=-2\sqrt{\pi p},\qquad\quad
I^{(r)}(p)=\frac{d^r}{dp^r}\,I(p).
\end{equation}
The first two components satisfy
\begin{equation}
g_0(x)+\frac{1}{\pi}\int_0^\infty\frac{e^{-y}g_0(y)}{x+y}dy=
\frac{1}{\pi}\int_0^\infty\frac{e^{-y}-1}{(x+y)y\sqrt{y}}dy,
\label{g0}
\end{equation}
\begin{equation}
g_1(x)+\frac{1}{\pi}\int_0^\infty\frac{e^{-y}g_1(y)}{x+y}dy=
\left(xg_0(x)-\frac{1}{\sqrt{x}}\right)L_1(\ln x).
\label{g1}
\end{equation}
These two integral equations are solved explicitly in appendix \ref{appA}.

The calculation of the density integral and the boundary value of $\chi_1$ is
analogous to the derivation of (\ref{hatQeq}). After a long calculation we get
\begin{equation}
O_{1c}=\frac{e^B G_+(i)}{2\pi}\left\{\frac{\delta_{N,3}}{e}\,\hat Q_+(i)\,\nu
+\frac{2}{\pi}{\cal P}\int_0^\infty\frac{d\xi}{1-2\xi}\left[
e^{-4B\xi}\beta(\xi)\hat Q_+(2i\xi)+\frac{1}{2\xi\sqrt{2\xi\pi\Delta}}\right]
\right\}
\label{O1c}
\end{equation}
and
\begin{equation}
\chi_1(B)=\frac{2}{\pi}{\cal P}\int_0^\infty\left\{
e^{-4B\xi}\beta(\xi)\hat Q_+(2i\xi)+\frac{1}{2\xi\sqrt{2\xi\pi\Delta}}\right\}
d\xi.
\end{equation}
For $N=3$ the trans-series for the density $O_{1c}$ contains all powers of
$\nu$. For $\chi_1(B)$, however, the trans-series is of the form
\begin{equation}
\chi_1(B)=\frac{1}{\sqrt{v\pi\Delta}}\sum_{m=0}^\infty u_m\,\nu^{2m\xi_o}  
\end{equation}
for all $N$. The first two coefficients are
\begin{equation}
u_0=\frac{1}{\pi}\inC \left\{
e^{-x}{\cal A}(x)\frac{K_0(x)}{x}+\frac{1}{x\sqrt{x}}\right\}dx=  
\frac{1}{\pi}\inC \left\{
e^{-x}\frac{\omega(x)}{x}+\frac{1}{x\sqrt{x}}\right\}dx,  
\end{equation}
\begin{equation}
u_1=\frac{1}{\pi}\inC
e^{-x}{\cal A}(x)\overline{K}_1(x)dx+\frac{i}{\xi_o}H_1y_{1,0}.  
\end{equation}

We now expand $u_0$ perturbatively:
\begin{equation}
u_0=\sum_{k=0}^\infty v^k u_{0,k}.  
\end{equation}
Here
\begin{equation}
u_{0,0}=\frac{1}{\pi}\int_0^\infty dx\left[e^{-x}g_0(x)+\frac{1}{x\sqrt{x}}\left(
1-e^{-x}\right)\right]=\bar\gamma_0+\frac{2}{\sqrt{\pi}},      
\end{equation}
\begin{equation}
u_{0,k}=\bar\gamma_k,\qquad k=1,2,\dots
\end{equation}
In appendix \ref{appA} we calculate
\begin{equation}
\bar\gamma_0=\frac{\sqrt{\pi}}{2}-\frac{2}{\sqrt{\pi}}  
\end{equation}
and so we can write the perturbative expansion of $u_0$ as
\begin{equation}
u_0=\frac{\sqrt{\pi}}{2}\left(1+\sum_{k=1}^\infty v^k r_k\right),\qquad\quad
r_k=\frac{2}{\sqrt{\pi}}\bar\gamma_k.  
\label{rk}
\end{equation}
The NLO coefficient $r_1$ is calculated in appendix \ref{appA}:
\begin{equation}
r_1=r_o+\frac{L}{2}.  
\label{r1}
\end{equation}

Next we turn to $O_{1c}$. We start with the trans-series expansion of
\begin{equation}
\hat Q_+(i)=\frac{1}{\sqrt{v\pi\Delta}}K(1/v)=\frac{1}{\sqrt{v\pi\Delta}}  
\sum_{m=0}^\infty \hat Q_{+,m}\,\nu^{2m\xi_o}
\end{equation}
and consider the leading coefficient
\begin{equation}
\begin{split}
\hat Q_{+,0}&=\frac{v}{\pi}\inC\frac{e^{-x}\omega(x)}{1+vx}dx
-\frac{1}{\pi}\inC\left[\frac{e^{-x}\omega(x)}{x}+\frac{1}{x\sqrt{x}}\right]dx\\
&=-\frac{\sqrt{\pi}}{2}\Bigg\{1+\frac{2}{\sqrt{\pi}}\Big[
\sum_{p+q\not=0}v^{p+q}(-1)^q\bar\gamma_{p,q}-\frac{1}{\pi}
\sum_{q=1}^\infty v^q\,I^{(q)}(1)\Big]\Bigg\}.  
\end{split}    
\end{equation}
Alternatively, $\hat Q_+(i)$ can be written as
\begin{equation}
\hat Q_+(i)=-\frac{1}{2\sqrt{v\Delta}}\left\{1+\sum_{k=1}^\infty \tilde R_k v^k
\right\}+O(\nu^{2\xi_o}),  
\end{equation}
where
\begin{equation}
\tilde R_k=\frac{2}{\sqrt{\pi}}\left[\sum_{p+q=k}(-1)^q
\bar\gamma_{p,q}-\frac{1}{\pi}I^{(k)}(1)\right].    
\end{equation}
The NLO result
\begin{equation}
\tilde R_1=r_1+\frac{3}{4}  
\end{equation}
is calculated in appendix \ref{appA}.

Next we concentrate on the second term in the curly bracket in (\ref{O1c}). It
is first rewritten as
\begin{equation}
\begin{split}
\frac{1}{\pi\sqrt{v\pi\Delta}}&\int_{{\cal C}_{+*}}\frac{dx}{1-vx}\Bigg[
e^{-x}{\cal A}(x)\frac{K(x)}{x}+\frac{1}{x\sqrt{x}}\Bigg]+
\frac{i}{\xi_o}\sum_{\ell=1}^\infty\frac{H_\ell y_\ell}
{\ell\sqrt{v\pi\Delta}(1-2\ell\xi_o)}\nu^{2\ell\xi_o}\\
&=\frac{1}{\sqrt{v\pi\Delta}}\sum_{m=0}^\infty U^{(II)}_m \nu^{2m\xi_o},
\end{split}    
\end{equation}
where the meaning of the ${\cal C}_{+*}$ integration is that the contour
goes slightly above the real line near the poles at $x=2\xi_\ell/v$ but it
remains a principal value integral around $x=1/v$.
The leading term in this trans-series expansion is
\begin{equation}
U^{(II)}_0=\frac{1}{\pi}\int_{{\cal C}_{+*}}\left[\frac{e^{-x}\omega(x)}{x(1-vx)}
+\frac{1}{x\sqrt{x}(1-vx)}\right]dx=\frac{\sqrt{\pi}}{2}\left(
1+\sum_{k=1}^\infty R_k v^k\right),
\end{equation}
where
\begin{equation}
R_k=\frac{2}{\sqrt{\pi}}\left[\sum_{p+q=k}\bar\gamma_{p,q}-\frac{1}{\pi}(-1)^k
I^{(k)}(1)\right].    
\end{equation}
The NLO result
\begin{equation}
R_1=r_1-\frac{3}{4}  
\end{equation}
is calculated in appendix \ref{appA}.

It is possible to calculate the first few perturbative coefficients of
$\hat u_1$ defined in (\ref{udefs}). The details of the calculation are given
in appendix~\ref{appA}.
For this purpose first one has to calculate the perturbative coefficients in
\begin{equation}
y_{1,0}=-\frac{\sqrt{\pi}}{2}\left(1+\sum_{k=1}^\infty a_k v^k\right).  
\label{y10exp}
\end{equation}
The leading ones are
\begin{equation}
a_1=r_1+\frac{3}{8\xi_o},\qquad\quad  
a_2=r_2+\frac{1}{2\xi_o}\left(\frac{5}{8}\gamma_1-\frac{3}{4}r_1\right)
-\frac{15}{128\xi_o^2}.
\label{a12}
\end{equation}
Next we introduce $\hat{\cal F}(x)$ and its moments by
\begin{equation}
{\cal A}(x)\overline{K}_1(x)
=\frac{\sqrt{\pi}iv}{4\xi_o^2}H_1\hat{\cal F}(x),\qquad
\hat{\cal F}(x)=\sum_{k=0}^\infty v^k\hat{\cal F}_k(x),\qquad f_k=\frac{1}{\pi}
\int_0^\infty e^{-x}\hat{\cal F}_k(x)dx.
\end{equation}
It satisfies
\begin{equation}
\frac{\hat{\cal F}(x)}{{\cal A}(x)}+\frac{1}{\pi}
\inC\frac{e^{-y}\hat{\cal F}(y)}
{x+y}dy=\frac{1+va_1+\cdots}{1+\frac{vx}{2\xi_o}}=1+v\left(a_1-\frac{x}{2\xi_o}
\right)+\cdots     
\end{equation}
The LO problem is solved by
\begin{equation}
\hat{\cal F}_0(x)=\psi_0(x),\qquad\quad f_0=\Omega_0=\frac{1}{4}  
\end{equation}
and the NLO order problem is 
\begin{equation}
\hat{\cal F}_1(x)+\frac{1}{\pi}\int_0^\infty \frac{e^{-y}\hat{\cal F}_1(y)}
{x+y}dy=a_1-\frac{x}{2\xi_0}+x\psi_0(x)L_1(\ln x),
\label{calF1}
\end{equation}
which is closely related to the NLO problem of $\psi(x)$ and the corresponding
moment is given by
\begin{equation}
f_1=\frac{a_1}{4}-\frac{9}{32}\left(1+\frac{1}{2\xi_o}\right)+\Omega_1.
\label{f1res}
\end{equation}
Finally the result for the boundary value of $\chi_1$ is 
\begin{equation}
\chi_1(B)=\frac{1}{2\sqrt{v\Delta}}\left\{\hat u_0-\frac{iH_1}{\xi_o}
\hat u_1\nu^{2\xi_o}+O(\nu^{4\xi_o})\right\},  
\end{equation}
where
\begin{equation}
\hat u_1=1+\left(r_1+\frac{1}{4\xi_o}\right)v+\left(a_2-\frac{f_1}{2\xi_o}
\right)v^2+O(v^3).  
\label{u1expansion}
\end{equation}

\section{The first two perturbative orders}
\label{appA}

In this appendix we present calculations which enable us to obtain the
first two terms in the perturbative expansion of the densities
$O_{cc}$, $O_{c1}$ and the boundary values $\chi_c(B)$, $\chi_1(B)$ analytically.
Originally these perturbative coefficients were determined numerically with high
precision \cite{Hasenfratz:1990zz,Hasenfratz:1990ab}. Later the numerical
calculations were confirmed also analytically \cite{Balog:1992cm}, although the
details of the calculations were not published. Recently, analytic calculations
based on the solution of the Airy kernel problem were published
\cite{Marino:2021dzn}. Here we reproduce the same results with an entirely
different method based on the Laplace transform of the Airy kernel. This method
was invented by the authors of \cite{Balog:1992cm}.

\subsection{Leading and subleading orders for the
$\chi_c$ and $\chi_1$ problems}

First we will explicitly solve the $\chi_c$ and $\chi_1$ problems
at leading (LO)
and subleading (NLO) order, the integral equations (\ref{psi0}) and
(\ref{psi1}), respectively. We recall that the linear function $L_1(u)$ on the
right hand side of the NLO equation is
\begin{equation}
L_1(u)=\gamma_1 u+\alpha_1-L,\qquad \alpha_1=-2r_o+\gamma_1(\gamma_E-1+2\ln2).
\end{equation}
We will divide the NLO problem into three partial NLO problems. First we write
\begin{equation}
\psi_1(x)=\psi^{(a)}_1(x)+(\alpha_1-L)\psi^{(b)}_1(x)+\gamma_1\psi^{(c)}_1(x),
\end{equation}
where $\psi_1^{(i)}(x)$, $i=a,b,c$, satisfy the integral equations
\begin{equation}
\psi_1^{(i)}(x)+\frac{1}{\pi}\int_0^\infty\frac{e^{-y}\psi_1^{(i)}(y)}{x+y}d y
=r_1^{(i)}(x),  
\label{psi1i}
\end{equation}
\begin{equation}
r_1^{(a)}(x)=x,\qquad r_1^{(b)}(x)=x\psi_0(x),\qquad r_1^{(c)}(x)=x\psi_0(x)\ln x.
\end{equation}
Analogously to (\ref{Omegaks}) we define the partial moments
\begin{equation}
\Omega^{(i)}_{1,r}=\frac{1}{\pi}\int_0^\infty e^{-x}x^r\psi^{(i)}_1(x)dx,\qquad\quad
\Omega^{(i)}_1=\Omega^{(i)}_{1,0},\qquad i=a,b,c.  
\label{Omega1i}
\end{equation}
The total $\Omega_{1,r}$ moments are given by
\begin{equation}
\Omega_{1,r}=\Omega^{(a)}_{1,r}+(\alpha_1-L)\Omega^{(b)}_{1,r}+
\gamma_1\Omega^{(c)}_{1,r}.
\end{equation}


For the $\chi_1$ case the LO and NLO equations are (\ref{g0}) and (\ref{g1}),
respectively. We again divide the NLO problem into parts:
\begin{equation}
g_1(x)=(\alpha_1-L)g_1^{(a)}(x)+\gamma_1 g_1^{(b)}(x),
\end{equation}
where $g_1^{(i)}(x)$, $i=a,b$, satisfy the integral equations
\begin{equation}
g_1^{(i)}(x)+\frac{1}{\pi}\int_0^\infty\frac{e^{-y} g_1^{(i)}(y)}{x+y}d y
=\tilde r_1^{(i)}(x),  
\label{g1i}
\end{equation}
\begin{equation}
\tilde r_1^{(a)}(x)=xg_0(x)-\frac{1}{\sqrt{x}},\qquad
\tilde r_1^{(b)}(x)=\tilde r_1^{(a)}(x)\ln x.
\end{equation}
The corresponding moments are
\begin{equation}
\bar\gamma_{1,r}=(\alpha_1-L)\bar\gamma^{(a)}_{1,r}+
\gamma_1\bar\gamma^{(b)}_{1,r},
\end{equation}
where
\begin{equation}
\bar\gamma^{(i)}_{1,r}=\frac{1}{\pi}\int_0^\infty e^{-x}x^r g^{(i)}_1(x)dx,
\qquad\quad \bar\gamma^{(i)}_1=\bar\gamma^{(i)}_{1,0},\qquad i=a,b.  
\label{gamma1i}
\end{equation}

We will need the leading small $x$ expansion of $g_0(x)$ and $g_1(x)$.
For this purpose we will use the following fact. Let us assume that the small
$x$ behaviour of a function $\phi(x)$ is
\begin{equation}
\phi(x)=\frac{1}{\sqrt{x}}G(\ln x)+H(x),
\end{equation}
where 
\begin{equation}
H(x)=O(\sqrt{x})
\end{equation}
and $G(u)$ is a polynomial of degree $D$. In this case
\begin{equation}
\begin{split}  
\frac{1}{\pi}\int_0^\infty\frac{e^{-y}\phi(y)}{x+y}dy
&=\frac{1}{\pi}\int_0^\infty\frac{G(\ln y)}{(x+y)\sqrt{y}}dy +O(1)\\ 
&=\frac{1}{\sqrt{x}}\sum_{k=0}^D \frac{1}{k!}G^{(k)}(\ln x)
\frac{1}{\pi}\int_0^\infty\frac{\ln^k u}{(1+u)\sqrt{u}}du +O(1). 
\end{split}
\end{equation}
From the above it follows that if the leading term of the polynomial is
$G(u)=G_ou^D+\dots$ then
\begin{equation}
\frac{1}{\pi}\int_0^\infty\frac{e^{-y}\phi(y)}{x+y}dy
=\frac{1}{\sqrt{x}}\hat G(\ln x)+O(\sqrt{x}),
\end{equation}
where $\hat G(u)=G_ou^D+\dots$ is also a polynomial of degree $D$ with the
same leading term. Since on the right hand side of (\ref{g0})
\begin{equation}
\frac{e^{-y}-1}{y\sqrt{y}}=-\frac{1}{\sqrt{y}}+O(\sqrt{y})
\end{equation}
we see that
\begin{equation}
g_0(x)=-\frac{1}{2\sqrt{x}}+O(\sqrt{x}).
\label{g0asy}
\end{equation}
From this it follows that the right hand side of (\ref{g1}) is
\begin{equation}
-\frac{1}{\sqrt{x}} L_1(\ln x)+O(\sqrt{x})  
\end{equation}
hence
\begin{equation}
g_1(x)=-\frac{1}{2\sqrt{x}} L_1(\ln x) +O(\sqrt{x}).
\label{g1asy}
\end{equation}

\subsection{Laplace transformation}

Our integral equations are generically of the form
\begin{equation}
\psi(x)+\frac{1}{\pi}\int_0^\infty\frac{e^{-y}\psi(y)}{x+y}dy=r(x).
\label{ai}
\end{equation}
It is easier to solve the problem for the Laplace transformed unknown function
\begin{equation}
{\cal F}(s)=\int_0^\infty e^{-(s+1)x} \psi(x)dx.  
\end{equation}
After Laplace transformation (\ref{ai}) takes the form
\begin{equation}
(1+K){\cal F}=R,  
\end{equation}
where the Laplace transformed integral operator is
\begin{equation}
K[{\cal F}](s)=\frac{1}{\pi}\int_0^\infty\frac{{\cal F}(p)}{s+p+1}dp  
\end{equation}
and
\begin{equation}
R(s)=\int_0^\infty e^{-(s+1)x} r(x)dx.  
\end{equation}
In this language the calculation of moments is also easier since
\begin{equation}
M_r=\frac{1}{\pi}\int_0^\infty e^{-x}x^r\psi(x)dx=\frac{1}{\pi}\left(
-\frac{d}{ds}\right)^r{\cal F}(s)\Bigg\vert_{s=0}.  
\end{equation}
We also note that if the leading small $x$ expansion of $\psi(x)$ is of the form
\begin{equation}
\psi(x)=\frac{A+B\ln x}{\sqrt{x}}+O(\sqrt{x})
\end{equation}
then the leading large $s$ behaviour of its Laplace transform becomes
\begin{equation}
{\cal F}(s)=\frac{\sqrt{\pi}}{\sqrt{s+1}}\left\{A-B[\ln(s+1)+\gamma_E+2\ln2]
\right\}+O\left(\frac{1}{\sqrt{s+1}^3}\right).  
\label{asy1}
\end{equation}
Similarly, for $\psi(x)$ behaving like
\begin{equation}
\psi(x)=\sqrt{x}[A+B\ln x]+O(\sqrt{x}^3)
\end{equation}
the asymptotics of its Laplace transform is
\begin{equation}
{\cal F}(s)=\frac{\sqrt{\pi}}{2\sqrt{s+1}^3}\left\{A-
B[\ln(s+1)+\gamma_E+2\ln2-2]
\right\}+O\left(\frac{1}{\sqrt{s+1}^5}\right).  
\label{asy2}
\end{equation}

\subsection{The Laplace transformed problems}

The Laplace transformed $\psi(x)$ function of the $\chi_c$ problem will be
denoted by ${\cal F}(s)$ and we similarly define the Laplace space
functions ${\cal F}_0$, ${\cal F}_1$, ${\cal F}^{(i)}_1$, corresponding to
(respectively) $\psi_0$, $\psi_1$, $\psi^{(i)}_1$. For later convenience we
will also use the ad hoc notation
\begin{equation}
{\cal F}_0(s)=U(s).  
\end{equation}
Using this notation the LO $\chi_c$ problem is
\begin{equation}
(1+K)U=\frac{1}{s+1}  
\end{equation}
and the three partial NLO problems are\footnote{In the third equation we used
the representation $\ln x=\int_0^\infty\frac{dp}{p}[e^{-p}-e^{-px}]$.}
\begin{equation}
\begin{split}
(1+K){\cal F}^{(a)}_1&=\frac{1}{(s+1)^2},\\
(1+K){\cal F}^{(b)}_1&=-U^\prime,\\
(1+K){\cal F}^{(c)}_1&=\int_0^\infty\frac{dp}{p}[U^\prime(p+s)-e^{-p}U^\prime(s)].
\end{split}    
\end{equation}
The small $x$ behaviour of $\psi(x)$, (\ref{psiasy}), is translated to
\begin{equation}
{\cal F}(s)=\frac{\ln(s+1)\hat\rho_o+{\rm const.}}{2\sqrt{s+1}^3}+
O\left(\frac{1}{\sqrt{s+1}^5}\right).
\label{asyF}
\end{equation}


For the $\chi_1$ problem the Laplace space functions are
${\cal G}_0$, ${\cal G}_1$,
${\cal G}^{(i)}_1$, corresponding to (repectively) $g_0$, $g_1$, $g^{(i)}_1$.
Again, it will turn out to be convenient to introduce the ad hoc notations
\begin{equation}
\rho(s)=\sqrt{s+1}-\sqrt{s},\qquad {\cal G}_0(s)=2\sqrt{\pi}(Y(s)-\sqrt{s+1}),
\qquad Q(s)=Y(s)-\sqrt{s},  
\end{equation}
and
\begin{equation}
{\cal G}_1^{(a)}(s)=2\sqrt{\pi}Z(s),\qquad\quad
{\cal G}_1^{(b)}(s)=2\sqrt{\pi}H(s).
\end{equation}
We start by calculating the right hand side of the Laplace transformed LO
equation (\ref{g0}):
\begin{equation}
\int_0^\infty dx e^{-(s+1)x}\frac{1}{\pi}\int_0^\infty\frac{e^{-y}-1}
{(x+y)y\sqrt{y}}dy=\frac{2}{\sqrt{\pi}}\int_0^\infty\frac{dp}{p+s+1}
(\sqrt{p}-\sqrt{p+1}).    
\end{equation}
It is easy to see that in this notation the LO problem becomes
\begin{equation}
(1+K)Q=\rho
\label{LOchi1}
\end{equation}
and the two partial NLO problems are
\begin{equation}
\begin{split}
(1+K)Z&=-Y^\prime,\\
(1+K)H&=\int_0^\infty\frac{dp}{p}[Y^\prime(p+s)-e^{-p}Y^\prime(s)].
\label{NLOchi1}
\end{split}    
\end{equation}

\subsection{Exchange relations}
\label{Xchange}

For the solution of the integral equations we will employ the following
identities, which are easily obtained by partial integration.
\begin{equation}
(K{\cal F})^\prime=-K[{\cal F}^\prime]-\frac{1}{\pi}\frac{{\cal F}(0)}{s+1},  
\label{ex1}
\end{equation}
\begin{equation}
(K{\cal F})^{\prime\prime}=K[{\cal F}^{\prime\prime}]
+\frac{1}{\pi}\frac{{\cal F}^\prime(0)}{s+1}  
+\frac{1}{\pi}\frac{{\cal F}(0)}{(s+1)^2},  
\label{ex2}
\end{equation}
\begin{equation}
K[s(s+1){\cal F}^{\prime\prime}]=s(s+1)(K{\cal F})^{\prime\prime}
+\frac{1}{\pi}\frac{{\cal F}(0)}{s+1},
\label{ex3}
\end{equation}
\begin{equation}
K[(2s+1){\cal F}^\prime]=(2s+1)(K{\cal F})^\prime
-\frac{1}{\pi}\frac{{\cal F}(0)}{s+1}.
\label{ex4}
\end{equation}
Introducing the second order differential operator
\begin{equation}
{\cal D}_a=s(s+1)\frac{d^2}{ds^2}+a(2s+1)\frac{d}{ds}  
\end{equation}
we can establish the exchange relation
\begin{equation}
K[{\cal D}_a{\cal F}]={\cal D}_a(K{\cal F})+\frac{(1-a){\cal F}(0)}{s+1}.  
\end{equation}

\subsection{Solution of the LO problems}

We start from the identity 
\begin{equation}
\hat{\cal D}_1\rho=-\frac{1}{4}(1+K)\left[\frac{1}{\sqrt{s}}\right],  
\end{equation}
where
\begin{equation}
\hat{\cal D}_1={\cal D}_1-\frac{3}{4}.
\end{equation}
The operator $\hat{\cal D}_1$ commutes with the integral operator $K$.

Applying $\hat{\cal D}_1$ to the LO equation we get
\begin{equation}
\hat{\cal D}_1(1+K)Q=(1+K)\hat{\cal D}_1Q
=-\frac{1}{4}(1+K)\left[\frac{1}{\sqrt{s}}\right],  
\end{equation}
which is equivalent to
\begin{equation}
\hat{\cal D}_1Q+\frac{1}{4\sqrt{s}}=\hat{\cal D}_1(Q+\sqrt{s})
=\hat{\cal D}_1Y=0.  
\end{equation}
The last differential equation is a slightly modified hypergeometric equation
with parameters $-1/2,3/2;1$ and argument $-s$. Since $Y(s)$ is regular at
$s=0$, $Y$ must be proportional to the hypergeometric function:
\begin{equation}
Y(s)=Y(0){}_2F_1\left(-\frac{1}{2},\frac{3}{2};1;-s\right).  
\end{equation}
From the known properties of the hypergeometric function we establish that for
large $s$
\begin{equation}
Y(s)=\frac{4Y(0)}{\pi}\sqrt{s+1}+O\left(\frac{1}{\sqrt{s+1}}\right).
\end{equation}
Since ${\cal G}_0(s)$ must vanish for large $s$, we conclude that $Y(0)=\pi/4$
and
\begin{equation}
Y(s)=\frac{\pi}{4}{}_2F_1\left(-\frac{1}{2},\frac{3}{2};1;-s\right).  
\end{equation}
For the leading moment we thus find
\begin{equation}
\bar\gamma_0=\frac{1}{\pi}{\cal G}_0(0)
=\frac{\sqrt{\pi}}{2}-\frac{2}{\sqrt{\pi}}.  
\end{equation}


The solution of the LO $\chi_c$ problem is completely analogous. Here we start
from
\begin{equation}
{\cal D}_2(1+K)U=(1+K)[{\cal D}_2U]+\frac{U(0)}{\pi(s+1)}=\frac{-2}{s+1},
\end{equation}
from which it follows that
\begin{equation}
(1+K)\left\{{\cal D}_2 U+\left[\frac{U(0)}{\pi}+2\right]U\right\}=0.  
\end{equation}
We again encounter a hypergeometric differential equation. In this case
the solution can be written in terms of a so far unknown parameter $\alpha$ as
\begin{equation}
U(s)=\pi\left(\frac{1}{4}-\alpha^2\right)
{}_2F_1\left(\frac{3}{2}-\alpha,\frac{3}{2}+\alpha;2;-s\right).  
\end{equation}
The asymptotic behaviour of the hypergeometric function implies that $U(s)$
for large $s$ behaves as
\begin{equation}
U(s)  \sim \frac{1}{(s+1)^{3/2-\vert\alpha\vert}}.
\end{equation}
On the other hand we know from (\ref{asy2}) that (up to logs)
\begin{equation}
U(s)  \sim \frac{1}{\sqrt{s+1}^3}.
\end{equation}
This fixes $\alpha=0$ and the final result is
\begin{equation}
U(s)=\frac{\pi}{4}{}_2F_1\left(\frac{3}{2},\frac{3}{2};2;-s\right).  
\end{equation}
The actual large $s$ behaviour of $U(s)$ is
\begin{equation}
U(s)=\frac{\ln(s+1)+4\ln2-2}{2\sqrt{s+1}^3}+
O\left(\frac{1}{\sqrt{s+1}^5}\right),
\end{equation}
which is consistent with (\ref{asyF}) at LO.

The leading moment here is given by
\begin{equation}
\Omega_0=\frac{1}{\pi}U(0)=\frac{1}{4}  
\end{equation}
and for later purposes we also compute
\begin{equation}
\Omega_{0,1}=-\frac{1}{\pi}U^\prime(0)=\frac{9}{32},\qquad\quad
\Omega_{0,2}=\frac{1}{\pi}U^{\prime\prime}(0)=\frac{75}{128}.
\end{equation}

\subsection{Solution of the NLO $\chi_1$ problem}

By taking the derivative of the LO equation (\ref{LOchi1}) and using (\ref{ex1})
we obtain
\begin{equation}
K[Y^\prime]=Y^\prime-\frac{1}{4(s+1)},
\end{equation}
which can be rewritten as
\begin{equation}
(1+K)[Y^\prime]=2Y^\prime-\frac{1}{4}(1+K)U.  
\end{equation}
Comparing this to the first equation in (\ref{NLOchi1}) we immediately see that
\begin{equation}
Z=-\frac{1}{2}\left(Y^\prime+\frac{U}{4}\right).  
\label{Zsol}
\end{equation}
To solve the second NLO equation we introduce
\begin{equation}
\hat{\cal D}_2={\cal D}_2+\frac{5}{4},\qquad\qquad \hat{\cal D}_2 Y^\prime=0.  
\end{equation}
Acting with $\hat{\cal D}_2$ on the second NLO equation we get
\begin{equation}
\hat{\cal D}_2(1+K)H=(1+K)\left\{\hat{\cal D}_2 H+\frac{H(0)}{\pi}U\right\}
=\int_0^\infty\frac{dp}{p}\hat{\cal D}_2 Y^\prime(p+s)=(2s+1)Y^{\prime\prime}+
3Y^\prime.
\end{equation}
Now it is easy to show that
\begin{equation}
(1+K)\left\{sY^{\prime\prime}-\frac{U}{4}\right\}=(2s+1)Y^{\prime\prime}  
\end{equation}
and this, together with (\ref{Zsol}) gives
\begin{equation}
\hat{\cal D}_2\left\{H+\left[\frac{1}{8}-\frac{H(0)}{\pi}\right]U\right\}=  
sY^{\prime\prime}+\frac{3}{2}Y^\prime=\frac{1}{2}\hat{\cal D}_2\left\{
\ln(s+1) Y^\prime+\frac{Y}{s+1}\right\}.
\end{equation}
Thus the solution for $H$ is of the form
\begin{equation}
H=\left[\frac{H(0)}{\pi}-\frac{1}{8}\right]U+\frac{1}{2}\ln(s+1)Y^\prime
+\frac{Y}{2(s+1)}+\xi Y^\prime,  
\end{equation}
where we have added the regular (at $s=0$) solution of the homogeneous equation
with a so far undetermined coefficient $\xi$. Self-consistency requires
\begin{equation}
H(0)=\frac{\pi}{8}+\frac{\pi\xi}{4}  
\end{equation}
and finally we get
\begin{equation}
H=\xi\left(Y^\prime+\frac{U}{4}\right)+\frac{1}{2}\ln(s+1)Y^\prime
+\frac{Y}{2(s+1)}.  
\end{equation}
The constant $\xi$ can be fixed as follows. Asymptotically
\begin{equation}
Z=-\frac{1}{4}\frac{1}{\sqrt{s+1}}+O\left(\frac{1}{\sqrt{s+1}^3}\right),\qquad  
H=\frac{1}{4}\frac{1}{\sqrt{s+1}}\left\{\ln(s+1)+2\xi+2\right\}
+O\left(\frac{1}{\sqrt{s+1}^3}\right)  
\end{equation}
and comparing this to (\ref{g1asy}) and (\ref{asy1}) we can fix
\begin{equation}
\xi=\frac{\gamma_E}{2}+\ln2-1.  
\end{equation}
This gives
\begin{equation}
\bar\gamma_1=\frac{\sqrt{\pi}}{4}\left\{L-\alpha_1
+\gamma_1(\gamma_E+2\ln2-1)\right\}=\frac{\sqrt{\pi}}{4}(L+2r_o)
\end{equation}
and using the definition (\ref{rk}) we reproduce (\ref{r1}).
Further we calculate
\begin{equation}
\bar\gamma_{0,1}=2\sqrt{\pi}\left(-Y^\prime(0)+\frac{1}{2}\right)\frac{1}{\pi}=  
\frac{1}{\sqrt{\pi}}-\frac{3\sqrt{\pi}}{8},\qquad\quad I^\prime(1)=-\sqrt{\pi}
\end{equation}
and find
\begin{equation}
R_1=\frac{2}{\sqrt{\pi}}\left\{\bar\gamma_1+\bar\gamma_{0,1}+\frac{1}{\pi}
I^\prime(1)\right\}=r_1-\frac{3}{4},\qquad
\tilde R_1=\frac{2}{\sqrt{\pi}}\left\{\bar\gamma_1-\bar\gamma_{0,1}
-\frac{1}{\pi}I^\prime(1)\right\}=r_1+\frac{3}{4}.
\end{equation}
For later use we also compute
\begin{equation}
\bar\gamma_{1,1}=\frac{2}{\sqrt{\pi}}\left\{(L-\alpha_1)Z^\prime(0)-\gamma_1  
H^\prime(0)\right\}=\frac{\sqrt{\pi}}{2}\left(\frac{3}{4}r_1-\frac{5}{8}
\gamma_1\right),
\end{equation}
\begin{equation}
\bar\gamma_{0,2}=\frac{2}{\sqrt{\pi}}\left(Y^{\prime\prime}(0)+\frac{1}{4}\right)
=-\frac{15}{64}\sqrt{\pi}+\frac{1}{2\sqrt{\pi}},\qquad
\bar\gamma_{0,2}-\frac{1}{\pi}I^{\prime\prime}(1)=-\frac{15\sqrt{\pi}}{64}.
\end{equation}

\subsection{Solution of the NLO $\chi_c$ problem}

Here we proceed analogously to the NLO $\chi_1$ case. Using the identities
of subsection \ref{Xchange} we easily find the results
\begin{equation}
{\cal F}_1^{(a)}=(2s+1)U^\prime+\frac{9}{4}U,\qquad\quad
{\cal F}_1^{(b)}=sU^\prime+U.
\end{equation}
To determine ${\cal F}_1^{(c)}$ we introduce
\begin{equation}
\hat{\cal D}_3={\cal D}_3+\frac{25}{4},\qquad\qquad \hat{\cal D}_3 U^\prime=0.  
\end{equation}
Acting with $\hat{\cal D}_3$ on the ${\cal F}_1^{(c)}$ equation we have
\begin{equation}
\hat{\cal D}_3\left\{(1+K){\cal F}_1^{(c)}\right\}=(1+K)\left\{  
\hat{\cal D}_3{\cal F}_1^{(c)}+\frac{2}{\pi}{\cal F}_1^{(c)}(0) U\right\}
=\int_0^\infty\frac{dp}{p}\hat{\cal D}_3 U^\prime(p+s)=(2s+1)U^{\prime\prime}
+5U^\prime.
\end{equation}
Using
\begin{equation}
(1+K)\left\{sU^{\prime\prime}-\frac{U}{4}+2{\cal F}_1^{(a)}\right\}
=(2s+1)U^{\prime\prime}
\end{equation}
we get
\begin{equation}
\hat{\cal D}_3{\cal F}_1^{(c)}=sU^{\prime\prime}+(2-s)U^\prime-\left[
\frac{3}{4}+\frac{2}{\pi}{\cal F}_1^{(c)}(0)\right]U.
\end{equation}
The solution of this differential equation is
\begin{equation}
{\cal F}_1^{(c)}=\frac{s}{2}U^\prime+\frac{\ln(s+1)}{2}U^\prime+(\gamma-3)
(U+sU^\prime)+\left[\frac{5}{s+1}-\frac{2}{(s+1)^2}\right]U
-\frac{2}{s+1}U^\prime+\eta U^\prime.
\end{equation}
Here we introduced the notation
\begin{equation}
\gamma=\frac{4}{\pi}{\cal F}_1^{(c)}(0)  
\end{equation}
and added the regular solution of the homogeneous equation with coefficient
$\eta$. Self-consistency requires $\eta=2$ and so we get
\begin{equation}
{\cal F}_1^{(c)}=\frac{s}{2}U^\prime+\frac{\ln(s+1)}{2}U^\prime+(\gamma-3)
(U+sU^\prime)+\left[\frac{5}{s+1}-\frac{2}{(s+1)^2}\right]U
+\frac{2s}{s+1}U^\prime.
\end{equation}
Next we study the asymptotics of the partial solutions ${\cal F}_1^{(i)}$. We
define the coefficients $\lambda^{(i)}$ by
\begin{equation}
{\cal F}_1^{(i)}(s)=\lambda^{(i)}\frac{\ln(s+1)+{\rm const.}}{2\sqrt{s+1}^3}
+O\left(\frac{1}{\sqrt{s+1}^5}\right),\qquad i=a,b,c.  
\end{equation}
We find
\begin{equation}
\lambda^{(a)}=-\frac{3}{4},\qquad \lambda^{(b)}=-\frac{1}{2},\qquad
\lambda^{(c)}=\frac{3}{4}-\frac{\gamma}{2}
\end{equation}
and so the total coefficient $\lambda$ is
\begin{equation}
\lambda=\lambda^{(a)}+(\alpha_1-L)\lambda^{(b)}+\gamma_1\lambda^{(c)}=  
R_1+\gamma_1\left[\frac{5}{4}-\frac{\gamma}{2}-\frac{\gamma_E}{2}-\ln2\right].
\end{equation}
From here we obtain
\begin{equation}
\gamma=\frac{5}{2}-\gamma_E-2\ln2.  
\end{equation}
We can now calculate the moments
\begin{equation}
\Omega_1=\frac{1}{\pi}\left\{{\cal F}_1^{(a)}(0)  
+(\alpha_1-L){\cal F}_1^{(b)}(0)+\gamma_1{\cal F}_1^{(c)}(0)\right\}=  
-\frac{3}{32}+\frac{3\Delta}{4}-\frac{r_o}{2}-\frac{L}{4},
\end{equation}
\begin{equation}
\overline{\Omega}_1=\Omega_1+\Omega_{0,1}=\frac{3}{16}+\frac{3\Delta}{4}
-\frac{r_o}{2}-\frac{L}{4}.
\end{equation}

\subsection{Calculation of $Z_{2x}$ and $a_1$, $a_2$, $f_1$}

We have all ingredients to calculate $Z_{2x}$ defined by (\ref{Zmx}-\ref{k1}):
\begin{equation}
\Sigma_{1,1}-\Omega_{1,1}=-2\Omega^{(a)}_{1,1}=\frac{2}{\pi}{\cal F}_1^{(a)\prime}  
(0)=-\frac{39}{32},
\end{equation}
\begin{equation}
Z_{2x}=\Sigma_{1,1}-\Omega_{1,1}+2\Omega_{0,2}=-\frac{3}{64}.
\end{equation}

$y_{1,0}$ is defined by (\ref{y10}) and can be expanded as
\begin{equation}
\begin{split}
y_{1,0}&=\frac{v}{\pi}\inC\frac{e^{-y}\omega(y)}{2\xi_o+vy}dy
-\frac{1}{\pi}\inC\Bigg[e^{-y}\frac{\omega(y)}{y}+\frac{1}{y\sqrt{y}}\Bigg]dy\\
&=\frac{v}{\pi}\frac{1}{2\xi_o}\int_0^\infty e^{-y}\Bigg[yg_0(y)
-\frac{1}{\sqrt{y}}\Bigg]dy
-\frac{v^2}{\pi}\frac{1}{(2\xi_o)^2}\int_0^\infty e^{-y}\left[y^2g_0(y)
-\sqrt{y}\right]dy\\
&+\frac{v^2}{\pi}\frac{1}{2\xi_o}\int_0^\infty e^{-y}yg_1(y)dy
-\frac{1}{\pi}\int_0^\infty \Bigg[e^{-y}g_0(y)+\frac{1-e^{-y}}{y\sqrt{y}}
\Bigg]dy\\
&-\frac{v}{\pi}\int_0^\infty e^{-y}g_1(y)dy
-\frac{v^2}{\pi}\int_0^\infty e^{-y}g_2(y)dy+O(v^3).
\end{split}  
\end{equation}
All integrals can be expressed in terms of moments already defined and we obtain
\begin{equation}
\begin{split}
y_{1,0}&=-\bar\gamma_0-\frac{2}{\sqrt{\pi}}-v\left[\bar\gamma_1-\frac{1}{2\xi_o}
\left(\bar\gamma_{0,1}-\frac{1}{\sqrt{\pi}}\right)\right]\\
&-v^2\left[\bar\gamma_2-\frac{1}{2\xi_o}\bar\gamma_{1,1}+\frac{1}{(2\xi_o)^2}
\left(\bar\gamma_{0,2}-\frac{1}{2\sqrt{\pi}}\right)\right]+O(v^3).
\end{split}    
\end{equation}
The first two expansion coefficients defined by (\ref{y10exp}) are
\begin{equation}
\begin{split}  
a_1&=\frac{2}{\sqrt{\pi}}\left[\bar\gamma_1
-\frac{1}{2\xi_o}\left(\bar\gamma_{0,1}-\frac{1}{\sqrt{\pi}}\right)\right]=
r_1+\frac{3}{8\xi_o},\\    
a_2&=\frac{2}{\sqrt{\pi}}\left[\bar\gamma_2
-\frac{1}{2\xi_o}\bar\gamma_{1,1}+\frac{1}{(2\xi_o)^2}\left(\bar\gamma_{0,2}
-\frac{1}{2\sqrt{\pi}}\right)\right]=
r_2+\frac{1}{2\xi_o}\left(\frac{5}{8}\gamma_1-\frac{3}{4}r_1\right)
-\frac{15}{128\xi_o^2}.
\end{split}
\end{equation}

The function $\hat{\cal F}_1$ is the solution of (\ref{calF1}) and can be
written as
\begin{equation}
\hat{\cal F}_1(x)=a_1\psi_0(x)-\left(1+\frac{1}{2\xi_o}\right)\psi_1^{(a)}(x)  
+\psi_1(x)
\end{equation}
and so its moment is given by (\ref{f1res}).

\section{Volin's $O(3)$ algorithm for general running coupling $v$}
\label{app:Volin}

In this section we briefly sketch how to calculate $\epsilon$ and $\rho$ in terms of $v$ (in particular in terms of
the special coupling $\beta$ (\ref{eq:specialL})) from the result of the algorithm introduced in Appendix E.3 of \cite{Volin:2010cq}.

Since from the ratio $\epsilon/\rho^{2}$ all $\ln B$-s drop out,
this algorithm generates the densities themselves without such terms,
as a plain power series in \textbf{$B^{-1}$}. As explained in \cite{Volin:2010cq}, this is done by formally treating $\ln B$
as a variable independent of $B^{-1}$, both inside the above quantities
and in the definition of the $\alpha$ running coupling:
\begin{equation}
	\frac{1}{\alpha}=B+\frac{1}{2}\ln B+2\ln2-\frac{1}{2}+\ln\hat{U}(B),\quad\hat{U}(B)\equiv\frac{4\pi e^{-\left(B+1/2\right)}}{\sqrt{B}}O_{1 c}.\label{eq:volinalpha}
\end{equation}
As $\ln B$ drops out from the final result $\epsilon/\rho^{2}$,
it was admissible to fix it to any value, $\ln B=\ln B_{0}$. Here
$B_{0}$ is arbitrary, and it was chosen to be $B_{0}=e^{-1}$ for
the $O(3)$ model to simplify the algorithm. As we need both $\epsilon$
and $\rho$ to calculate the capacitance $C$, we must either restore
their $\ln B$ dependence or try to reuse the data generated in the
above conventions. The idea is that since the densities $W_{0}$ and
$\hat{\rho}_{0}$ are also a power series in the running coupling
$v$, we might fix $\ln v$ (as an independent variable) to a constant
as well, which must also correspond to the above $\ln B=\ln e^{-1}=-1$
choice. Taking the logarithm of (\ref{eq:runningcouplingwithL}) gives
us 
\begin{equation}
\ln v + \ln(2B) =\ln\left(1+v(\ln v + L)\right),
\end{equation}
thus if we fix $\ln(B)=-1$, $L=\ln\left(\frac{2}{e}\right)$, and
$\ln v =-L$ simultaneously, we can satisfy the above equation. This
basically means, we may simply relate the $B^{-1}$ expansion coming
from the algorithm to our $v$ expansion by 
\begin{align*}
2B & =\frac{1}{v},
\end{align*}
that is, what remains from (\ref{eq:runningcouplingwithL}). What
we get is the result in terms of a specific running coupling - let us denote this particular choice of the coupling as $\gamma$: 
\begin{equation}
2B=\frac{1}{\gamma}+\ln \gamma+\ln\left(\frac{2}{e}\right).\label{eq:technicalcoupling}
\end{equation}
Now the problem is that in the original algorithm we have also got
rid of all the $\ln 2$-s. However in our special running coupling
(\ref{eq:specialL}) the quantities are also free of $\ln 2$. It is
then admissible to switch from (\ref{eq:technicalcoupling}) to
(\ref{eq:specialL}) :
\begin{equation}
2B=\frac{1}{\gamma}+\ln \gamma+\ln 2 -1\quad\Rightarrow\quad2B=\frac{1}{\beta}+\ln \beta- 3\ln 2+1,
\label{eq:vcouplingchange}
\end{equation}
while putting $\ln 2\to 0$ everywhere in the above equations, i.e. omitting the $\ln 2$  term in the definition of $\gamma$, and the $-3\ln 2$ term in that of $\beta$. In the end, we may switch from the $\beta$ coupling to any $v$ coupling with arbitrary $L$ value (while of course keeping the $\ln 2$-s in both the definition of $\beta$ and the generic $v$).

To see an example, let us show this procedure step-by-step for $W_0$ in (\ref{W0pert}). In the upper line, we start from the (incomplete) log-free result given by the algorithm in $B$ and substitute $B^{-1} \to 2\gamma$ simply:
\begin{equation}
	\begin{array}{lll}
		 1 + \frac{1}{4}B^{-1} + \frac{19}{32}B^{-2}+\mathcal{O}(B^{-3})  & & 1+ \frac{1}{2} \gamma + \frac{19}{8}\gamma^2+\mathcal{O}(\gamma^{3})    \\ 
		 &\;\Rightarrow \;&  \\
		 1 + \frac{1}{4}B^{-1}+ \left[\frac{15}{32} -\frac{1}{8}\ln(16 B)\right] B^{-2}+\mathcal{O}(B^{-3}) & & 1+ \frac{1}{2} \gamma + \left(\frac{19}{8} -2 \ln 2\right)\gamma^2 +\mathcal{O}(\gamma^3),   \\
	\end{array} 		
\end{equation}
while in the line underneath we start from the correct result and use (\ref{eq:technicalcoupling}). Then, for the log-free case, we may replace $ \gamma^{-1}+\ln \gamma-1  \to \beta^{-1}+\ln \beta +1 $, while for the exact case we must use (\ref{eq:vcouplingchange}) without dropping $\ln2$-s. One can achieve the change of couplings by calculating the series expansion of $\gamma$ in terms of $\beta$ iteratively in both cases:
\begin{equation}
	\begin{array}{rlll}
		\gamma(\beta)\vert_\text{no logs} & = \beta -2 \beta^2+\mathcal{O}(\beta^3) & & \\
		&&\;\Rightarrow\;&  W_0 = 1 + \frac{1}{2}\beta +\frac{11}{8}\beta^2+ \mathcal{O}(\beta^3),
\\
		\gamma(\beta) & = \beta - \left(2-4\ln 2\right)\beta^2 + \mathcal{O}(\beta^3)&&\\
	\end{array}
\end{equation}
and then realize that in the end, both methods give the same result in $\beta$.

To summarize, we can directly use the coefficients of the original algorithm, which drops $\ln B$ and $\ln 2$ from the start by a simple change of variables (\ref{eq:vcouplingchange}) - without $\ln(2)$-s, and get the perturbative coefficients of the densities in terms of any running coupling $v$.

\bibliographystyle{JHEP}
\bibliography{Running}

\end{document}